\documentstyle[12pt]{article}
\catcode`@=11
\def\@cite#1#2{$^{\scriptscriptstyle
\mbox{\rm\scriptsize#1\if@tempswa , #2\fi}}$}

\def\mite{\@ifnextchar [{\@tempswatrue\@mitex}{\@tempswafalse\@mitex[]}}

\def\@mitex[#1]#2{\if@filesw\immediate\write\@auxout{\string\citation{#2}}\fi
  \def\@mitea{}\@mite{\@for\@miteb:=#2\do
    {\@mitea\def\@mitea{,}\@ifundefined
       {b@\@miteb}{{\bf ?}\@warning
       {Red alert: `\@miteb' found on deck \thepage \space }}
\hbox{\csname b@\@miteb\endcsname}}}{#1}}

\def\@mite#1#2{$\mbox{\rm#1\if@tempswa , #2\fi}$}
\catcode`@=12

\renewcommand{\theequation}{\thesection.\arabic{equation}}

\def\onward{\addtocounter{section}{1} \setcounter{equation}{0} }

\def\tetra#1#2#3#4#5#6{\setlength{\unitlength}{1.0cm}
\begin{picture}(2,2.0)(1.8,2.5)
\put (2.4,2.5){$#1$}
\put (3.9,2.82){$#2$}
\put (3.5,0.77){$#3$}
\put (3.1,1.75){$#4$}
\put (3.5,4.1){$#5$}
\put (3.1,3.22){$#6$}
\end{picture}
{}~~~~~~~~~}

\def\cpstetra#1#2#3#4#5#6{\setlength{\unitlength}{1.0cm}
\begin{picture}(2,2.0)(1.8,2.5)
\put (2.4,2.5){$#1$}
\put (3.9,2.82){$#2$}
\put (3.5,0.77){$#3$}
\put (3.1,1.75){$#4$}
\put (3.5,4.1){$#5$}
\put (3.1,3.22){$#6$}
\put (3.6,2.2){${\sct \cK^{#2 #4}_{#6}}$}
\put (2.1,3.7){${\sct \cK^{#5 #6}_{#1}}$}
\put (2.1,1.4){${\sct \cK_{#3 #4}^{#1}}$}
\end{picture}
{}~~~~~~~~~}

\def\cprtstetra#1#2#3#4#5#6{\setlength{\unitlength}{1.0cm}
\begin{picture}(2,2.0)(1.8,2.5)
\put (2.4,2.5){$#1$}
\put (3.9,2.82){$#2$}
\put (3.5,0.77){$#3$}
\put (3.1,1.75){$#4$}
\put (3.5,4.1){$#5$}
\put (3.1,3.22){$#6$}
\put (3.6,2.2){${\sct \cK^{#4}_{#2 #6}}$}
\put (2.1,3.7){${\sct \cK^{#5 #6}_{#1}}$}
\put (2.1,1.4){${\sct \cK_{#3 #4}^{#1}}$}
\end{picture}
{}~~~~~~~~~}

\def\cpastetra#1#2#3#4#5#6{\setlength{\unitlength}{1.0cm}
\begin{picture}(2,2.0)(1.8,2.5)
\put (2.4,2.5){$#1$}
\put (3.9,2.82){$#2$}
\put (3.5,0.77){$#3$}
\put (3.1,1.75){$#4$}
\put (3.5,4.1){$#5$}
\put (3.1,3.22){$#6$}
\put (3.6,2.2){${\sct \cK^{#6 #2}_{#4}(i)}$}
\put (2.1,3.7){${\sct \cK^{#5}_{#1 #6}}$}
\put (2.1,1.4){${\sct \cK^{#1 #4}_{#3}}$}
\end{picture}
{}~~~~~~~~~}

\def\cpbstetra#1#2#3#4#5#6{\setlength{\unitlength}{1.0cm}
\begin{picture}(2,2.0)(1.8,2.5)
\put (2.4,2.5){$#1$}
\put (3.9,2.82){$#2$}
\put (3.5,0.77){$#3$}
\put (3.1,1.75){$#4$}
\put (3.5,4.1){$#5$}
\put (3.1,3.22){$#6$}
\put (3.6,2.2){${\sct \cK^{#6 #2}_{#4}(j)}$}
\put (2.1,3.7){${\sct \cK^{#1 #5}_{#6}}$}
\put (2.1,1.4){${\sct \cK^{#4}_{#1 #3}}$}
\end{picture}
{}~~~~~~~~~}

\def\cpcstetra#1#2#3#4#5#6{\setlength{\unitlength}{1.0cm}
\begin{picture}(2,2.0)(1.8,2.5)
\put (2.4,2.5){$#1$}
\put (3.9,2.82){$#2$}
\put (3.5,0.77){$#3$}
\put (3.1,1.75){$#4$}
\put (3.5,4.1){$#5$}
\put (3.1,3.22){$#6$}
\put (3.6,2.2){${\sct \cK^{#6}_{#4 #2}(i)}$}
\put (2.1,3.7){${\sct \cK^{#5}_{#1 #6}}$}
\put (2.1,1.4){${\sct \cK^{#1 #4}_{#3}}$}
\end{picture}
{}~~~~~~~~~}

\def\cpdstetra#1#2#3#4#5#6{\setlength{\unitlength}{1.0cm}
\begin{picture}(2,2.0)(1.8,2.5)
\put (2.4,2.5){$#1$}
\put (3.9,2.82){$#2$}
\put (3.5,0.77){$#3$}
\put (3.1,1.75){$#4$}
\put (3.5,4.1){$#5$}
\put (3.1,3.22){$#6$}
\put (3.6,2.2){${\sct \cK^{#6}_{#4 #2}(k)}$}
\put (2.1,3.7){${\sct \cK^{#1 #5}_{#6}}$}
\put (2.1,1.4){${\sct \cK^{#4}_{#1 #3}}$}
\end{picture}
{}~~~~~~~~~}

\def\sptetra#1#2#3#4#5#6{\setlength{\unitlength}{1.0cm}
\begin{picture}(2,2.0)(1.8,2.5)
\put (2.4,2.5){$#1$}
\put (3.9,2.82){$#2$}
\put (3.5,0.77){$#3$}
\put (3.1,1.75){$#4$}
\put (3.5,4.1){$#5$}
\put (3.1,3.22){$#6$}
\end{picture}
{}~~~~~~~~~}

\def\spotetra#1#2#3#4#5#6{\setlength{\unitlength}{1.0cm}
\begin{picture}(2,2.0)(1.8,2.5)
\put (2.4,2.5){$#1$}
\put (3.9,2.82){$#2$}
\put (3.5,0.77){$#3$}
\put (3.1,1.75){$#4$}
\put (3.5,4.1){$#5$}
\put (3.1,3.22){$#6$}
\end{picture}
{}~~~~~~~~~}

\def\revtetra#1#2#3#4#5#6{\setlength{\unitlength}{1.0cm}
\begin{picture}(2,2.0)(1.8,2.5)
\put (2.4,2.5){$#1$}
\put (3.9,2.82){$#2$}
\put (3.5,0.77){$#3$}
\put (3.1,1.75){$#4$}
\put (3.5,4.1){$#5$}
\put (3.1,3.22){$#6$}
\end{picture}
{}~~~~~~~~~}

\def\trefoil#1{\setlength{\unitlength}{1.0cm}
\begin{picture}(3.0,2.9)(1.8,2.5)
\put (2.0,2.5){$#1$}
\end{picture}
{}~~~~~~~~~~~~~~~~}

\def\figeight#1{\setlength{\unitlength}{1.0cm}
\begin{picture}(2.0,1.2)(1.8,2.5)
\put (2.2,2.5){$#1$}
\end{picture}
{}~~~~~~~~~~~~}

\def\mfigeight#1{\setlength{\unitlength}{1.0cm}
\begin{picture}(2.0,1.2)(1.8,2.5)
\put (2.2,2.5){$#1$}
\end{picture}
{}~~~~~~~~~~~~}

\def\yfancy#1#2#3{\setlength{\unitlength}{1.0cm}
\begin{picture}(1.5,1.0)(2.0,2.5)
\put (1.95,3.2){$#1$}
\put (3.65,3.2){$#2$}
\put (3.1,1.7){$#3$}
\end{picture}
{}~~~~}

\def\gbrtwis#1#2#3{\setlength{\unitlength}{1.0cm}
\begin{picture}(1.5,1.0)(2.0,2.5)
\put (2.1,3.2){$#1$}
\put (3.7,3.2){$#2$}
\put (3.2,1.7){$#3$}
\end{picture}
{}~~~~}

\def\fourlineq#1#2#3{\setlength{\unitlength}{1.0cm}
\begin{picture}(5.0,3.0)(2.0,2.5)
\put (2.5,2.5){$#1$}
\put (5.55,5.4){$#2$}
\put (3.8,4.65){$#3$}
\end{picture}
{}~~~~~~~~~~~}

\def\bilinkeq#1#2#3{\setlength{\unitlength}{1.0cm}
\begin{picture}(5.5,2.5)(1.8,2.5)
\put (2.4,2.5){$#3$}
\put (5.5,4.1){$#1$}
\put (5.5,0.82){$#2$}
\end{picture}
{}~~~~~~~~~~~~~~}

\def\bilink#1#2{\setlength{\unitlength}{1.0cm}
\begin{picture}(5.8,4.0)(1.8,2.5)
\put (6.2,5.2){$#1$}
\put (6.2,-0.2){$#2$}
\put (2.3,2.5){$n_{1}$}
\put (6.2,3.5){$n_{3}$}
\put (9.8,2.5){$n_{2}$}
\end{picture}
{}~~~~~~~~~~~~~~~~}

\def\npovtet#1#2#3#4#5#6{\setlength{\unitlength}{1.0cm}
\begin{picture}(1.9,1.7)(1.8,2.5)
\put (1.85,2.5){$#1$}
\put (4.45,2.5){$#2$}
\put (3.05,1.1){$#3$}
\put (2.85,3.1){$#4$}
\put (3.1,3.75){$#5$}
\put (2.85,1.8){$#6$}
\end{picture}
{}~~~~~~~~~}

\def\ltetra#1#2#3#4#5#6{\setlength{\unitlength}{1.0cm}
\begin{picture}(3,2.3)(1.4,2.5)
\put (2.0,2.5){$#1$}
\put (3.95,2.84){$#2$}
\put (3.5,0.4){$#3$}
\put (2.9,1.5){$#4$}
\put (3.5,4.6){$#5$}
\put (2.9,3.5){$#6$}
\end{picture}
{}~~~~~~~~~}

\def\fork#1#2#3#4#5{\setlength{\unitlength}{1.0cm}
\begin{picture}(3,2.3)(1.4,2.5)
\put (1.5,2.5){$#1$}
\put (2.95,3.9){$#2$}
\put (4.05,3.9){$#3$}
\put (5.5,2.5){$#4$}
\put (3.5,2.1){$#5$}
\end{picture}
{}~~~~~~~~~}

\def\tfork#1#2#3#4#5{\setlength{\unitlength}{1.0cm}
\begin{picture}(3,2.3)(1.4,2.5)
\put (1.5,2.5){$#1$}
\put (2.95,3.9){$#2$}
\put (4.05,3.9){$#3$}
\put (5.5,2.5){$#4$}
\put (3.5,2.1){$#5$}
\end{picture}
{}~~~~~~~~~}

\def\fusfork#1#2#3#4#5{\setlength{\unitlength}{1.0cm}
\begin{picture}(3,2.3)(1.4,2.5)
\put (1.5,2.5){$#1$}
\put (2.7,3.9){$#2$}
\put (4.4,3.9){$#3$}
\put (5.5,2.5){$#4$}
\put (3.8,2.9){$#5$}
\end{picture}
{}~~~~~~~~~}

\def\cptetra#1#2#3#4#5#6{\setlength{\unitlength}{1.0cm}
\begin{picture}(3.4,2.3)(1.4,2.5)
\put (2.0,2.5){$#1$}
\put (1.7,3.9){$\cK^{ba}_{s}$}
\put (1.7,1.1){$\cK_{ab}^{s}$}
\put (3.95,2.84){$#2$}
\put (5.5,2.5){$\cK_{bt}^{a}~~~$}
\put (3.6,2.2){$\cK^{tb}_{a}$}
\put (3.5,0.4){$#3$}
\put (2.9,1.5){$#4$}
\put (3.5,4.6){$#5$}
\put (2.9,3.5){$#6$}
\end{picture}
{}~~~~~~~~~}

\def\tmtetra#1#2#3#4#5#6{\setlength{\unitlength}{1.0cm}
\begin{picture}(3.4,2.3)(1.4,2.5)
\put (1.88,2.58){$#1$}
\put (3.85,2.84){$#2$}
\put (3.4,0.3){$#3$}
\put (2.9,1.5){$#4$}
\put (3.3,4.6){$#5$}
\put (2.9,3.5){$#6$}
\end{picture}
{}~~~~~~~~~}

\def\dutetra#1#2#3#4#5#6{\setlength{\unitlength}{1.0cm}
\begin{picture}(3.4,2.3)(1.4,2.5)
\put (1.88,2.58){$#1$}
\put (3.85,2.84){$#2$}
\put (3.5,0.3){$#3$}
\put (2.9,1.5){$#4$}
\put (3.5,4.6){$#5$}
\put (2.9,3.5){$#6$}
\end{picture}
{}~~~~~~~~~}

\def\gfour#1#2#3#4#5#6{\setlength{\unitlength}{1.0cm}
\begin{picture}(2,2.5)(1.8,2.5)
\put (2.5,2.5){$#1$}
\put (3.5,2.5){$#2$}
\put (5.5,2.5){$#2$}
\put (6.6,2.5){$#1$}
\put (4.35,3.85){$#1$}
\put (4.35,4.9){$#2$}
\put (4.35,1.8){$#1$}
\put (4.35,0.73){$#2$}
\put (3.0,1.0){$#3$}
\put (3.0,4.2){$#4$}
\put (5.4,4.2){$#5$}
\put (5.4,1.0){$#6$}
\end{picture}
{}~~~~~~~~~~~~~~~~~~~~~}

\def\yperm#1#2#3{\setlength{\unitlength}{1.0cm}
\begin{picture}(1.5,1.0)(2.0,2.3)
\put (2.2,2.65){$#1$}
\put (3.5,2.65){$#2$}
\put (3.1,1.4){$#3$}
\end{picture}
{}~~~~}

\def\openy#1#2#3{\setlength{\unitlength}{1.0cm}
\begin{picture}(1.5,1.0)(2.0,2.5)
\put (2.1,3.2){$#1$}
\put (3.7,3.2){$#2$}
\put (3.2,1.7){$#3$}
\end{picture}
{}~~~~}

\def\curchan#1#2#3#4#5{\setlength{\unitlength}{1.0cm}
\begin{picture}(2.5,1.9)(1.4,2.5)
\put (2.2,1.35){$#1$}
\put (3.5,1.35){$#2$}
\put (2.2,3.6){$#3$}
\put (3.5,3.6){$#4$}
\put (3.2,2.5){$#5$}
\end{picture}
{}~~~~}

\def\dcurchan#1#2#3#4#5{\setlength{\unitlength}{1.0cm}
\begin{picture}(2.5,1.9)(1.4,2.5)
\put (2.2,1.35){$#1$}
\put (3.5,1.35){$#2$}
\put (2.2,3.6){$#3$}
\put (3.5,3.6){$#4$}
\put (3.2,2.5){$#5$}
\end{picture}
{}~~~~}

\def\rcurchan#1#2#3#4#5{\setlength{\unitlength}{1.0cm}
\begin{picture}(2.5,1.9)(1.4,2.5)
\put (2.2,1.35){$#1$}
\put (3.5,1.35){$#2$}
\put (2.2,3.6){$#3$}
\put (3.5,3.6){$#4$}
\put (3.2,2.5){$#5$}
\end{picture}
{}~~~~}

\def\curid#1#2#3#4{\setlength{\unitlength}{1.0cm}
\begin{picture}(2.5,1.9)(1.4,2.5)
\put (2.2,1.35){$#1$}
\put (3.5,1.35){$#2$}
\put (2.2,3.6){$#3$}
\put (3.5,3.6){$#4$}
\end{picture}
{}~~~~}

\def\rcurid#1#2#3#4{\setlength{\unitlength}{1.0cm}
\begin{picture}(2.5,1.9)(1.4,2.5)
\put (2.2,1.35){$#1$}
\put (3.5,1.35){$#2$}
\put (2.2,3.6){$#3$}
\put (3.5,3.6){$#4$}
\end{picture}
{}~~~~}

\def\baryon#1#2#3{\setlength{\unitlength}{1.0cm}
\begin{picture}(2,2.0)(1.8,2.5)
\put (2.3,2.5){$#1$}
\put (3.6,2.5){$#2$}
\put (5.05,2.5){$#3$}
\end{picture}
{}~~~~~~~~~~~~}

\def\cpbaryon#1#2#3{\setlength{\unitlength}{1.0cm}
\begin{picture}(2,2.0)(1.8,2.5)
\put (2.3,2.5){$#1$}
\put (3.6,2.5){$#2$}
\put (5.05,2.5){$#3$}
\put (3.1,4.25){$\cK_{#1 #2 #3}$}
\put (3.1,0.8){$\cK^{#1 #2 #3}$}
\end{picture}
{}~~~~~~~~~~~~}

\def\duubaryon#1#2#3{\setlength{\unitlength}{1.0cm}
\begin{picture}(2,2.0)(1.8,2.5)
\put (2.3,2.5){$#1$}
\put (3.6,2.5){$#2$}
\put (5.05,2.5){$#3$}
\end{picture}
{}~~~~~~~~~~~~}

\def\udubaryon#1#2#3{\setlength{\unitlength}{1.0cm}
\begin{picture}(2,2.0)(1.8,2.5)
\put (2.3,2.5){$#1$}
\put (3.6,2.5){$#2$}
\put (5.05,2.5){$#3$}
\end{picture}
{}~~~~~~~~~~~~}

\def\smduubaryon#1#2#3{\setlength{\unitlength}{1.0cm}
\begin{picture}(1,1)(2.3,2.5)
\put (2.4,2.58){$\sct #1$}
\put (3.55,2.58){$\sct #2$}
\put (4.25,2.58){$\sct #3$}
\end{picture}
{}~~~~~~~~~}

\def\smddubaryon#1#2#3{\setlength{\unitlength}{1.0cm}
\begin{picture}(1,1)(2.3,2.5)
\put (2.4,2.58){$\sct #1$}
\put (3.55,2.58){$\sct #2$}
\put (4.25,2.58){$\sct #3$}
\end{picture}
{}~~~~~~~~~}

\def\smuudbaryon#1#2#3{\setlength{\unitlength}{1.0cm}
\begin{picture}(1,1)(2.3,2.5)
\put (2.4,2.58){$\sct #1$}
\put (3.55,2.58){$\sct #2$}
\put (4.25,2.58){$\sct #3$}
\end{picture}
{}~~~~~~~~~}

\def\smdudbaryon#1#2#3{\setlength{\unitlength}{1.0cm}
\begin{picture}(1,1)(2.3,2.5)
\put (2.4,2.58){$\sct #1$}
\put (3.55,2.58){$\sct #2$}
\put (4.25,2.58){$\sct #3$}
\end{picture}
{}~~~~~~~~~}

\def\hsmdudbaryon#1#2#3{\setlength{\unitlength}{1.0cm}
\begin{picture}(1,1)(2.3,2.0)
\put (2.4,2.58){$\sct #1$}
\put (3.55,2.58){$\sct #2$}
\put (4.25,2.58){$\sct #3$}
\end{picture}
{}~~~~~~~~~}

\def\bigraph#1#2#3#4#5{\setlength{\unitlength}{1.0cm}
\begin{picture}(2,2.5)(1.8,2.5)
\put (3.1,1.8){$#1$}
\put (3.1,3.3){$#2$}
\put (6.1,1.8){$#1$}
\put (6.1,3.3){$#2$}
\put (2.4,2.5){$#3$}
\put (7.4,2.5){$#4$}
\put (4.5,2.8){$#5$}
\put (4.5,4.1){$#1$}
\put (4.5,1.45){$#2$}
\end{picture}
{}~~~~~~~~~~~~~~~~~~~~~~~~~~~~~~~}

\def\punknot#1{\setlength{\unitlength}{1.0cm}
\begin{picture}(1,1)(2.7,2.0)
\put (3.45,3.4){$#1$}
\end{picture}
{}~~~~~~}

\def\idrop#1{\setlength{\unitlength}{1.0cm}
\begin{picture}(1,1.5)(2.1,2.0)
\put (2.15,2.5){$#1$}
\put (3.2,1.3){$0$}
\end{picture}
{}~~~~~~}

\def\twist#1#2{\setlength{\unitlength}{1.0cm}
\begin{picture}(5.8,3.3)(1.8,2.5)
\put (2.5,2.5){$#1$}
\put (6.35,-1.5){$#2$}
\put (5.2,0.4){$n_{1}$}
\put (5.2,4.6){$n_{2}$}
\put (7.3,4.6){$n_{3}$}
\put (7.3,0.4){$n_{4}$}
\end{picture}
{}~~~~~~~~~~~~~~~~}

\def\smultab#1#2#3#4#5#6#7{\put (0,4){\line(1,0){#1}}
                    \multiput(0,3)(1,0){#1}{\line(1,0){1}}
                    \multiput(1,3)(1,0){#1}{\line(0,1){1}}
                    \multiput(0,2)(1,0){#2}{\line(1,0){1}}
                    \multiput(1,2)(1,0){#2}{\line(0,1){1}}
                    \multiput(0,1)(1,0){#3}{\line(1,0){1}}
                    \multiput(1,1)(1,0){#3}{\line(0,1){1}}
                    \multiput(0,0)(1,0){#4}{\line(1,0){1}}
                    \multiput(1,0)(1,0){#4}{\line(0,1){1}}
                    \multiput(0,-1)(1,0){#5}{\line(1,0){1}}
                    \multiput(1,-1)(1,0){#5}{\line(0,1){1}}
                    \multiput(0,-2)(1,0){#6}{\line(1,0){1}}
                    \multiput(1,-2)(1,0){#6}{\line(0,1){1}}
                          \put (0,4){\line(0,-1){#7}}}

\def\young#1#2#3#4#5#6#7{\begin{picture}(#1,#7)(0,3)
                \thicklines \smultab#1#2#3#4#5#6#7 \end{picture}}

\def\fund{\young1000001}

\newcommand{\eq}{\begin{equation}}
\newcommand{\en}{\end{equation}}
\newcommand{\ie}{{\it i.e.}}

\def\ZZ{Z\!\!\!Z}

\def\col{c}

\def\row{l}

\def\CS{{\sst {\rm CS}}}
\def\cU{{\cal U}}

\def\cR{{\cal R}}

\def\cH{{\cal H}}

\def\cK{{\cal K}}
\def\cO{{\cal O}}
\def\cL{{\cal L}}

\def\half{{\scriptstyle \frac{1}{2}}}
\def\third{\frac{1}{3}}
\def\sst{\scriptscriptstyle}
\def\sct{\scriptstyle}
\def\ds{\displaystyle}

\def\ta{\widetilde{a}}
\def\tb{\widetilde{b}}
\def\tc{\widetilde{c}}

\def\vev#1{\langle #1 \rangle}

\def\bra#1{\langle #1 |}
\def\ket#1{| #1\rangle}

\def\sun{{\rm SU}(N)}

\def\spn{{\rm Sp}(N)}
\def\gnk{{\rm G}(N)_{K}}
\def\gkn{{\rm G}(K)_{N}}
\def\gln{{\rm GL}(N)}
\def\glk{{\rm GL}(K)}
\def\sunk{{\rm SU}(N)_{K}}
\def\sukn{{\rm SU}(K)_{N}}
\def\spnk{{\rm Sp}(N)_{K}}
\def\spkn{{\rm Sp}(K)_{N}}

\def\su{{\rm su}}
\def\smlef{\hbox{$\sct ($}}
\def\smrig{\hbox{$\sct )$}}
\def\smbar{\hbox{$\sct |$}}
\def\asupmn{\su{\raise 0.25mm\smlef}m{\raise 0.25mm\smbar}n
                                             {\raise 0.25mm \smrig}}
\def\asupmnk{\su{\raise 0.25mm\smlef}m{\raise 0.25mm\smbar}n
                                             {\raise 0.25mm \smrig}_{{\sst K}}}
\def\asummnn{\su{\raise 0.25mm\smlef}m-n{\raise 0.25mm \smrig}}
\def\basupmn{\bf {\rm\bf su}{\raise 0.5mm\hbox{{$\sct ($}}} m {\sct |} n
{\raise 0.25mm \smrig}}

\def\asum{\su {\raise 0.25mm\smlef}  m {\raise 0.25mm\smrig}}
\def\asun{\su {\raise 0.25mm\smlef}  n {\raise 0.25mm\smrig}}
\def\asuN{\su {\raise 0.25mm\smlef} {\sct N} {\raise 0.25mm\smrig}}
\def\asumn{\su {\raise 0.25mm\smlef}  m-n {\raise 0.25mm\smrig}}
\def\asumnk{\su {\raise 0.25mm\smlef} m-n {\raise 0.25mm\smrig}_{K}}

\def\asupno{\su{\raise 0.25mm\smlef}n+{\sct 1}{\raise 0.25mm\smbar}n
                                               {\raise 0.25mm\smrig} }
\def\asupo{\su{\raise 0.25mm\smlef} n {\raise 0.25mm\smbar} {\sct 1}
                                               {\raise 0.25mm\smrig} }

\def\uone{{\rm u} {\raise 0.25mm\smlef} {\sct 1} {\raise 0.25mm\smrig}}

\def\Tr{{\rm Tr}}
\def\gn{{\rm G}(N)}
\def\spnk{{\rm Sp}(N)_{K}}
\def\spkn{{\rm Sp}(K)_{N}}
\def\gnk{{\rm G}(N)_{K}}
\def\gkn{{\rm G}(K)_{N}}
\def\gk{{\rm G}(K)}

\def\sun{{\rm SU}(N)}
\def\spn{{\rm Sp}(N)}
\def\son{{\rm SO}(N)}

\def\sodn{{\rm SO}(2n+1)}
\def\sodnk{{\rm SO}(2n+1)_{2k+1}}

\def\ason{{\rm so}(N)}
\def\asonk{{\rm so}(N)_{K}}

\def\asodn{{\rm so}(2n+1)}
\def\asodnk{{\rm so}(2n+1)_{2k+1}}
\def\asodkn{{\rm so}(2k+1)_{2n+1}}
\def\asodnK{{\rm so}(2n+1)_{K}}

\def\cO{{\cal O}}
\def\cK{{\cal K}}
\def\cC{{\cal C}}
\def\cW{{\cal W}}

\def\cL{{\cal L}}
\def\cF{{\cal F}}

\def\rb{\rho (b)}
\def\rc{\rho (c)}
\def\rd{\rho (d)}

\def\rtb{\rho (\widetilde{b})}
\def\rtc{\rho (\widetilde{c})}

\def\ta{\widetilde{a}}
\def\tb{\widetilde{b}}
\def\tc{\widetilde{c}}
\def\td{\widetilde{d}}
\def\ts{\widetilde{s}}
\def\tt{\widetilde{t}}
\def\trr{\widetilde{r}}
\def\tu{\widetilde{u}}
\def\CS{{\rm CS}}

\begin{document}
\setlength{\unitlength}{0.25cm}
\thispagestyle{empty}

\noindent{\small April 1992, hep-th/9205082, {\bf No Figures}}\hfill
                      \begin{tabular}{l} {\sc BRX-TH}-302 \\
                                                 {\sc JHU-TIPAC}-920013
                                       \end{tabular}
\vspace{0.4cm}

\begin{center}
{\large\bf   Simple-Current Symmetries, Rank-Level Duality,
    and Linear Skein Relations for Chern-Simons Graphs\footnote{Supported
           in part by  the DOE under grant
               DE-FG02-92ER40706}}\\[1.0cm]

\setcounter{footnote}{1}
{\large \sl Stephen G. Naculich,\footnote{Supported in part by
the NSF under grant PHY-90-96198} Harold A. Riggs, and Howard J. Schnitzer}

\vspace{0.5cm}
{ \sl
\begin{tabular}{c}
   Department of Physics, Brandeis University \\
   Waltham, MA 02254
\end{tabular}  }

$^{\dagger}${\normalsize \sl Dept. of
Physics and Astronomy, The Johns Hopkins University\\
 Baltimore, MD 21218}\\[0.6cm]

\end{center}

\vfill
\begin{center}
{\sc Abstract}
\end{center}

\begin{quotation}
A previously proposed two-step
algorithm for calculating the expectation values of
arbitrary Chern-Simons graphs fails to determine certain crucial signs.
The step which involves calculating tetrahedra by solving
certain non-linear equations is repaired by introducing
additional linear equations.
The step which involves reducing arbitrary graphs to
sums of products of tetrahedra remains seriously disabled, apart
from a few exceptional cases.
As a first step towards a new algorithm
for general graphs we find useful linear equations for
those special graphs which support knots and links.

Using the improved set of
equations for tetrahedra we examine the symmetries between tetrahedra
generated by arbitrary simple currents.
Along the way we
uncover the simple, classical origin of simple-current charges.
The improved skein relations also lead to exact
identities between planar tetrahedra
in level $K$ $\gn$ and level $N$ $\gk$ Chern-Simons theories, where $\gn$
denotes a classical group. These results
are recast as WZW braid matrix identities
and as identities between quantum $6j$-symbols at appropriate roots of
unity.

We also obtain the transformation properties
of arbitrary graphs, knots, and links
under simple-current symmetries and rank-level duality.
For links with knotted components
this requires precise control of the braid eigenvalue
permutation signs, which we obtain from plethysm and
an explicit expression for the (multiplicity-free) signs, valid
for all compact gauge groups and
all fusion products.

\end{quotation}
\vfill

\setcounter{page}{0}
\newpage
\setcounter{page}{1}

\vspace{1.0cm}
\noindent {\bf 1. Introduction}
\vspace{0.3cm}
\setcounter{section}{1}
\setcounter{footnote}{1}

Topologically invariant Chern-Simons gauge theories in $2+1$ dimensions are
interacting---yet completely soluble---quantum field theories.\cite{witten}
Quantization of such a theory with compact gauge group $G$
forces the coupling constant $K$ to be an integer.
The fixing of these integer values of $K$ leads
to the appearance of discrete symmetries
associated with automorphisms of
the extended Dynkin diagram for $G$,\cite{tame}
as well as remarkable relations between those models with
a classical group $\gn$ as gauge group and coupling constant $K$
and those with gauge group $G(K)$ and coupling
constant $N$.\cite{nrs}  While the
presence of these symmetries in several contexts has long been
known,\cite{earlydd}$^{-}$\cite{earlyrl}
only recently have the pervasive implications of these
symmetries for  Chern-Simons theories,\cite{nrs,recentus}
integrable lattice models,\cite{recentjapan,recentfrance}
and quantum groups\cite{recentfrance}
begun to be studied.\footnote{The general idea
that Wilson lines with representations related by the Dynkin diagram
symmetries are equivalent up to phases has long been
known,\cite{tame} however,
and has been used to understand Chern-Simons theories
with gauge groups of the form
$G/({\rm discrete~subgroup})$.}
In much of this existing work the effects of these discrete symmetries
have been derived only for restricted cases. For
the extended Dynkin diagram automorphisms
(which signal the presence of simple-current symmetries in the
associated WZW model\cite{earlydd}) the properties of
the modular transformation matrix have been of central
interest.\cite{earlydd2,schellekens}$^{-}\!$\cite{abc} This
corresponds in Chern-Simons theory only to the expectation values of
the simplest knot, the unknot,
and the simplest link, two linked unknots (the Hopf link).\cite{witten}
Similarly, although the rank-level
duality of the characteristic polynomial of the braid matrix or its
spectral decomposition holds for many tensor representations of the
groups involved, this result only leads to a duality of expectation values
for special classes of knots or links.\cite{nrs,recentus}
The analogous results for
the associated quantum groups and lattice
models,\cite{recentjapan,recentfrance} which have proceeded
by explicit construction of the quantities involved, have only been obtained
for restricted classes of representations: the completely symmetric
and antisymmetric representations of $\sun$.
Our goal in this paper is to attain an exact, general result, without
restrictions on the representations or classes of links involved.

The advantage of examining this question in the context of
three-dimensional Chern-Simons theory is that, in addition to providing a
powerful and unifying approach to almost all the different areas in which
these discrete symmetries appear, it provides the tools
with which to demonstrate
the general implications of these symmetries
without having to explicitly
solve for the quantities being related.

Of special interest are
certain gauge invariant Chern-Simons observables: the planar
tetrahedra. This follows since all Chern-Simons
observables can be expressed as sums of products of these
tetrahedra.\cite{witten,martin} They are also the $q$-$6j$-symbols
of the related quantum groups evaluated at the associated roots of unity.
In addition, the planar tetrahedra are related by simple
phases to limiting values
of the Boltzmann weights of integrable lattice models and to
the braid matrices of WZW theory.\cite{witten}

Our strategy will be to examine sets of skein relations that
completely determine the expectation values of
arbitrary tetrahedra. A previously proposed set of non-linear
skein relations\cite{martin} suffers from a sign ambiguity
that renders them ineffective for the exact determination of all
tetrahedra.
We will remove this ambiguity
by supplementing these non-linear equations with a
set of inhomogeneous linear equations.
We are then able to show that given any tetrahedron there exists
a class of tetrahedra, related to this one by the symmetries of
the extended Dynkin diagram of the gauge group,
whose expectation values only differ
from that of the original tetrahedron
by (in general, vertex-normalization dependent) signs.
If the tetrahedral expectation values are
intrinsic (\ie, independent of the sign convention for
vertex normalization), then the relative sign is
given by a certain product of braid eigenvalues.
We further show that given a tetrahedron in one
theory, there are tetrahedra in the rank-level dual theory with
the same expectation values (up to sign). If the signs
of the tetrahedra so related are intrinsic, the relative sign
is again given by a simple product of braid eigenvalues.

Since arbitrary observables in a $\gnk$ Chern-Simons theory
(including knots, links, and
graphs) can be reduced to sums of products of tetrahedra,
we expect that all such observables will
fall into sets related by the
Dynkin diagram automorphisms, and have, in addition,
rank-level duals in the $\gkn$ theory.
A general proof along these lines is stymied
by the presence of certain undetermined,
{\it normalization-independent} signs in most such reductions.
Nevertheless, on the basis of cases where an
unambiguous  reduction is possible and on the basis of the
properties of the linear equations which we obtain for
any link-type graph, we state the expected general result
for such graphs for both types of symmetry.
Identities between knot and link expectation values defined
with representations related by the diagram automorphisms can
be proved via a direct cabling argument;  the aforementioned
identities for the underlying link-type graphs then follow.
The identities that relate knot and link expectation values
in rank-level dual theories are shown to follow from the
conjectured identities between dual graphs.

Our approach only deals with the local, Lie-algebraic structure of
Chern-Simons theory; to avoid conflicts with global constraints\cite{DikgWitt}
we assume that each theory is defined with a compact,
connected, and simply-connected gauge group. This is appropriate
since such Lie groups are
in one-to-one correspondence with (complex)
Lie algebras (and so with the standard set of Dynkin diagrams).
For example, in order to obtain a theory with the local
structure of $B_{n}$ or $D_{n}$
with arbitrary coupling constant $K$,
and in order to include their spinor representations,
it must be defined via the simply-connected
covering group of $\son$,
${\rm Spin}(N)$.
Since it will be useful (in section 4, essential)
to characterize these theories using the tensorial
language of the orthogonal group we will refer to these
theories as level $K$ ${\rm so}(N)$ theories (by which
we mean level $2K$ ${\rm Spin}(N)$
\setcounter{footnote}{1}
theories).\footnote{The (non-simply-connected) level $K$ $\son$
theory is defined by the level $2K$ ${\rm Spin}(N)/\ZZ_{2}$ theory,
given appropriate restrictions on $K$.}

In section 2 we describe the known non-linear equations for
tetrahedra and find the supplemental equations that remove
the sign ambiguity of the non-linear set.
We pay particular attention to the permutation signs appearing
in the diagonal action of the braid matrix
on the legs of trivalent vertices and the
normalization of these vertices, in order to deal uniformly with
real, pseudoreal, and complex representations.
We adopt a natural system of permutation signs different from
that commonly prescribed in the literature.
In section 3 we show that replacing the representations
around a closed loop of edges of a tetrahedron with
representations related to these by simple-current
symmetries leaves the tetrahedral expectation value essentially invariant.
We will call the tetrahedra (or representations) so related
{\em cominimally equivalent}.\cite{nrs,recentus}
In section 4 we find that, given a tetrahedron
in a $\gnk$ theory, with $\gn$ denoting $\sun$, $\spn$,
or the double-cover of
$\sodn$,\footnote{Similar but more complicated results
hold for dual pairs involving
${\rm so}(2n)$; we will not, however, present the details here.
As a result we only examine rank-level duality for $\asodnK$ with $K$ odd.}
and with the representations on its edges
specified by Young tableaux, the tetrahedra in the $\gkn$ theory with
edge representations specified by the
transposes (or certain cominimal
equivalents of the transposes) of these Young tableaux have
essentially the same expectation value.
In section 5 these results are recast as
identities for WZW fusion and braid matrices
as well as for the $q$-$6j$-symbols appearing in
quantum group\cite{qg} theory.
In section 6, we examine more general graphs and the knots and
links based on them.
We give
a simple, graph-independent, and completely general derivation of the
phases that relate links (or knots) with cominimally
equivalent representations on corresponding components.
We state (and give evidence for, but
do not prove) identities that
we expect to hold between arbitrary Chern-Simons observables
related by rank-level duality.

The technical note following the conclusion describes certain subtleties
of baryon normalization, a proof of a useful identity by means of plethysm,
and several other results on
permutation signs needed in the text, including an explicit formula
for these signs in the multiplicity-free case, valid for all compact
gauge groups and all fusion products.

\newpage

\vspace{0.8cm}
\noindent {\bf 2.  Tetrahedral Skein Relations: A Sign Problem and its
Solution}
\vspace{0.3cm}
\onward

We shall see later that a previously proposed algorithm for calculating
arbitrary Chern-Simons observables\cite{martin}
suffers from sign ambiguities that render it ineffective for
evaluating general graphs and links.
However, the part of this algorithm that involves solving certain
non-linear (associativity or pentagon) equations for
planar tetrahedra is repairable, which is the task of the
present section. Along the way we give a more comprehensive
account of the permutation signs that arise in the diagonal
action of the braid matrix than has previously appeared, and
take account of certain subtleties of baryon normalization.

We begin with a level $K$ Chern-Simons theory with compact, connected,
and simply-connected gauge group $G$ defined on the $3$-manifold $M=S^{3}$.
Since $G$ is simply-connected, $G$-bundles over $M$ are trivial
and the theory is well-defined by the action\cite{DikgWitt}
\eq
    S_{\CS} = \frac{\sct K}{4\pi} \Tr \int (A \wedge dA
                + \frac{2}{3} A \wedge A \wedge A)
\en
where $A$ is the $G$ gauge connection, and quantization forces $K$ to
be an integer.  In order for the partition function
\eq
       Z(S^{3}) = \int {\cal D}A ~e^{i S_{\CS}}
\en
to be unambiguously defined it must be regularized by specifying
a framing\cite{ccw}
of $S^{3}$.  We choose this framing\cite{ccw} so that $Z(S^{3}) = S_{00}$,
where $S_{00}$ is the identity-identity component of the modular transformation
matrix of the level $K$ characters of the affinization of $G$.
The gauge invariant observables
$\cO$, whose expectation values are given by
\eq
        Z(\cO; S^{3}) = \int {\cal D}A ~e^{i S_{\CS}} ~\cO  ~~~,
\en
correspond to linked---often knotted---Wilson lines
and graphs. To obtain an unambiguous definition of these expectation
values we will assume throughout a
vertical framing of the Wilson lines and graphs.\cite{witten}
We will be concerned exclusively with the normalized
expectation values,
\eq
  \vev{\cO} =
\frac{Z(\cO; S^{3})}{Z(S^{3})}  ~~.
\en

Among the graph observables are those
specified by single-component planar graphs with four trivalent
gauge-invariant vertices---the planar tetrahedra.
Any tetrahedron corresponds to a pair of
fusion rule channels, with each channel defined by
a pair of compatible fusion rules.
These channels come in natural sets of three
which, following ref.~\mite{martin}, we call the $S$, $T$, and $U$ channels.
Each of these channels corresponds to a basis of the Hilbert
space on which some braid matrix acts diagonally. The compatibility of
these bases gives a set of equations that constrain
the basis change coefficients, which are, essentially, the
expectation values of planar tetrahedra.

\vspace{0.6cm}
\noindent {\it 2.1 ~~S, T, and U channel bases}
\vspace{0.3cm}

Choose a surface $S^{2}$ that divides $S^{3}$ into
two halves (call one the interior half; the other, the exterior),
in such a way that it is punctured at exactly four points by static
charges, corresponding to the four representations
$a$, $b$, $\rho(c)$, and $\rho(d)$ of $\gn$. (For the groups we
are considering, the dual representation $\rho(r)$ of $r$ is just
the representation conjugate to $r$.)
The Hilbert space $\cal H$  associated with this surface, considered
as the boundary of the interior half of $S^{3}$, exists
and is $f= \sum_{s} {N_{ab}}^{s} {N_{cd}}^{s} $ dimensional
whenever the pair of fusion rules
\eq
\begin{array}{ccl}
  \phi_{a} \cdot \phi_{b} & = & \ds ~\sum_{s} {N_{ab}}^{s} \phi_{s} \\[0.4cm]
   \phi_{c} \cdot \phi_{d} & = & \ds ~\sum_{s} {N_{cd}}^{s} \phi_{s}
\end{array}
\en
has some representation in common (\ie, whenever $f\neq 0$).
Such a pair of fusion rules will be called {\em compatible}.
The path integral on the interior half
can produce a variety of states in ${\cal H}$ depending on how
the Wilson lines or graphs intertwine in the interior.
We will consider the relation between three different
planar bases of this space,  corresponding to the
three sets of compatible fusion rule pairs
\eq
\begin{array}{ccc}
  S{\rm -channel} &    T{\rm -channel}  &  U{\rm -channel} \\[0.3cm]
\ds
\phi_{a} \cdot \phi_{b}   =  \sum_{s} {N_{ab}}^{s} \phi_{s}  &
\ds
{}~~\phi_{a} \cdot \phi_{\rc} =  \sum_{t} {N_{a\rc}}^{t} \phi_{t}  &
\ds
{}~~\phi_{a} \cdot \phi_{\rd} =  \sum_{u} {N_{a\rd}}^{u} \phi_{u} \\[0.4cm]
\ds
\phi_{c} \cdot \phi_{d}   =  \sum_{s} {N_{cd}}^{s} \phi_{s}   &
\ds
{}~~\phi_{\rb} \cdot \phi_{d} =  \sum_{t} {N_{\rb d}}^{t} \phi_{t} &
\ds
{}~~\phi_{\rb} \cdot \phi_{c} =  \sum_{u} {N_{\rb c}}^{u} \phi_{u}
\end{array}
\label{threechan}
\en
Each non-zero fusion coefficient ${N_{r_{1} r_{2}}}^{r_{3}}$
corresponds to a set of ${N_{r_{1} r_{2}}}^{r_{3}}$ gauge invariant
couplings, each of which permits construction of a gauge invariant
trivalent vertex. In order to describe the properly framed graphs that
specify the bases of ${\cal H}$
constructed with these vertices, we must keep track of
the (here vertical) framing.  This is done by a generic projection of the graph
onto a plane and a restriction to certain allowed
moves\cite{kauffman} in manipulating the
resulting diagrams---the standard Chern-Simons link and graph
moves.\cite{witten}

Then, with the bases defined by

{}~
\vspace{1.3in}
\setlength{\unitlength}{1.0cm}
\begin{picture}(1,2)(0,2.5)
\put (0,2.1){$\ket{s} =$}
\put (1,3.3){$c$}
\put (1,1){$a$}
\put (3,3.3){$d$}
\put (3,1){$b$}
\put (1.6,2.1){$s$}
\put (2.33,1.6){$\sct \cK_{ab}^{s}$}
\put (2.33,2.75){$\sct \cK^{cd}_{s}$}
\put (5,2.1){$\ket{t} =$}
\put (6,3.3){$c$}
\put (6,1){$a$}
\put (7.4,2.3){$t$}
\put (8.7,3.3){$d$}
\put (8.7,1){$b$}
\put (10,2.1){$\ket{u} =$}
\put (11.2,3.3){$c$}
\put (11,1){$a$}
\put (12.3,1.7){$u$}
\put (13.4,3.3){$d$}
\put (13.6,1){$b$}
\end{picture}

\noindent each of these channels corresponds to an $f$-dimensional
basis of the Hilbert space $\cH$.
We will denote the basis vectors
in the various channels by a single label
$\ket{s}, \ket{t}, {\rm ~or~} \ket{u}$ where, for example,
$\ket{s} = \ket{ab \rightarrow cd; s, \cK_{ab}^{s}, \cK^{cd}_{s}}$
denotes the vector created by the path integral in the
presence of the $S$-channel graph shown above, with
$\cK_{ab}^{s}$ and $\cK^{cd}_{s}$ used to couple the representations
associated with the graph edges.
Upper indices on a coupling indicate outgoing edge arrows at the vertex;
lower indices, incoming edge arrows. (For visual clarity,
we will mostly omit explicit display of these couplings on graph vertices.)

We assume that a basis of couplings can be chosen so that the relevant
braid matrices act diagonally even in the case of multiplicities.
The bases defined above then have the dual bases

{}~
\vspace{1.4in}
\setlength{\unitlength}{1.0cm}
\begin{picture}(1,2)(0.5,2.5)
\put (0,2.1){$\bra{s} =$}
\put (1,3.3){$a$}
\put (1,1){$c$}
\put (3,3.3){$b$}
\put (3,1){$d$}
\put (1.6,2.1){$s$}
\put (2.33,1.6){$\sct \cK_{cd}^{s}$}
\put (2.33,2.75){$\sct \cK^{ab}_{s}$}
\put (3.5,2.1){$\bra{t} = \eta^{a\rho(a)}_{0} \eta^{d\rho(d)}_{0}$}
\put (6.0,3.3){$a$}
\put (6,1){$c$}
\put (7.4,2.4){$t$}
\put (8.7,3.3){$b$}
\put (8.7,1){$d$}
\put (8.6,2.1){$\bra{u} = \eta^{a\rho(a)}_{0} \eta^{c\rho(c)}_{0}$}
\put (11.0,3.3){$a$}
\put (11,1){$c$}
\put (12.3,2.4){$u$}
\put (13.6,3.3){$b$}
\put (13.6,1){$d$}
\end{picture}

\noindent The permutation signs $\eta^{r\rho(r)}_{0} = \pm 1$
appearing here result from the signs of certain baryon expectation
values (as described in the technical note and the next subsection)
and insure that these states satisfy the orthonormalizations
\eq
\vev{ s^{\prime} | s}
= \delta_{s^{\prime} s} ~\chi_{q}(a) \chi_{q}(b) \chi_{q}(c)\chi_{q}(d) > 0 ~~,
\en
where $\chi_{q}(r)$ denotes the $q$-dimension of the representation $r$.
The delta function means that the {\it entire} intermediate channel,
as specified by a representation and a pair of couplings, must be
dual.  Thus $\ket{s}$ is orthogonal to $\ket{s^{\prime}}$ unless
the intermediate representation is the same, {\em and} unless the couplings
\setcounter{footnote}{1}
at each vertex are the same, in the basis chosen above.\footnote{\ie,
since dual couplings satisfy $\cK^{ab}_{s}(i) \cK_{ab}^{s}(i^{\prime})
\propto \delta_{i i^{\prime}}$ with $i,i^{\prime}= 1,\ldots, {N_{ab}}^{s}$,
this means, unless $i=i^{\prime}$.}

\vspace{0.6cm}
\noindent {\it 2.2 Permutation Signs and Baryons}
\vspace{0.3cm}

Associated with each planar basis of ${\cal H}$ is a pair of braid
operators that act diagonally on the basis vectors
by interchanging adjacent vertex legs
of the basis vector graphs. For example, the action of the
braid operator $B_{ab}$ on the dual $S$-channel vectors
is specified by
\eq
       B_{ab} \yperm{a}{b}{c} = \varrho^{ab}_{c} \yperm{b}{a}{c} ~~~,
\label{mainact}
\en
\vspace{1.1cm}

\noindent where the diagonal matrix elements
(which we will loosely call `eigenvalues') are
\eq
       \varrho^{ab}_{c} = \eta^{ab}_{c}
       q^{{\sct {1 \over 4}} (Q(a) + Q(b) - Q(c))} ~~{\rm with~~}
q = e^{2\pi i /(K + \overline{g})} ~~~.
\label{breigdef}
\en
Here $\eta^{ab}_{c} = \pm 1$ is
the Chern-Simons permutation sign,
$\overline{g}$ is the dual Coxeter number of $G$, and
$Q(r)= 2(K+\overline{g}) h(r)$ with $h(r)$ the conformal weight
of the level $K$ representation of the affinization of $G$ specified
by $r$.
For the simple, compact gauge groups considered in depth here
$Q(r)$ is the quadratic Casimir of the representation $r$.
We will specify all braid eigenvalues with reference to the specific
edge orientations shown in~\ref{mainact} so that $\varrho^{ab}_{c}$
always corresponds to $c \in a\otimes b$ and the quadratic Casimir of
the lower-index representation always enters
with a minus sign (as in~\ref{breigdef}).
On the other hand, we will adopt an index convention for
$\eta^{ab}_{c}$ that keeps track of edge orientations:
the same index convention
as that of the coupling $\cK^{ab}_{c}$;
namely, that upper (lower) indices imply outgoing (incoming) arrows.

Several of the results in section $3$ hold for
Chern-Simons theories with more general compact gauge groups, such
as those corresponding to various rational conformal field theories.
In these cases, $Q(\phi) = 2 (K + \overline{g}) h(\phi)$ with $h(\phi)$ the
conformal weight of the primary field $\phi$ of the associated
conformal field theory.
In addition, eqs.~\ref{blineq}-\ref{lasord} hold as identities between
quantum $6j$-symbols (using the connection with planar tetrahedra
given in section $5$) with the braid eigenvalues
given by eq.~\ref{breigdef} but with $q$ {\em not}
necessarily a root of unity.

The permutation signs require careful consideration in order
to give a consistent treatment in all cases, including that of groups
with pseudoreal representations. They satisfy the identity
\eq
      {\cal P}^{ab} \cK^{ab}_{c} = \eta^{ab}_{c} \cK^{ba}_{c} ~~,
\label{epdef}
\en
where ${\cal P}^{ab}: V^{a} \otimes V^{b} \rightarrow V^{b} \otimes V^{a}$
is the permutation operator.
If $a=b$ the sign $\eta^{aa}_{c}$
is independent of the normalization sign $\omega$
of the coupling $\cK^{aa}_{c}$ and only
depends on the embedding of the representation $c$ in the
tensor product $a \otimes a$.
If $a\neq b$, however,
$\eta^{ab}_{c}$ is fixed by the choice of the relative sign of the
normalization signs  $\omega_{1}$ of the coupling $\cK^{ab}_{c}$ and
$\omega_{2}$ of the permuted coupling $\cK^{ba}_{c}$.
We will refer to the normalization independent
signs $\eta^{aa}_{s}$ as the {\em intrinsic} permutation signs.
Once a consistent system of permutation signs is chosen there still
remains a single undetermined sign $\omega$ for each triple $\{a,b;c\}$
related by $a \otimes b \ni c$ . Baryon normalization
fixes the normalization $\overline{\omega}$
of the dual couplings $\cK_{ab}^{c}$ in terms of the sign
$\omega$ chosen for $\cK^{ab}_{c}$, so
that for each set $\{ \cK^{ab}_{c}, \cK^{ba}_{c},
\cK_{ab}^{c}, \cK_{ba}^{c} \}$ there remains just
one undetermined residual sign $\omega$ (whether $a=b$ or not).

Throughout this paper, the sign of any expectation value will be called
{\em intrinsic} or vertex-normalization independent,
if, given a system of permutation signs and baryon normalizations, it
is independent of the choice of these residual normalization signs.
Note that the word `intrinsic' is being
used for two slightly different concepts.

We can make any choice of the relative sign of
the normalizations $\omega_{1}$ and $\omega_{2}$
for a particular set of couplings, and so any choice for $\eta^{ab}_{c}$
with $a\neq b$, that we wish.  However, in order to
obtain a consistent set of topologically meaningful
graphical moves with which to manipulate the planar projections of knot,
link, and graph observables with {\em arbitrary} representations,
certain constraints on these signs must be satisfied.
In the technical note we find that, given the
standard normalization of baryons, the following three conditions
must be satisfied if we are to retain the standard graphical moves
for graphs with arbitrary representations on their edges.
\eq
\begin{array}{clcll}
1. & \eta^{ab}_{c} & = &  \eta^{a\rho(c)}_{\rho(b)} & {\rm crossing} \\
2. & \eta^{\rho(a)\rho(b)}_{\rho(c)} & = &
                 \eta^{ab}_{c} & {\rm conjugation}   \\
3. & \eta^{c \rho(c)}_{0} & = &
           \eta^{a\rho(a)}_{0} \eta^{b\rho(b)}_{0}
                    ~~~~{\rm for}~c\in a\otimes b~~~ &   {\rm fusion}
\end{array}
\label{chernsign}
\en
(Here, and throughout, $0$ denotes the identity representation.)
While we have singled out these three constraints,
the use of eq.~\ref{arrowreversal} to reduce
vertices with orientations other
than that of eq.~\ref{mainact} to this standard form
(before acting with $B_{ab}$) leads to
various other crossing constraints of the form
$\eta^{ab}_{c} = \eta^{ab\rho(c)} = \eta_{\rho(a) \rho(b) c}
= \ldots $, so that we need only (and will only) refer to the
standard form $\eta^{ab}_{c}$.

An immediate consequence of the fusion constraint
in~\ref{chernsign}
is that
the charge conjugation signs   must satisfy
(with $r(a)$ denoting the number of boxes of the tableau
  $a$)
\setlength{\unitlength}{0.15cm}
\eq
\begin{array}{rcll}
\eta^{a\rho(a)}_{0} & = & (\eta^{\fund \rho(\fund )}_{0})^{r(a)}
    &  ~~~{\rm all~classical~groups~} \\
                      ~                     \\
\eta^{ \{\psi;a\} \rho(\{\psi;a\})}_{0} & = &
             \eta^{\psi \rho(\psi)}_{0}
\eta^{a\rho(a)}_{0}   &  ~~~\ason
\end{array}
\label{fusimp}
\en
for all tensor representations $a$ and
spin representations $\{\psi;a\}$ with tensor part $a$. The fundamental
spinor is denoted by $\psi$, and
\setlength{\unitlength}{0.23cm} $\fund$ denotes
the fundamental tensor representation.
A further useful consequence of the fusion
constraint in~\ref{chernsign} for a compatible fusion rule channel (as in
eq.~\ref{threechan}) is that $\eta^{a\rho(a)}_{0} \eta^{b\rho(b)}_{0}
= \eta^{c\rho(c)}_{0} \eta^{d\rho(d)}_{0}$.

The common (manifestly crossing-symmetric)
proposal that $\eta^{ab}_{c}$ be uniformly chosen
positive in the cases with  $a\neq b$, $\rho(c) \neq a$,
and $\rho(c) \neq b$ (simultaneously), cannot be
adopted here, since it runs afoul of the fusion
constraint in~\ref{chernsign} as follows.
Consider $\sun$ with $N=(2\times ~{\rm odd~number})$,
\ie, the unitary groups with pseudoreal representations.
Then, with $a$ denoting the pseudoreal representation specified
by a single column of $N/2$ boxes, we necessarily have, from the first
equation in~\ref{fusimp}, that
\setlength{\unitlength}{0.15cm}
\eq
       \eta^{\fund \rho(\fund)}_{0} = \eta^{aa}_{0} = -1  ~~,
\en
even though $\eta^{\fund \rho(\fund)}_{0}$
is not an intrinsic sign, since
\setlength{\unitlength}{0.23cm}
$\fund  \neq \rho(\fund )$ unless $N=2$.
The first identity in eq.~\ref{fusimp} then fixes the charge
conjugation signs of all representations whose
tableaux have an odd number of boxes
(only a fraction of which
are connected by crossing to an intrinsic sign) to be
negative.
Therefore, the first identity in~\ref{fusimp} leads to a
series of counter-examples to
the `positive if not connected by crossing to an
intrinsic sign' prescription.

The other well-known system of permutation signs, the
standard ${\rm SU}(2)$ group theory signs
\eq
     \epsilon^{j_{1} j_{2}}_{j_{3}} = (-1)^{j_{1} + j_{2} - j_{3}} ~,
\label{sutwosign}
\en
cannot be  directly used in ${\rm SU}(2)$ Chern-Simons theory since they
do not, in fact, obey the crossing constraint
listed in eq.~\ref{chernsign}.

A crossing-symmetric system of signs that also obeys the fusion
constraint can, however, be directly
obtained in terms of a certain, natural system of group theory signs.
The group theory signs are defined by
\eq
      {\cal P}^{ab} \cC^{ab}_{c} = \epsilon^{ab}_{c} \cC^{ba}_{c} ~~,
\label{epdeftwo}
\en
where $\cC^{ab}_{c}$ is the transpose of
the matrix of Clebsch-Gordan
coefficients $\cC_{ab}^{c} : V_{a} \otimes V_{b} \longrightarrow V_{c}$.
Again, if $a=b$ these signs are intrinsic but if $a\neq b$ we may
choose them at will.  There is, however, a natural choice for
these latter signs.
In ref.~\mite{newp} this natural
sign system is obtained for all compact, simple
Lie groups and all tensor products.
The centerpiece for this system
is a general formula for $\epsilon^{ab}_{c}$ (in terms of the highest
weight vectors of the representations $a$, $b$, and $c$ alone)
in the case where
$c$ occurs in the tensor product $a\otimes b$ with no multiplicity
\eq
{}~~~~~\epsilon^{ab}_{c} =
(-1)^{\half (\lambda  | a + b - c) }
           ~~~~~~~~({\rm no~multiplicity}).
\label{classepdef}
\en
Here $\lambda$ is the level vector for the group $G$ (defined by
the condition that $(\lambda | \alpha) = 1$ for all simple
\setcounter{footnote}{1}
roots $\alpha$ of $G$).\footnote{If $\alpha$ is long, we set
$(\alpha | \alpha) = 2$, in which case the level vector differs
from the sum of positive roots only if $G$ is non-simply-laced.}
(Since $\lambda = \alpha$ for ${\rm SU}(2)$, eq.~\ref{sutwosign}
turns out to be a particular case of eq.~\ref{classepdef}.)
The problems that arise in the case of
multiplicities are discussed in the technical note.
This is a {\em natural} system of signs
in the sense that $\epsilon^{ab}_{c}$ originates in the structure of
the embedding of $c$ in the tensor product $a \otimes b$ for $a\neq b$
in exactly the same way that the sign $\epsilon^{aa}_{c}$
originates in the embedding of $c$ in
$a \otimes a$ (as shown in ref.~\mite{newp} and illustrated in
the technical note), and corresponds to the
choice of normalization $\omega_{1} = \omega_{2} = \omega$ (with
$\omega$ a residual, undetermined sign).

This natural system of group theory signs satisfies the identities
\eq
\begin{array}{clcll}
1. & \epsilon^{ab}_{c} & = &  \epsilon^{a\rho(c)}_{\rho(b)}
                \epsilon^{a \rho(a)}_{0} & {\rm crossing} \\
2. & \epsilon^{\rho(a)\rho(b)}_{\rho(c)} & = &
                 \epsilon^{ab}_{c} & {\rm conjugation}   \\
3. & \epsilon^{c \rho(c)}_{0} & = &
           \epsilon^{a\rho(a)}_{0} \epsilon^{b\rho(b)}_{0}
                    ~~~~{\rm for}~c\in a\otimes b~~~ &   {\rm fusion}
\end{array}
\label{fuscon}
\en
The first identity shows that these signs do not satisfy
the crossing constraint
on the Chern-Simons signs in eq.~\ref{chernsign} when
$\epsilon^{a\rho(a)}_{0}$ is
negative.

While this natural system of permutation signs is not crossing
symmetric, the modified system given by
\eq
{}~~~~~\eta^{ab}_{c} = \epsilon^{ab}_{c} \epsilon^{c\rho(c)}_{0}
= (-1)^{\half ( \lambda | a + b + c)}  ~~~~~~~~({\rm no~multiplicity})
\label{cstoclass}
\en
does provide a consistent crossing-symmetric sign system for
Chern-Simons theory that satisfies all the
constraints of~\ref{chernsign}.

Since the fusion identity in eq.~\ref{fuscon}
implies that $\epsilon^{c\rho(c)}_{0} = 1$ if $a=b$
(\ie, if $c \in a\otimes a$), the
{\em intrinsic} Chern-Simons permutation signs equal the
{\em intrinsic} group theory  signs, as is well known.
Since this fusion identity
also implies that $\epsilon^{00}_{0} =1$, eq.~\ref{cstoclass} also
equates the Chern-Simons charge conjugation signs $\eta^{a\rho(a)}_{0}$
with the (natural) group theory signs $\epsilon^{a\rho(a)}_{0}$.

\vspace{0.6cm}
\noindent {\it The Natural Charge Conjugation Signs}
\vspace{0.3cm}

Since the natural charge conjugation
signs $\eta^{a\rho(a)}_{0}$ appear pervasively,
and since they often necessarily differ from
the common expectation that only pseudoreal representations require
a negative charge conjugation sign, we exhibit their values in detail.
These signs are given---in all cases since the
identity always appears with multiplicity one---by
eq.~\ref{cstoclass}, which reduces to
\eq
      \eta^{a\rho(a)}_{0} = (-1)^{ ( \lambda | a ) }
\label{ccexp}
\en
in this case
since $(\lambda | a ) = (\lambda | \rho(a) )$.
In most cases these signs are
actually completely determined by the fusion
constraint (or in any case fixed by being either
intrinsic or related by the fusion constraint to an
intrinsic sign)
so that there is actually little
freedom of choice in Chern-Simons theory for these signs.
\setlength{\unitlength}{0.2cm}

Since ~$\fund = \rho(\fund)$ for $\ason$ and $\spn$, we
\setlength{\unitlength}{0.15cm}
must have $\eta^{\fund ~\!\rho(\fund)}_{0} =1$ for $\ason$ (which also
follows necessarily from the fusion constraint)
and $\eta^{\fund ~\!\rho(\fund)}_{0} =-1$ for $\spn$ (which does not).
Then eq.~\ref{fusimp} requires, in agreement with eq.~\ref{ccexp}, that
\eq
  \eta^{a\rho(a)}_{0} =  \left\{ \begin{array}{cc}
            1 &  \ason   \\
          (-1)^{r(a)} & \spn
             \end{array}    \right.
\label{ccperm}
\en
for all tensor representations $a$.\footnote{
Only for the (complex) tensor representations of
${\rm so}(4n+2)$ with self-associate Young
tableaux is eq.~\ref{ccperm} a constraint
beyond that required by pure group theory.}
For ${\rm so}(8n+4 \pm 1)$ (${\rm so}(8n+4)$) with $n=0, 1, \ldots$
the fundamental spinor(s) is(are) pseudoreal.
Since $\eta^{\psi \rho(\psi)}_{0} = - 1$ in these cases, eqs.~\ref{fusimp}
and~\ref{ccperm} then imply that the charge conjugation
sign is necessarily negative for all spin-tensor representations $\{\psi;a\}$
of these groups. However, for ${\rm so}(8n + 6)$ the fundamental
spinors are complex and we choose (though this
is not required by eq.~\ref{chernsign}) the charge conjugation sign
to be $-1$ in accordance with eq.~\ref{ccexp}.

For $G_{2}$, $F_{4}$, $E_{6}$, and $E_{8}$, the Chern-Simons constraints
in eq.~\ref{chernsign} alone fix all charge conjugation signs to be positive.
For $E_{6}$ this represents a constraint beyond pure
group theory, since the fundamental representation
is complex. For $E_{7}$ the
charge conjugation sign of the fundamental is necessarily negative since
this representation is pseudoreal. While this fixes all other charge
conjugation
signs via eqs.~\ref{chernsign}, these are all self-conjugate and
so intrinsic.

For $\sun$  the natural sign
for the fundamental is
\setlength{\unitlength}{0.15cm}
$\eta^{\fund \rho(\fund )}_{0} = (-1)^{N-1}$.
In fact, if $N = ({\rm odd~number})$ the fusion constraint forces
$\eta^{\fund \rho(\fund )}_{0} = 1$, since the number of boxes
modulo two is not conserved by the tensor ring.
As noted previously, if $N=(2\times ~{\rm odd~number})$
then we must set
\setlength{\unitlength}{0.15cm}
$\eta^{\fund \rho(\fund)}_{0} = -1 $
due to the fusion constraint in~\ref{chernsign}.
On the other hand, for $N = (2\times ~{\rm even~number})$
the natural sign for the fundamental (which is negative)
is not intrinsic and not required by the fusion constraint.
We will adopt---for all values of $N$---the natural system
for the charge conjugation signs, so that
\eq
  \begin{array}{cccc}
     \eta^{a\rho(a)}_{0} & = &  (-1)^{(N-1)r(a)}  & ~~~~\sun
 \end{array}
\label{ccpermsu}
\en
for all representations $a$.

In the following table we summarize these results. In all cases
the representations $a$ with $\eta^{a \rho(a)}_{0} =1$
form a closed (sub)ring. The fusion constraint alone
actually forces the positive value on this (sub)ring.
In most other cases the fusion constraint connects the
charge conjugation sign of
the remaining representations to an intrinsic sign.
The (remaining free) choice of the signs for the fundamental
representation of ${\rm SU}(4n)$
and the fundamental spinors of ${\rm so}(4n+2)$ then leaves all signs fixed.
Note that the appearance of minus signs is
(necessarily) not restricted to the pseudoreal case.

\begin{center}
\begin{tabular}{|l|cc|}    \hline
 Group   & \multicolumn{2}{c|}{$\eta^{a\rho(a)}_{0}$}
                                   \\ \hline\hline
${\rm SU}(2n)$, ${\rm Sp}(N)$   &\multicolumn{2}{c|}{$(-1)^{r(a)}$}
                                     \\ \hline
${\rm SU}(2n+1)$ & \multicolumn{2}{c|}{$1 $}             \\ \hline
$G_{2}$, $F_{4}$, $E_{6}$, $E_{8}$
        &\multicolumn{2}{c|}{$ 1$}   \\ \hline
$   E_{7}$       &$\begin{array}{lr}
                       {\rm ~real} &  ~~~~1 \\
                      {\rm  ~pseudoreal} & ~~~~-1
                          \end{array}$
                         &   ~                     \\ \hline
${\rm SO}(N_{+})$ &\multicolumn{2}{c|}{$1$}  \\ \hline
${\rm SO}(N_{-})$   & $\begin{array}{lr}
                       {\rm ~tensors} &  ~~~~1 \\
                      {\rm  ~spinors~~~} & ~~~~-1
                          \end{array}$
                       &  ~    \\ \hline
\multicolumn{3}{|l|}{$N_{+} = \{7,8,9,10\} + 8n$ and} \\
\multicolumn{3}{|l|}{$N_{-} = \{3,4,5,~~\! 6\} + 8n$ with
                                                 $n=0,1, \ldots$} \\ \hline
\end{tabular}
\end{center}

\vspace{0.6cm}
\noindent {\it
2.3 Non-linear Skein Relations for Planar Tetrahedra}
\vspace{0.3cm}

Since the vectors $\ket{s}, \ket{t}, \ket{u}$ (defined in {\it 2.1})
form different bases of the same space, we can
expand any given basis vector in terms of the other bases.
\eq
\begin{array}{ccc}
   \ket{s}  =  \ds \sum_{t} F_{st} \ket{t} ~, &
   ~\ket{s}  =  \ds \sum_{u} (\varrho^{cd}_{s})^{-1} ~G_{su} \ket{u} ~, &
   ~\ket{t}  =  \ds \sum_{u} (\varrho^{tc}_{a})^{-1}
                \varrho^{\rho(c)b}_{\rho(u)}  ~H_{tu} \ket{u}
\end{array}
\label{tetradef}
\en
The entries in the $f \times f$ matrices
$F$, $G$, and $H$ are just the expectation values of planar tetrahedra
\eq
  \begin{array}{ccc}
F_{st} & = & {\punknot{t} \over \smduubaryon{a}{t}{c} \smddubaryon{b}{t}{d} }
                \tetra{s}{t}{a}{b}{c}{d}   \\
         &   &          \\
         &   &          \\
  G_{su} & = &  {\punknot{u} \over \smduubaryon{a}{u}{d} \smddubaryon{b}{u}{c}
}
                 \tetra{s}{u}{a}{b}{d}{c}  \\
          &  &                  \\
         &   &          \\
  H_{tu} & = &  {\punknot{u} \over \smduubaryon{a}{u}{d} \smddubaryon{b}{u}{c}
}
                 \revtetra{t}{u}{a}{c}{d}{b}
  \end{array}
\label{tetradisp}
\en
\vspace{1.0in}

\noindent Note that the edge orientations on the baryons
imply (via eq.~\ref{arrbarynorm}) the possible presence of minus signs
in the baryon expectation values.

Explicit calculation of the inverse transformations along with the
orthogonality and completeness of the three bases
shows that the matrices of tetrahedra satisfy\cite{martin}
\eq
\begin{array}{rcr}
F F^{T} = I &   ~~~~ &
{}~F F^{\dagger} = I \\
G G^{T} = I &   ~~~~ &
  ~G G^{\dagger} = I \\
H H^{T} =  I &   ~~~~ &
  ~H H^{\dagger} = I
\end{array}
\label{orthounit}
\en
{}From these equations one concludes that $F$, $G$, and $H$ are real matrices.
Written out, the remaining independent equations for these real matrices are
\eq
\begin{array}{ccc}
  \ds  \sum_{s} F_{s t} F_{s t^{\prime}} = \delta_{t t^{\prime}}~~~
 & \ds  \sum_{u} G_{s u} G_{s^{\prime} u} = \delta_{s s^{\prime}}~~~
 &  \ds  \sum_{t} H_{t u} H_{t u^{\prime}} = \delta_{u u^{\prime}}
\end{array}  ~~.
\label{inveq}
\en
These constitute $3f(f+1)/2$ equations for $3f^{2}$ real unknowns and so
do not by themselves contain enough information to solve for the tetrahedra.
The associativity of the basis change
operations (\ref{tetradef}) gives the further set of equations\cite{martin}
\eq
     F_{st} = \sum_{u}
(\varrho^{cd}_{s} \varrho^{a\rc}_{t}
            \varrho^{\rho(b) c}_{u}  \cF^{-1}_{c})  G_{su} H_{tu}
\label{keyeq}
\en
where $\cF_{c}= q^{{\sct {1 \over 2}} Q(c)}$ is the vertical-framing factor
incurred in undoing a self-crossing of a Wilson line
carrying the representation $c$.
This equation and its complex conjugate provide $2f^2$ further
constraints, so that we have in general more constraints than unknowns.
These conditions ostensibly overdetermine the tetrahedra, given that no
unforeseen degeneracies in the braid eigenvalues occur.\cite{martin}

However, eqs.~\ref{inveq} and~\ref{keyeq} have certain discrete
symmetries, a fact which  is both necessary and problematic.
For if we have generic tetrahedra ($a$, $b$, $c$, and $d$ all different),
then changing the residual normalization $\omega$ of the coupling
$\cK^{ab}_{s}$ at the common vertex connecting
$a$, $b$, and $s$ in $F_{st}$ and $G_{su}$, for example,
will change the sign of  a row of $F_{st}$ and of $G_{su}$, but
the non-linear equations will remain unchanged. Since these
equations hold for arbitrary tetrahedra, they must allow for
and not determine this sign ambiguity, since this is entirely
a matter of the arbitrary choice of sign of
the vertex normalization.
However, certain tetrahedra have
intrinsic signs which are then not determined by these
equations even though their signs are independent of vertex normalization.
Primary among these latter tetrahedra are the
{\it link-type} tetrahedra, which have $d=a$ and $c=b$:
\eq
            \cptetra{s}{t}{a}{b}{b}{a}
\en
\vspace{1.3cm}

\noindent  Here the vertices occur in dual pairs whose relative
normalization is fixed once the permutation signs and
baryon normalization are fixed. Therefore a change of residual vertex
normalization cancels and does not affect the overall sign of such
link-type tetrahedra. Since this sign is intrinsic and
since the non-linear equations cannot determine it
we need a further prescription
that will enable the calculation of these intrinsic signs.

\vspace{0.6cm}
\noindent {\it 2.4 Linear Skein Relations for Planar Tetrahedra}
\vspace{0.3cm}

The problem of determining the signs of link-type tetrahedra
that remain unfixed by the (non-linear) orthogonality and
associativity equations is solved by appeal to
the following inhomogeneous linear equations.  While they are
special cases of the general construction valid for all
link-type graphs described in section~6, they also follow
directly from the definitions of the tetrahedra
in eq.~\ref{tetradef}.  By acting with
$B_{ab}$ and $B^{-1}_{ab}$ on the expansion
of $\ket{s}$ in terms of $\ket{t}$, and vice
versa, we obtain, from the braid closure of the resulting
diagrams, the $4f$ equations for the $f^2$
quantities $F_{st}$
\eq
\begin{array}{ccc}
\ds    \smdudbaryon{a}{s}{b} ~ ({\varrho^{ab}_{s}})^{\pm 1} & =
& \ds \sum_{t} F_{st} ~({\varrho^{a\rho(b)}_{t}})^{\mp 1}
                                ~\smuudbaryon{b}{t}{a} \\
          &          &                    \\
         &         &            \\
    \smuudbaryon{b}{t}{a}  ~ ({\varrho^{a\rho(b)}_{t}})^{\pm 1} &
         = & \ds \sum_{s} F_{st}
           ~({\varrho^{a b}_{s}})^{\mp 1} ~ \smdudbaryon{a}{s}{b}
\end{array}
\label{blineq}
\en
\vspace{1cm}

\noindent From the results of section {\it 2.2} and the
technical note we find that

\eq
{\hsmdudbaryon{a}{s}{b} \over \smuudbaryon{b}{t}{a}}
 = \eta^{b\rho(b)}_{0} \sqrt{ {\chi_{q}(s) \over \chi_{q}(t) }}
\en
\vspace{1.0cm}

\noindent Then the equations in~\ref{blineq} can be written in
the form
\eq
\begin{array}{rcl}
    ({\varrho^{ab}_{s}})^{2} \sqrt{\chi_{q}(s)} & =
& \ds \sum_{t} ~(\eta^{b\rho(b)}_{0}
              \varrho^{ab}_{s} F_{s t} \varrho^{a \rho(b)}_{t})~
  \sqrt{\chi_{q}(t)} ({\varrho^{a\rho(b)}_{t}})^{- 2}  \\[0.3cm]
    ({\varrho^{a\rho(b)}_{t}})^{2} \sqrt{\chi_{q}(t)} & =
& \ds \sum_{s} ~(\eta^{b\rho(b)}_{0}
  \varrho^{ab}_{s} F_{s t} \varrho^{a \rho(b)}_{t})~
       \sqrt{\chi_{q}(s)} ({\varrho^{ab}_{s}})^{- 2}
\end{array}
\label{lineq}
\en
(or, using the fact that the $F_{st}$ are
real, in the form of the complex conjugates of these equations).
For the following, the quantity
\eq
\eta^{b\rho(b)}_{0}
\varrho^{ab}_{s_{i}} F_{s_{i} t_{j}} \varrho^{a \rho(b)}_{t_{j}}
\label{keyquant}
\en
will then be of prime interest.  The non-linear equations
(\ref{inveq} \&~\ref{keyeq}) determine the absolute value of $F_{st}$;
the linear equations (\ref{blineq}) fix the remaining sign ambiguity.
The signs of the link-type tetrahedra
$F_{st}$ clearly depend on the choice of a system
of permutation signs and baryon normalizations.
However, from eq.~\ref{lineq} it is seen that the quantity in
eq.~\ref{keyquant} only depends on the {\em squares} of braid
eigenvalues and on the $q$-dimensions. It not only does not
depend on the residual vertex normalization signs;
it also does not depend on any choice of a particular system of
permutation signs.

If we think of the freedom to choose the vertex normalization
(and so the permutation signs) as a
local, discrete gauge symmetry (in which guise it does appear
in integrable lattice models) then the product of tetrahedra and
eigenvalues in eq.~\ref{keyquant} is a natural gauge invariant quantity.

In section 5 we shall see that
the expressions in~\ref{keyquant} are
essentially the non-planar tetrahedra that equal the matrix elements
of a class of WZW braid matrices.

For $f=1, 2, 3,$ and $4$ the equations
in~\ref{blineq} alone determine $F_{st}$, and an explicit
general solution for all link-type tetrahedra
is possible in these cases without appeal to the non-linear equations.

For $f=1$ one finds that
\eq
 F_{st} =
\eta^{b\rho(b)}_{0} \sqrt{ {\chi_{s} \over \chi_{t} }}
                   \varrho^{ab}_{s} \varrho^{a\rho(b)}_{t}  ~~.
\label{fone}
\en
Since $\chi_{t} = \chi_{s}$ when $f=1$, this simplifies to
\eq
{}~~~~F_{st} =
\eta^{b \rho(b)}_{0} \varrho^{ab}_{s} \varrho^{a \rho(b)}_{t}
{}~~~~~~~~~~(f=1).
\en
In addition, the non-linear, orthogonality condition
(\ref{inveq}) yields $F_{st}= \pm 1$ for all tetrahedra if $f=1$.
Therefore, we must have $\varrho^{ab}_{s} \varrho^{a\rho(b)}_{t}
= (\varrho^{ab}_{s} \varrho^{a\rho(b)}_{t})^{-1} = \pm 1$, which
constrains the values of the quadratic Casimirs appearing in
an $f=1$ fusion rule.

For $f=2$, with an arbitrary ordering of the two terms in each of
the fusion rules
$\phi_a \cdot \phi_b = \phi_{s_{1}} + \phi_{s_{2}}$
and $\phi_a \cdot \phi_{\rho(b)} = \phi_{t_{1}} + \phi_{t_{2}}$, we find that
\eq
     F_{s_{1} t_{1}} =
\eta^{b\rho(b)}_{0} \sqrt{ {\chi_{s_{1}} \over \chi_{t_{1}} }}
  \varrho^{ab}_{s_{1}} \varrho^{a \rho(b)}_{t_{1}}
{ 1- (\varrho^{ab}_{s_{1}})^{-2} (\varrho^{a \rho(b)}_{t_{2}})^{-2}
   \over 1- (\varrho^{a\rho(b)}_{t_{1}})^{2} (\varrho^{a\rho(b)}_{t_{2}})^{-2}}
{}~~~~~~(f=2)
\label{lasord}
\en

In these solvable cases we see explicitly that $F_{st}$
is a rational function of braid eigenvalues with an overall sign
that depends in a complicated way on the permutation signs,
baryon normalization, and the form of this function.
In contrast, the quantity in
eq.~\ref{keyquant} is a rational function of the {\em squares}
of the braid eigenvalues and does not depend on the system of
permutation signs or the baryon normalization.  While
its value depends on the values of the Casimirs and the structure
of this function, this information is just that encoded in the
structure of eq.~\ref{lineq}.
The same will be true for any $f$ by appeal to the whole system
of non-linear and linear equations (except that an explicit solution of these
equations will not be generally available).
This means that  the symmetries of this combined set of equations will
be exact symmetries of the combination $\eta^{b\rho(b)}_{0}
\varrho^{ab}_{s_{i}} F_{s_{i} t_{j}} \varrho^{a \rho(b)}_{t_{j}}$
for all consistent choices of permutation signs and baryon
normalization.

Although a special case, the link-type tetrahedra are important
because they are the only tetrahedra that directly support knot
and link invariants. They will also be important in section 6
where their special properties lead to proofs of certain
identities for more complex graphs.

Since the combined skein relations in eqs.~\ref{inveq},~\ref{keyeq}
and~\ref{lineq} are maximally effective for the exact determination
of all tetrahedra, we can now explore the symmetries between
tetrahedra by examining the transformation properties of the
fusion coefficients, quadratic Casimirs, and permutation signs
appearing as coefficients of eqs.~\ref{keyeq} and~\ref{lineq}.

\vspace{2.0cm}
\noindent {\bf 3. Simple Currents, Cominimal Equivalence, and Planar
Tetrahedra}
\vspace{0.4cm}
\onward

The fusion ring of a level $K$ WZW model based on any simple, compact Lie
group $G$ arises as a quotient of the classical tensor ring of $G$ by a
certain ideal. Such a fusion ring has automorphisms of the form
\eq
 {N_{\sigma^{m}(a) \sigma^{n}(b)}}^{\sigma^{m+n}(c)}  =  {N_{a b}}^{c}
  ~~~{\rm with}~~ \sigma^{p}(r) = r  ~~{\rm for~all~} r
\label{comfusion}
\en
for some positive integer $p$
if and only if the center $Z$ of $G$ is nontrivial---with
exactly one exception (the $E_{8}$ level $2$ fusion ring).\cite{earlydd,fuchs}
Such automorphisms correspond to the presence of a
discrete (integer) charge\cite{schellekens} $\gamma$ that is conserved mod $p$
by the fusion product:
\eq
      {N_{a b}}^{c} \neq 0  \Rightarrow
       \gamma(c) = \gamma(a) + \gamma(b) ~~~~~~~~{\rm mod} ~p.
\label{comcharge}
\en
Curiously, these charges have---as we shall see---a completely
classical origin: the classical tensor ring has exactly
this mod $p$ conservation law, so that
the fusion ring (as a quotient) necessarily
inherits the same additive conservation law.
Since all known rational fusion rings (those with a finite
number of elements) are obtained from these WZW fusion rings
by forming further products and quotients and
since these automorphisms have profound implications\cite{schellekens} for
coset field identifications,\cite{tame,fieldident}
simple-current fixed-point resolution, and the
construction of modular invariant partition functions,\cite{sy} it
is of interest to study their exact consequences in
the associated Chern-Simons theories.
\newpage

Given rather general properties of any fusion ring,
eq.~\ref{comfusion} implies the following constraint
on the conformal dimensions\cite{schellekens}
for an automorphism $\sigma$ of order $p$:
\eq
   h(\sigma(a)) = h(a) + h(\sigma(0)) - k(a)/p
\label{simpcur}
\en
for some integer $k(a)$ (unknown at this level
of generality). If we define
\eq
    \gamma(a) = k(a) ~~~~~~{\rm mod}~p
\en
then eq.~\ref{comcharge} is satisfied. It will
be useful to define the fractional charge
\setcounter{footnote}{1}
$q(a) = \gamma(a)/p$ which is conserved
mod~one.\footnote{This charge $q(a)$ is
the same fraction ($|q(a)|<1$) defined in ref.~\mite{schellekens}
via the leading  pole in the conformal block of $\phi_{a}$ appearing
in an operator product expansion, although there the integer
ambiguity  is compounded with the question of which
field in the conformal block occurs as the leading pole.}
Since $q(\rho(a)) = - q(a)$ we have the braid eigenvalue
identities (from eq.~\ref{breigdef})
\eq
  \begin{array}{lcl}
\varrho^{\sigma(a) b}_{\sigma(s)} & = & (\pm)_{s} ~e^{i\pi q(b) }
                                           ~~\varrho^{a b}_{s} \\[0.3cm]
\varrho^{\sigma(a) \rho(b)}_{\sigma(t)} & = &
   (\pm)_{t} ~e^{-i\pi q(b)} \varrho^{a \rho(b) }_{t}
   \end{array}
\label{comgcfdim}
\en
where the undetermined sign depends on $s$ (or $t$).
In addition, it follows from
general constraints and eq.~\ref{comfusion} that the $q$-dimensions (given
that they are real, linear functions on the fusion ring) must
satisfy\cite{schellekens}
\eq
      \chi_{q}(\sigma(a)) = \chi_{q}(a)    ~~.
\label{comgqdim}
\en

Precise results for the transformation properties of
the tetrahedral quantity in eq.~\ref{keyquant},
as well as those of linked unknots, will only require the three
constraints, eqs.~\ref{comfusion},~\ref{comgcfdim}, and~\ref{comgqdim},
and will not depend on any choice of a system of permutation signs.
Therefore, these results will hold in {\em any} Chern-Simons theory
with simple current symmetries.

In order to obtain analogous results for (linked) {\em knots},
or to calculate the braid eigenvalues and permutation sign in
eq.~\ref{keyquant}, we will need to know the integers $k(a)$ exactly.
In addition we will need some understanding of how the permutation signs
transform under the fusion rule automorphisms. We provide this level
of precision only for Chern-Simons theories with compact, simply-connected
gauge groups.
In the cases $\sun$, $\spn$, $\asodn$, ${\rm so}(4n+2)$,
$E_{6}$, and $E_{7}$ the center is  cyclic and
isomorphic to the group generated by a single
automorphism $\sigma$ (of order $p = {\rm ~dim}\{ {\rm center}\}$) of the
extended Dynkin diagram of $G$ which permutes the affine vertex
with a vertex of the ordinary Dynkin diagram. For ${\rm so}(4n)$
the center is isomorphic to the group generated by two independent
automorphisms $\sigma_{1}$ and $\sigma_{2}$ each of order two.
The elements of the fusion ring are divided into
equivalence classes---we
call them\cite{nrs,recentus}
{\it cominimal equivalence classes}---by the map
between representations associated with the action of
the aforementioned diagram automorphism groups.

For the classical groups $\sigma$ has a natural interpretation in terms
of Young tableaux:
\begin{list}{label}{\setlength{\leftmargin}{1.8cm}
                    \setlength{\labelsep}{0.5cm}}
\item[$\sunk$]    ~

   $\sigma$ acts on a reduced tableau $a$ by adding a row
   of length $K$ to the top of $a$. (A tableau is reduced if it has no
   columns of length $N$.)
\item[$\spnk$]   ~

   $\sigma(a)$ denotes the complement of the tableau $a$ in an
   $N \times K$ rectangle.
\item[$\asodnK$] ~

   $\sigma$ maps a tableau with first row length $\row_{1}$ to a tableau
   with first row length $K-\row_{1}$ (but otherwise identical).
\item[${\rm so}(4n)_{K}$]  ~

   $\sigma_{1}$ has the same definition
          as the (tableau) map $\sigma$ for $\asodnK$.

   $\sigma_{2}$ denotes the complement of a tableau in an
   ${1 \over 2} N \times {1\over 2} K$ rectangle (with $N=4n$)
   if $\row_{1} \leq {1\over 2} K$ and if $\row_{N/2} \geq 0$; in general,
   $\row_{i}(\sigma_{2}(a)) +  \row_{N/2 +1- i}(a)  = {1\over 2} K $.
   If $K$ is odd $\sigma_{2}$ interchanges spinors and tensors.
\item[${\rm so}(4n+2)_{K}$] ~

   $\sigma = \sigma_{2} \circ \sigma_{1}$ is the composition of the two
   operations just defined for ${\rm so}(4n)$ (except that now
   $N=4n+2$ in the definition of $\sigma_{2}$); its order is four.
\end{list}
For the spin-tensors of ${\rm so}(2n+1)$ or ${\rm so}(2n)$ we add a column
of $n$ half-boxes to the Young tableaux in order to implement
the operations just described. (Consult the initial
paragraphs of the technical note for the translation between
Young tableau row lengths, the labels $\row_{i}$,
and Dynkin indices.)

Using the Dynkin numbering for the $E_{6}$ diagram, the
Dynkin indices of $\sigma(a)$, $a_{i}^{\sigma}$, are such that
$a_{1}^{\sigma} = a_{0}$; the $\ZZ_{2}$ generator
$\sigma$ for $E_{7}$ is unambiguous.

The representations related by the action of the groups generated by
these maps are usefully termed `cominimally equivalent'
since the quadratic Casimirs (so also the conformal dimensions),
fusion coefficients, and $q$-dimensions
have (by explicit
verification)
the transformation properties:
\eq
\begin{array}{ccl}
 h(\sigma(a))    & = &
\cases{ h(a) + \frac{(N-1)K}{2N} - \frac{r(a)}{N} & for $\sun$ \cr
                      ~         &          ~ \cr
        h(a) + \frac{1}{4} NK  - \frac{1}{2} r(a) & ~~~~$\spn$ \cr
                        ~     &       ~ \cr
        h(a) + \frac{1}{2} K  -  \row_{1}(a) & ~~~~$\asodn$   \cr
                ~         &   ~~\& ${\rm so}(4n)$ if $\sigma =\sigma_{1}$ \cr
        h(a) +  \frac{1}{16} NK - \frac{1}{2} r(a) &  ~~~~${\rm so}(4n+2)$ \cr
            ~        &        ~~\& ${\rm so}(4n)$ if $\sigma=\sigma_{2}$ \cr
   h(a) + \third 2K  - \third (2 ( \kappa | a) + 3 I_{1})
                              & ~~~~$E_{6}$ \cr
                            ~ &   ~  \cr
   h(a) + \frac{1}{4} 3K  -  \frac{1}{2}
(3 (\kappa | a) + 2 I_{2})
                                        & ~~~~$E_{7}$} \\
     &  &   \\
    {N_{\sigma^{m}(a) \sigma^{n}(b)}}^{\sigma^{m+n}(c)} & = &
      {N_{a b}}^{c}  \\
     &  &   \\
    \chi_{q}(\sigma(a)) & = & \chi_{q}(a)
\end{array}
\label{propeq}
\en
Here
$$
  r(a) = \sum_{i=1}^{{\rm rank}\{G\}} \row_{i}(a) ,
$$
and the $\row_{i}$ are given in terms of the Dynkin indices in the
technical note.
For all representations $a$ of $\sun$ or $\spn$,
$r(a)$ equals the number of boxes in the associated Young tableaux.
For any (tensor or spinor) representation of ${\rm so}(2n)$ the quantity
$r(a)$ and the number of boxes $r^{y}(a)$ of
the diagram of $a$ are related by
\eq
     r(a) = r^{y}(a) - 2 \nu |\row_{N/2}(a)| ~~.
\en
Here $\nu$ equals zero (one) if $a_{n} > a_{n-1}$ ($a_{n} < a_{n-1}$).
The diagram of a spin-tensor
$\{\psi;a\}$ is formed by adjoining a column of $n$ half-boxes
to the Young tableau for $a$, so that
$r^{y}(\{\psi;a\}) = r^{y}(a) + n/2$.
While for tensors of ${\rm so}(N)$ the label $\row_{1}$ is
an integer (the first Young tableau
row length), for spin-tensors $\{\psi;a\}$ it is not:
$\row_{1}(\{\psi;a\}) = \row_{1}(a) + \half$.
With Dynkin numbering for the $E_{6}$ and $E_{7}$ Dynkin indices, we
have set
$$
\begin{array}{cc}
   I_{1} =   2 a_{3} + a_{4} + a_{6},~~    &
  {\rm and~~}I_{2} = a_{1} + 2 a_{2} + 3 a_{3}  + a_{4} + a_{5} ~.
\end{array}
$$
In addition, the product of the (congruence) vector\cite{patera} $\kappa$ with
the highest weight of $a$ is explicitly given
by $ (\kappa | a) = a_{1} + 2 a_{2} + a_{4} + 2 a_{5}$ for $E_{6}$,
and by $(\kappa | a) = a_{4} + a_{6} + a_{7}$ for $E_{7}$.
\newpage

\vspace{0.6cm}
\noindent {\it The Origin of Simple--Current Charges}
\vspace{0.3cm}

The formulae for $h(\sigma(a))$ provide explicit expressions for
the simple current
charges. Inspection of these expressions shows that, in all cases
of WZW simple currents (apart from $E_{8}$ level $2$),
the simple-current charges $\gamma$ coincide exactly with
the congruence classes of the group $G$. The congruence class of
a representation $a$ of $G$ is given by $(\kappa | a)$ mod $p$ (where
$p$ is the order of the relevant automorphism generator discussed above
and $\kappa$ is the associated congruence vector). The importance of
these classes stems from the fact that, for all $c \in a \otimes b$,
\eq
   (\kappa | c ) =  (\kappa | a ) + (\kappa | b )
               ~~~~{\rm mod}~p ~.
\en
This is the conservation law of the tensor ring that is inherited
by the fusion ring.
In the case of the classical groups these
congruence classes have simple Young tableau interpretations,
which  clearly indicates their origin in the $\gln$ tensor ring.
\begin{list}{label}{\setlength{\leftmargin}{1.8cm}
                    \setlength{\labelsep}{0.5cm}}
\item[$\sun$]  ~

   $\gamma(a)= r(a)$ mod $N$ ~distinguishes the well-known $N$-ality classes.
\item[$\spn$] ~

   $\gamma(a)= r(a)$ mod $2$ ~distinguishes real ($r(a)$ even) from pseudoreal
   ($r(a)$ odd) representations.
\item[$\asodn$] ~

   $\gamma(a) = 2\row_{1}(a)$ mod 2  ~indicates
   whether $a$ is a tensor ($\row_{1}(a)$ integer) or
   spinor ($\row_{1}(a)$ half-integer) representation.
\item[${\rm so}(4n)$] ~

   $\gamma_{1}(a) = 2\row_{1}(a)$ mod $2$  ~again distinguishes
   tensor and spinor representations.

   $\gamma_{2}(a) = r(a)$ mod $2$  ~indicates whether
   the number of boxes associated with a tensor representation
   is even or odd. For spinor representations it indicates
   whether the number of diagram boxes plus $\nu$ is even or odd.
   (The relation to the standard definition of $\kappa$
   is  that $r(a) {\rm ~mod~} 2 = \half [ (\kappa | a) {\rm ~mod~} 4]$
   with $(\kappa | a) = 2 r(a)$.)
\item[${\rm so}(4n+2)$] ~

   $\gamma(a) = 2 r(a)$ mod $4$  ~differentiates tensors ($\gamma = 0,2$)
   with an even ($\gamma=0$) or odd ($\gamma=2$) number of boxes
   from spinors ($\gamma = 1,3$) with the number of boxes
   plus $\nu$ even or odd.
\item[$E_{6}$] ~

   $\gamma(a) = 2 (\kappa | a)$ mod $3$ coincides with the
                  standard triality classes $\{0, 1, 2\}$,
                except that the labels $1$ and $2$ are interchanged.
\item[$E_{7}$] ~

   $\gamma(a) = 3 (\kappa | a)$ mod $2$ coincides with the duality which
   distinguishes real and pseudo-real representations (except that
   the labels $0$ and $1$ are exchanged).
\end{list}
If we define new congruence vectors for $E_{6}$ ($\kappa^{\prime}=2\kappa$),
$E_{7}$ ($\kappa^{\prime}=3\kappa$) and $\sigma_{2}$ of
${\rm so}(4n)$ ($\kappa_{2}^{\prime} = \half \kappa_{2}$)
then in all cases
\eq
           \gamma(a) = (\kappa | a) ~~~~{\rm mod~} p.
\en
For the classical groups we see that all the tensorial charges originate
(ultimately) in the exact conservation of the number of Young tableau boxes
(\ie, in the fact that $c\in a\otimes b \longrightarrow r(a) + r(b) = r(c)$)
exhibited by the $\gln$ or ${\rm U}(N)$ tensor rings.  In each case the
classical
tensor ring of $G$ conserves the charge mod a number that reflects how
$G$ is defined as a subgroup of $\gln$ or ${\rm U}(N)$. For example, the
various mod $2$ quantities result from the existence of
an invariant tensor that implements contractions of tensor indices
two at a time.
For the orthogonal groups the charge $\nu$ distinguishes the two
irreducible representations of ${\rm so}(N)$ that branch from
self-associate representations of ${\rm O}(N)$. The
distinction between tensors and spinors is the
only additional concept (which can be seen as exact conservation of half
boxes).

\vspace{0.6cm}
\noindent {\it Permutation Signs and Cominimal Equivalence}
\vspace{0.3cm}

The remaining ambiguity in eq.~\ref{comgcfdim} arises from the
permutation signs. For any given sign $\eta^{ab}_{c}$
eq.~\ref{cstoclass} rapidly yields its value if $c$
appears in the {\em tensor} product $a \otimes b$ with multiplicity one.
Difficulties can arise if one tries to obtain general formulae from
this equation alone because it is difficult to implement the condition
of no multiplicities. In the case of $\sun$, for example, with
\eq
      \Delta^{ab}_{c} = {r(a) + r(b) - r(c) \over N}  ~~,
\label{sundelta}
\en
we find (from eq.~\ref{cstoclass}) that
\eq
\begin{array}{ccl}
  \eta^{\sigma(a) b}_{\sigma(c)} &  = &
  e^{i\pi (K(N+1)+r(b) + N\Delta^{ab}_{c})} \eta^{ab}_{c} ~~,
\end{array}
\label{eppropeq}
\en
{\em if} $c$ appears with multiplicity one in the decomposition
of $a \otimes b$, and the same for $\sigma(c)$ in $\sigma(a) \otimes b$.
The iteration of eq.~\ref{eppropeq} (or direct calculation)
in the case $b=\sigma(a)$ and $c=\sigma(s)$ then leads to
\eq
\eta^{\sigma(a) \sigma(a)}_{\sigma^{2}(s)} =
e^{i\pi (r(a)+r(\sigma(a)) +
N\Delta^{aa}_{s} + N\Delta^{a\sigma(a)}_{\sigma(s)})} \eta^{aa}_{s}
{}~~~~~~(\sunk)
\label{sunsign}
\en
which holds exactly if both $s$ is multiplicity free in $a \otimes a$ and
$\sigma^2(s)$ is multiplicity free in $\sigma(a) \otimes \sigma(a)$.
If $N$ is odd, then eq.~\ref{sunsign}  depends on the intermediate channel
\eq
   \eta^{\sigma(a) \sigma(a)}_{\sigma^{2}(s)} =
     e^{i\pi (K+\row_{N-1}(s))} \eta^{aa}_{s}
        ~~~~~~(N~{\rm odd;~multiplicity~free})
\en
However, exactly this dependence also arises from the transformation
of the conformal dimensions (\ref{propeq}) in the braid eigenvalue so that
(in agreement with eq.~\ref{comgcfdim})
\eq
     \varrho^{\sigma(a) \sigma(a)}_{\sigma^2(s)}
=  e^{i\pi { N+1 \over N} (r(\sigma(a)) + r(a))} ~\varrho^{aa}_{s}
{}~~~~~~(N~{\rm odd})
\label{mfeigsun}
\en
holds if both multiplicity free conditions hold.
For $N$ even, eq.~\ref{sunsign} reduces to
\eq
\eta^{\sigma(a) \sigma(a)}_{\sigma^{2}(s)} =
e^{i\pi K} \eta^{aa}_{s}    ~~~~~~(N~{\rm even;~multiplicity~free})
\en
and there is no intermediate channel dependence to cancel
that coming from the conformal dimensions. Then, given the
multiplicity free conditions,
\eq
     \varrho^{\sigma(a) \sigma(a)}_{\sigma^2(s)}
=  e^{i\pi (\row_{N-1}(a) + \row_{N-1}(s) )}
e^{i\pi { N+1 \over N}(r(\sigma(a)) + r(a))} ~\varrho^{aa}_{s}
{}~~~~~~(N~{\rm even})
\en
seems to contain some intermediate channel dependence.
Despite this appearance, this dependence is spurious.
In order to demonstrate this, and to get at the cases with
multiplicity, consider the following.
{}From the (independently proved) result in~\ref{comlinres} for a
trefoil-type knot with an odd number of
twists $t$ (with $t=\pm 3$ being the case of
the trefoil proper),
\eq
\trefoil{\!\!\!\!\!\! \sigma(a) ~~~~t~\left\{
                 {\raise 1.4cm \hbox{$~$}} \right. } =
   (\varrho^{\sigma(0) \sigma (0)}_{\sigma^2(0)})^{t}
   (\varrho^{\sigma(0) a}_{\sigma(a)})^{2t}
\trefoil{~a ~~~~t~\left\{ {\raise 1.4cm \hbox{$~$}} \right. }
\label{sigtref}
\en
\vspace{1.3cm}

\noindent we obtain
\eq
 \sum_{s} {N_{\sigma(a)\sigma(a)}}^{\sigma^2(s)}~
(\varrho^{\sigma(a)\sigma(a)}_{\sigma^2(s)})^{t} ~\chi_{q}(\sigma^2(s))
 = (\varrho^{\sigma(0) \sigma (0)}_{\sigma^2(0)})^{t}
   (\varrho^{\sigma(0) a}_{\sigma(a)})^{2t}
\sum_{s}  {N_{aa}}^{s}~
(\varrho^{a a}_{s})^{t} ~\chi_{q}(s)   ~~,
\label{firsteq}
\en
upon insertion of an $S$-channel spectral decomposition on each side
of eq.~\ref{sigtref}.
Since $\sigma^2(0)$ always appears in
$\sigma(0) \otimes \sigma(0)$ with multiplicity one we can use
eq.~\ref{sunsign} to find that
\eq
     \eta^{\sigma(0) \sigma(0)}_{\sigma^2(0)} = (-1)^{K} ~~,
\label{speceig}
\en
so that (using eq.~\ref{propeq})
\eq
\varrho^{\sigma(0) \sigma (0)}_{\sigma^2(0)}
(\varrho^{\sigma(0) a}_{\sigma(a)})^2
= e^{i\pi (N+1) \row_{N-1}(a)}
e^{i\pi {N+1 \over N} (r(\sigma(a)) + r(a))}  ~~.
\en
Using this result, eq.~\ref{firsteq} becomes
\eq
\sum_{s} \chi_{q}(s)  (e^{-i \pi h(s)})^{t} {N_{aa}}^{s}
   [\eta^{aa}_{s} - e^{-i\pi (r(a) + r(\sigma(a))+\Delta^{aa}_{s} +
             \Delta^{a\sigma(a)}_{\sigma(s)}+ (N+1)\row_{N-1}(a) )}
   \eta^{\sigma(a)\sigma(a)}_{\sigma^{2}(s)}] = 0.
\label{multproof}
\en
(Note that, unlike eq.~\ref{sunsign}, there is no factor of $N$
multiplying either $\Delta$.)
Consider first the terms with single-multiplicity representations $s$.
If $N$ is odd these vanish identically (due to eq.~\ref{sunsign}), while
if $N$ is even the expression in brackets reduces to
\eq
 \eta^{aa}_{s} [ 1 - e^{i\pi \row_{N-1}(s)} ] ~.
\label{curious}
\en
Now consider the terms where $s$ occurs with multiplicity
in the tensor product.
It is important to realize that, although
${N_{\sigma(a)\sigma(a)}}^{\sigma^2(s)} ={N_{aa}}^{s}$ implies
that (cf. eq.~\ref{lastlab})
$$
{N_{\sigma(a)\sigma(a)}}^{\sigma^2(s)} =
{N_{\sigma(a)\sigma(a)}^{+}}^{\sigma^2(s)} +
{N_{\sigma(a)\sigma(a)}^{-}}^{\sigma^2(s)} =
{N_{aa}^{+}}^{s} + {N_{aa}^{-}}^{s} = {N_{aa}}^{s}
$$
the map $\sigma$ has not been defined for symmetric versus
anti-symmetric copies in an intermediate channel. We would like
to define it so that the cases of multiplicities and no
multiplicities agree.
If eq.~\ref{multproof} is written with the
multiplicities of symmetric and anti-symmetric terms explicitly
displayed, it becomes an equation for the differences
${\Delta N_{ab}}^{s} = {N_{ab}^{+}}^{s} - {N_{ab}^{-}}^{s}$
\eq
 \begin{array}{ll}
\ds \sum_{s} \chi_{q}(s)  (e^{-\pi i h(s)})^{t}   \\
{}~~~~~~~~   \times ~[ {\Delta N_{aa}}^{s}
  - e^{-i\pi (r(a) + r(\sigma(a))+\Delta^{aa}_{s} +
             \Delta^{a\sigma(a)}_{\sigma(s)}+ (N+1)\row_{N-1}(a)  )}
   ~{\Delta N_{\sigma(a)\sigma(a)}}^{\sigma^2(s)}  ] = 0.
\end{array}
\label{endproof}
\en
If $N$ is odd we can allow the sum to just run over the terms
with tensor multiplicity greater than one,
while if $N$ is even we must include
the single-multiplicity terms.
Since this equation  must hold
for all integers $t$, it will hold  only if
both the bracketed expression in eq.~\ref{curious} vanishes
(for $N$ even) and if
$$
 {N_{\sigma(a)\sigma(a)}^{+}}^{\sigma^2(s)} -
{N_{\sigma(a)\sigma(a)}^{-}}^{\sigma^2(s)} = \pm
({N_{aa}^{+}}^{s} - {N_{aa}^{-}}^{s})
$$
(where the $\pm$ sign is just that in eq.~\ref{endproof}).
Then the map $\sigma$ can be extended so that
\eq
     \varrho^{\sigma(a) \sigma(a)}_{\sigma^2(s)}
=  e^{i\pi (N+1) \row_{N-1}(a)}
e^{i\pi { N+1 \over N}(r(\sigma(a)) + r(a))} ~\varrho^{aa}_{s}
=e^{i\pi {N+1 \over N}K }  e^{2\pi i  {r(a) \over N}  } ~\varrho^{aa}_{s}
{}~~~~~~(\sun, ~{\rm all~} N)
\label{newpat}
\en
holds, even in the case of fusion multiplicities.
While this agrees transparently
with the direct calculation in the case of
$N$ odd (\ref{mfeigsun}), for $N$ even
it implies the
\begin{quotation}
\noindent Proposition: $s$ multiplicity free in $a \otimes a$ and
$\sigma^2(s)$ multiplicity free in $\sigma(a) \otimes \sigma(a)$
imply that
$\row_{N-1}(s)$ and $\row_{N-1}(\sigma^2(s))$ are even.
\end{quotation}
The interesting contrapositive,
\begin{quotation}
  \noindent With $s\in a \otimes a$  and
      $\sigma^2(s) \in \sigma(a) \otimes \sigma(a)$ in ${\rm SU}(2n)$, if
$\row_{2n-1}(s)$ or $\row_{2n-1}(\sigma^2(s))$ is odd,
 then either $s$ occurs with multiplicity in $a \otimes a$,
      or $\sigma^2(s)$ occurs with multiplicity in
                            $\sigma(a) \otimes \sigma(a)$,
\end{quotation}
is a useful (though incomplete) diagnostic for the presence of $\sun$ tensor
ring multiplicities.
For the remaining groups we obtain the
relatively uncomplicated formulas
\eq
   \begin{array}{ccc}
\eta^{\sigma(a) \sigma(a)}_{\sigma^{2}(s)} & = & \eta^{aa}_{s} \times
   \cases{
  e^{i\pi NK} & for $\spn$  \cr
  1           & ~~~~$E_{6}$, $\asodn$ \& $\sigma_{1}$ of ${\rm so}(4n)$ \cr
  (-1)^{nK}   & ~~~~$\sigma_{2}$ of ${\rm so}(4n)$ \cr
  (-1)^{K} & ~~~~$E_{7}$ \& ${\rm so}(4n+2)$ ~~~.}
\end{array}
\label{orthsign}
\en
in the multiplicity free case. The orthogonal group results hold for
$a$ either a tensor or spinor. Examination of these values shows that
\eq
\eta^{\sigma(a) \sigma(a)}_{\sigma^{2}(s)} =
     \eta^{aa}_{s} (-1)^{(\kappa | \sigma(0))}  \times
  \cases{ (-1)^{\row_{2n}(s)}  &  ${\rm SU}(2n+1)$ \cr
           1                   &   otherwise}
\en
is correct in all cases. Explicitly,
\eq
    (\kappa | \sigma(0)) = K \times \cases{ 1 &  $\sun$ \cr
2  & $E_{6}$, ${\rm so}(2n+1)$, \& ${\rm so}(4n)$ if $\sigma = \sigma_{1}$ \cr
              3    &   $E_{7}$  \cr
                2n-1 & ${\rm so}(4n+2)$  \cr
               n   &   ${\rm so}(4n)$ if $\sigma = \sigma_{2}$  \cr
                           N &  $\spn$~~~.}
\label{sigintval}
\en
The trefoil-type-knot-based argument then yields, with
$p$ the order of the automorphism $\sigma$,
\eq
    \begin{array}{ccl}
   \varrho^{\sigma(a) \sigma(a)}_{\sigma^{2}(s)} & = &
e^{2\pi i q(a)}
e^{i\pi {p+1 \over p} (\kappa|\sigma(0))}
\varrho^{aa}_{s}
\end{array}
\label{braidfin}
\en
for all groups and all representations.
It also implies that if $s$ and $\sigma^2(s)$ are multiplicity
free in ${\rm so}(4n+2)$
Kronecker squares, then $\row_{1}(s)$ and $\row_{1}(\sigma^2(s))$
are even, and if multiplicity free in Kronecker squares of $E_{6}$, then
$a_{2}(s) + a_{6}(s)$ and $a_{2}(\sigma^{2}(s)) + a_{6}(\sigma^{2}(s))$
are even.

Only eqs.~\ref{comfusion}-\ref{comgqdim} will be needed for the arguments
in the rest of this section. This means that the following arguments
hold for {\em all} Chern-Simons theories that display simple-current
symmetries. Equation~\ref{braidfin} will, however,
be relevant for the results of section~6 that
involve knot expectation values.
\newpage

\vspace{0.6cm}
\noindent {\it Cominimal Equivalence of Tetrahedra}
\vspace{0.3cm}

We will henceforth display compatible
sets of $S$, $T$, and $U$ fusion rule channels without the fusion
coefficients written explicitly by assuming that the sums only
run over the representations with non-zero coefficients
(but with multiplicity).
For example, the fusion rules in eq.~\ref{threechan} can be
written compactly (with a slight abuse of notation) as
\eq
\begin{array}{ccc}
  S{\rm -channel} &    T{\rm -channel}  &  U{\rm -channel} \\
                     &                    &              \\
\ds
a \cdot b   =  \sum_{s} s~~~  &
\ds
a \cdot \rc =  \sum_{t} t~~~  &
\ds
a \cdot \rd =  \sum_{u} u \\
 &   &    \\
\ds
c \cdot d   =  \sum_{s} s~~~   &
\ds
\rb \cdot d =  \sum_{t} t~~~ &
\ds
\rb \cdot c =  \sum_{u} u
\end{array}
\label{nofuseq}
\en
The associated matrices of expectation values of tetrahedra
are real and satisfy non-linear equations that only depend on the coefficients
appearing in eq.~\ref{keyeq} (and on those in eq.~\ref{lineq}
for link-type tetrahedra).
We now consider the cominimally equivalent set of fusion rules
\eq
\begin{array}{ccc}
  S{\rm -channel} &    T{\rm -channel}  &  U{\rm -channel} \\
   &    &  \\
\ds
\sigma(a) \cdot b   =  \sum_{s} \sigma(s)  &
\ds
\sigma(a) \cdot \rc =  \sum_{t} \sigma(t)  &
\ds
\sigma(a) \cdot \rho(\sigma(d)) =  \sum_{u} u \\
   &    &  \\
\ds
c \cdot \sigma(d)   =  \sum_{s} \sigma(s)   &
\ds
\rb \cdot \sigma(d) =  \sum_{t} \sigma(t) &
\ds
{}~~~~~~\rb \cdot c =  \sum_{u} u
\end{array}
\en
The sums here run over the same representations that
appear in eq.~\ref{nofuseq}.
The pairs of fusion rules displayed here exist
due to eq.~\ref{comfusion}
(or the second equality in~\ref{propeq}) and due to the fact that
\eq
   \rho(\sigma(a))= \sigma^{-1}(\rho(a))  ~~.
\label{sigrho}
\en
Each channel provides a basis that spans a new
Hilbert space ${\cal H}_{\sigma}$ of the same dimensionality $f$
as ${\cal H}$. The tetrahedral expectation values satisfy eqs.~\ref{inveq}
and~\ref{keyeq}, but with the coefficients in~\ref{keyeq} now given by
{\large
\eq
          \varrho^{c\sigma(d)}_{\sigma(s)}
 \varrho^{\sigma(a) \rc}_{\sigma(t)} \varrho^{\rho(b) c}_{u}  \cF^{-1}_{c}
\label{prodeq}
\en  }

\noindent Insertion of the identities in eq.~\ref{comgcfdim} shows that
the complex part of the relative phases cancel so that
{\large
\eq
       \varrho^{c\sigma(d)}_{\sigma(s)}
 \varrho^{\sigma(a) \rc}_{\sigma(t)} \varrho^{\rho(b)c}_{u} \cF^{-1}_{c} =  \pm
\varrho^{cd}_{s}
 \varrho^{a \rc}_{t} \varrho^{\rho(b)c}_{u} \cF^{-1}_{c}
\en  }

\noindent This means that the coefficients in the equations that determine the
two sets of tetrahedra are equal up to sign, so that corresponding
tetrahedra appearing in the two cases are equal up to sign.
Pictorially, by going to the cominimally equivalent set of fusion rules
we have changed the representations around a particular
closed oriented loop uniformly by $r \rightarrow \sigma(r)$.
The same result holds in general for all other loops and for
iterations around each loop.
In all cases the channels that appear produce exactly the same set of
constraining equations up to sign.  The general result is that
\vspace{1cm}
\eq
 \tmtetra{\sigma^{n+m}(s)}{\sigma^{l-m}(t)}{\sigma^{n}(a)}{\sigma^{m}(b)}
            {\!\!\! \sigma^{n+m-l}(c)}{\sigma^{l}(d)}
    ~~~~~~~~~=~ \pm~ \ltetra{s}{t}{a}{b}{c}{d}
\label{sigres}
\en
\vspace{1.8cm}

This is all that one can say about the general case of
non-link-type tetrahedra, since the sign of each tetrahedron
is a matter of residual vertex normalization. For link-type tetrahedra
note that the coefficients in the equations
in eq.~\ref{lineq} are squares and so do not
depend on any of the undetermined signs in eq.~\ref{comgcfdim}.
Therefore, using eq.~\ref{sigres}, as well as eq.~\ref{comgcfdim}
in eq.~\ref{lineq}, we obtain the exact result
\eq
\varrho^{\sigma(a)b}_{\sigma(s)}~  \varrho^{\sigma(a) \rho(b)}_{\sigma(t)}~
\ltetra{\sigma(s)}{\sigma(t)}{\sigma(a)}{b}{b}{\sigma(a)}
  =  ~\varrho^{ab}_{s}~ \varrho^{a\rho(b)}_{t} ~\ltetra{s}{t}{a}{b}{b}{a}
\label{exacres}
\en
\vspace{1.9cm}

\noindent While the permutation sign $\eta^{b\rho(b)}_{0}$
cancels trivially in this particular case, the general result
$$
\begin{array}{c}
\!\!\!\!\!\!\!\!\!\!\!\!\!\!
\!\!\!\!\!\!\!\!\!\!\!\!\!\!\!\! \eta^{\sigma^{m}(b) \rho(\sigma^{m}(b))}_{0}
\varrho^{\sigma^{n}(a)\sigma^{m}(b)}_{\sigma^{n+m}(s)}~\!
\varrho^{\sigma^{n}(a) \rho(\sigma^{m}(b))}_{\sigma^{n-m}(t)}~
\tmtetra{\sigma^{n+m}(s)}{\sigma^{n-m}(t)}{\sigma^{n}(a)}{\sigma^{m}(b)}{\sigma^{m}(b)}{\sigma^{n}(a)}
  =  \eta^{b\rho(b)}_{0}
 ~\varrho^{ab}_{s} \varrho^{a\rho(b)}_{t} ~\ltetra{s}{t}{a}{b}{b}{a}  \\
  ~
\end{array}
$$
\vspace{1.3cm}
\eq
 ~~
\label{genexacres}
\en
contains contributions from these signs.
The product of these charge conjugation signs is given by
\setcounter{footnote}{1}
\eq
\eta^{\sigma^{m}(b) \rho(\sigma^{m}(b))}_{0}
     \eta^{b \rho(b)}_{0} = (-1)^{m (p-1) (\kappa |\sigma(0)) }  ~~.
\en

The complex parts of all the phases in eqs.~\ref{exacres} \&~\ref{genexacres}
cancel leaving only a calculable sign.
This relative sign results both from
the transformation property of the permutation signs and from
that of the conformal dimensions involved. While one might hope
that the permutation signs could be chosen in such a way that
the signs of tetrahedra are uniform within cominimal equivalence
classes, simple counter-examples exist
in which it is impossible to arrange this, even using all
possible freedoms of normalizations.

\vspace{2.0cm}
\noindent {\bf 4. Rank-level Duality of Planar Tetrahedra}
\vspace{0.4cm}
\onward

The aim of this section is to show that each tetrahedron
in an $\sunk$, $\spnk$, or $\asodnk$ theory
has at least one partner in $\sukn$, $\spkn$, or $\asodkn$, respectively,
with the same expectation value up to sign.
In the case of $\asodnk$ this only applies to tetrahedra
with tensor representations on all edges.
The results of section 3 then show that each tetrahedron is also dual to
entire cominimal equivalence classes of
tetrahedra that all have the same expectation value (up to sign).

The map between integrable representations of
$\gnk$ and its rank-level dual $\gkn$ given by tableau
transposition
\eq
           a \in \gnk \rightarrow \widetilde{a} \in \gkn
\label{transmap}
\en
connects representations with closely related
quadratic Casimirs, fusion coefficients, and $q$-dimensions,
as follows.
In order to treat $\sun$, $\spn$, and $\asodn$ in parallel,
we extend eq.~\ref{sundelta} by defining
\eq
    \Delta^{ab}_{c} = \left\{ \begin{array}{ll}
                         (r(a) + r(b) - r(c))/N &   \sun \\
                         r(a) + r(b) - r(c)     & \asodn  \\
                                 0              & \spn
                             \end{array}    \right.    ~~.
\label{newdeltdef}
\en
For all three groups the non-zero fusion coefficients of $\gnk$
are related to those of $\gkn$ by\cite{recentus,recentjapan}
\eq
        {N_{a b}}^{c} = {N_{\ta \tb}}^{\sigma^{\Delta^{ab}_{c}}(\tc)}
\label{fusdual}
\en
(Note that the na\"{\i}ve relation
${N_{a b}}^{c} = {N_{\ta \tb}}^{\tc}$ does {\em not}
hold in general for
the fusion coefficients.)
For any tensor representation\cite{nrs,abc,nacschnit} $a$
\eq
 h(a)_{{\sct \gnk}} + h(\tilde{a})_{{\sct \gkn}} =
 \cases{ {r(a) \over 2} (1- {r(a) \over NK}) & for $\sunk$ \cr
               {r(a) \over 2} & for $\asonk$ and $\spnk$  }
\en
In addition,\cite{recentus} if $s \in a \otimes b$,
\eq
   h(s)_{{\sct \gnk}} + h(\sigma^{\Delta^{ab}_{s}}(\tilde{s}))_{{\sct \gkn}}
=   \cases{ {r(a) + r(b) \over 2} (1- {r(a) + r(b) \over NK})
   + \Omega^{ab}(s) & for $\sunk$ \cr
               {r(a) + r(b) \over 2} - \Gamma^{ab}(s) & for $\asodnk$ \cr
                           & ~~~~\& $\spnk$  }
\label{greekdef}
\en
where $\Gamma^{ab}(s)$ denotes the number of contractions
(of tensor indices) needed to
obtain $s$ in the tensor product $a\otimes b$ and
\eq
  \Omega^{ab}(s)
= \sum_{i= K- \Delta^{ab}_{s}+1}^{K} \col_{i}(s)   ~~.
\label{realgrdef}
\en
where the $\col_{i}(s)$ are the column lengths
of the reduced tableau $s$.
Both quantities are
integers.
In addition, the $q$-dimensions satisfy\cite{nrs,modint}
\eq
    \{ \chi_{q}(a)\}_{\gnk} =  \{ \chi_{q}(\tilde{a})\}_{\gkn} ~~.
\label{rlprop}
\en
Without considering
the exact values of the permutation signs, these equations yield
the braid eigenvalue relations
\eq
       \varrho^{ab}_{s} \varrho^{\ta\tb}_{\sigma^{\Delta^{ab}_{s}}(\ts)}
               =  (\pm)_{s}  ~e^{\pi i \Phi(a,b)} ~~.
\label{confopdual}
\en
where the sign $(\pm)_{s}$ depends on the intermediate channel $s$ in a
complicated way.
Here, and henceforth,
\eq
       \Phi(a,b) = \left\{ \begin{array}{ll}
                   ~r(a) r(b)/NK~~~~~ & \sunk \\
                    ~0                 & \asonk ~\&~ \spnk ~~.
                 \end{array}
             \right.
\label{fivethree}
\en
The dual identities for the special
tetrahedral quantity in eq.~\ref{keyquant}
will only require these identities.

For the case of (linked) knots it will also be important to understand
how the permutation signs $\eta^{aa}_{s}$ and
$\eta^{\ta \ta}_{\sigma^{\Delta^{aa}_{s}}(\ts)}$ are related. We find that
\eq
\eta^{aa}_{s} \eta^{\ta\ta}_{\sigma^{\Delta^{aa}_{s}}(\ts)}  =
         \cases{ e^{i\pi r(a)} e^{i\pi \Omega^{aa}(s)} & for $\sunk$ \cr
                        e^{i\pi r(a)} e^{i\pi \Gamma^{aa}(s)}
                                   & for $\spnk$ \& $\asonk$}
\label{dualeps}
\en
In the technical note we obtain a proof
of~\ref{dualeps} for $\spnk$, $\asodnk$,  and
for $\sunk$ in the special case
$\Omega^{aa}(s)=0$ (which often occurs
for $\Delta^{aa}_{s} \neq 0$), if $s$ does not appear with
reduced multiplicity. While eq.~\ref{duallinres} implies
eq.~\ref{dualeps} without any such restrictions via an argument using the
$t$ twisted trefoil-type knot analogous to that of the
previous section, we do not have an independent
proof of eq.~\ref{duallinres} unless $t=\pm 1$, the case of
the twisted unknot,\cite{nrs}
\eq
  \left<\!\!{\raise 0.5cm \hbox{$~$}}  \!\!\figeight{a} \!~\right>_{\gnk}
    =  e^{i\pi r(a)} e^{\pi i \Phi(a,a)}
  \left<\!\!{\raise 0.5cm \hbox{$~$}} \!\!\mfigeight{\ta} \!~\right>_{\gkn}
{}~~.
\en
\vspace{0.8cm}

\noindent Insertion of an $S$-channel spectral decomposition on both sides
yields (by using eqs.~\ref{fusdual}-\ref{rlprop})
\eq
  \sum_{s} {N_{aa}}^{s} \chi_{q}(s) e^{-i\pi h(s)}
  [ \eta^{aa}_{s} - e^{-i\pi (r(a)+ \Omega^{aa}(s))}
    \eta^{\ta\ta}_{\sigma^{\Delta^{aa}_{s}}(\ts)} ] = 0
\en
which constitute two real equations for the difference between
$\eta^{aa}_{s}$ in
$\sunk$ and $\eta^{\ta\ta}_{\sigma^{\Delta^{aa}_{s}}(\ts)}$ in $\sukn$
(replace $\Omega$ with $\Gamma$ for the other groups).
If we use eq.~\ref{dualeps} for the terms
that appear with no reduced multiplicity (or with
$\Omega^{aa}(s) = 0$ for $\sun$), then we obtain two
equations for the cases with reduced multiplicity (or for those with
$\Omega^{aa}(s) \neq 0$). This yields
a proof of a restricted but infinite set of cases where
eq.~\ref{dualeps} holds for all $s$ and with $\Omega^{aa}(s) \neq 0$
in the case of $\sun$.
(That the map in
eq.~\ref{transmap} is undefined between symmetric
and anti-symmetric copies of the same tableau in an intermediate
channel gives one
the freedom to define it so that eq.~\ref{dualeps} will
continue to hold in the case of multiplicity.)

While the product of permutation signs {\em does} depend on certain
details of the intermediate channel,
the product of braid eigenvalues does not:
\eq
\varrho^{aa}_{s} \varrho^{\ta\ta}_{\sigma^{\Delta^{aa}_{s}}(\ts)}  =
            e^{i\pi r(a)} e^{\pi i \Phi(a,a)}
\label{exdualbr}
\en
That the product does not depend on the intermediate channel
is the result of a remarkable cancellation between the
permutation signs and a contribution from the conformal dimensions.
\newpage

\vspace{0.6cm}
\noindent {\it Tetrahedral Duality}
\vspace{0.3cm}

Given the set of compatible fusion rule channels
specified in eq.~\ref{nofuseq} in the level $K$ $\gn$
theory, we now consider a dual set of fusion rules in the
level $N$ ${\rm G}(K)$ theory,
\eq
\begin{array}{ccc}
  S{\rm -channel} &    T{\rm -channel}  &  U{\rm -channel} \\
\ds
 \ta \cdot \tb = \sum_{s} \sigma^{\Delta^{ab}_{s}}(\ts)  &
\ds
\ta \cdot  \rtc =\sum_{t} \sigma^{\Delta^{a}_{ct}}(\tt~\!) &
\ds
\ta \cdot
 \rho(\sigma^{\delta}(\td~\!)) =  \sum_{u} \sigma^{\Delta^{c}_{bu}}(\tu) \\
\ds
\tc  \cdot \sigma^{\delta}(\td~\!)   =
                 \sum_{s}  \sigma^{\Delta^{ab}_{s}}(\ts) &
\ds
\rtb \cdot \sigma^{\delta}(\td~\!) =  \sum_{t}
                              \sigma^{\Delta^{a}_{ct}}(\tt~\!) &
\ds
\rtb \cdot  \tc =  \sum_{u}  \sigma^{\Delta^{c}_{bu}}(\tu)
\end{array}
\label{dualfusion}
\en
The letters $a,  \ldots, t, u$  denote $\gnk$ representations and
the `tilde' symbol again denotes the map from $\gnk$
integrable representations to
$\gkn$ integrable representations given by tableau transposition.
In addition, for $\spnk$ and $\sodnk$, $\rho(a) = a$ for all
representations $a$.
The integer $\Delta^{ab}_{c}$ is defined in
eq.~\ref{newdeltdef} and
the quantity
\eq
\delta \equiv \Delta^{ab}_{s} - \Delta^{cd}_{s}
\label{smdeltadef}
\en
measures the failure of exact conservation of the
number of boxes across an intermediate (here the $S$-) channel.
We have also adopted a generalization of the
index convention implicit in $\Delta^{ab}_{c}$:
an upper (lower) index indicates that $r$ comes in with a plus (minus)
sign.  For example,
\eq
\Delta^{a}_{bc} =
 \cases{(r(a) - r(b) - r(c))/N & for $\sunk$ \cr
        r(a) - r(b) - r(c)   & for $\sodnk$ \cr
                    0        & for $\spnk$  }
\label{deltadef}
\en
In all cases in which they appear these will be integers; in the
example just given this quantity  would
only appear for $c\in a \otimes \rho(b)$.

This dual set of fusion rules consistently defines three
$f$-dimensional bases of a $\gkn$ Hilbert space, with each
channel corresponding to a pair of compatible fusion rules.
This follows from the
dual and cominimal properties of the fusion coefficients in
eqs.~\ref{fusdual} and~\ref{propeq}, respectively, as well as
eq.~\ref{sigrho}. For $\sunk$ the identity
\eq
\rho(\widetilde{a}) = \sigma^{K-\row_{1}(a)}(\widetilde{\rho(a)})
\en
is also needed. This last identity is readily demonstrated by implementing
the operations on each side of the equation diagrammatically
(\ie, via a series of Young tableaux). The
existence of this dual set of fusion channels establishes a well-defined
map from tetrahedra in one theory to those of the other.

Corresponding to these two sets of fusion rules is
the following simple relation between the $S$ and $T$ channel
products of braid matrix eigenvalues:
\eq
\begin{array}{ccc}
  \varrho^{ab}_{s} \varrho^{\ta ~\!\tb}_{\sigma^{\Delta^{ab}_{s}}(\ts~\!)}
               & = &  \pm e^{\pi i \Phi(a,b)}  \\[0.3cm]
     \varrho^{a\rho(c)}_{t}
\varrho^{\ta\rho(\tc)}_{\sigma^{\Delta^{a}_{ct}}(\tt~\!)}
               & = &  \pm e^{-\pi i \Phi(a,c)}
\end{array}
\label{opchan}
\en
(The analogous $U$ channel identities are obtained as
cases of the $T$ channel
identity by setting $c \rightarrow d$, {\it etc.})

Using these equations and eq.~\ref{comgcfdim} we find that
the products of eigenvalues that appear as coefficients in
the two sets of non-linear constraint equations
corresponding to the two sets of fusion rules in~\ref{nofuseq}
and~\ref{dualfusion}
are the same up to sign for general tetrahedra
{\large
\eq
\varrho^{cd}_{s}  \varrho^{a \rc}_{t} \varrho^{\rho(b) c}_{u}\cF^{-1}_{c} =
    \pm    ~\cF_{\tc}~
(\varrho^{\tc\sigma^{\delta}(\td~\!)}_{\sigma^{\Delta^{ab}_{s}}(\ts~\!)}~
\varrho^{\ta \rho( \tc )}_{\sigma^{\Delta^{a}_{ct}}(\tt~\!)} ~
\varrho^{\rtb\tc}_{\sigma^{\Delta^{c}_{bu}}(\tu)} )^{-1}
\en
}

\noindent Given this result it immediately follows from the nonlinear
set of equations (eqs.~\ref{inveq} \&~\ref{keyeq}) that
the tetrahedra of one theory and the dual tetrahedra of the dual theory
satisfy exactly the same set of equations up to sign.
The result for general tetrahedra is that
\eq
 \ltetra{s}{t}{a}{b}{c}{d} ~~=~ \pm~~~~~~
\ltetra{\sigma^{\Delta^{ab}_{s}}(\ts)}{\sigma^{\Delta^{a}_{ct}}(\tt~\!)}
    {\ta}{\tb}{\tc}{\sigma^{\delta}(\td~\!)}
\en
\vspace{1.4cm}

\noindent
For link-type tetrahedra (for which $\delta = 0$ always) the exact result
$$
\!\!\!\! \eta^{b\rho(b)}_{0}   \varrho^{ab}_{s} ~\varrho^{a\rho(b)}_{t}
        \ltetra{s}{t}{a}{b}{b}{a} ~~ =
\eta^{\tb\rho(\tb)}_{0}
(\varrho^{\ta \tb}_{\sigma^{\Delta^{ab}_{s}}(\ts~\!)}
{}~\varrho^{\ta \rho(\tb)}_{\sigma^{\Delta^{a}_{bt}}(\tt~\!)})^{-1}
\dutetra{\sigma^{\Delta^{ab}_{s}}(\ts)}
{\!\!\! \sigma^{\Delta^{a}_{bt}}(\tt~\!)}{\ta}{\tb}{\tb}{\ta}
$$
\vspace{1.0cm}
\eq
{}~
\label{dualexres}
\en
follows from the supplemented skein relations
(eqs.~\ref{keyeq} \&~\ref{lineq}),
showing that the same product of tetrahedra and
braid eigenvalues is an exact invariant under both types of
discrete symmetries.
The product of charge conjugation signs is given by
\eq
\eta^{b\rho(b)}_{0}     \eta^{\tb\rho(\tb)}_{0} = \cases{
                        (-1)^{(N+K)r(b)}  & for $\sunk$ \cr
                              1   &   for $\spnk$ and $\asodnk$}
\en

\vspace{0.8cm}
\noindent {\bf 5. Two Applications: WZW Models and Quantum Groups}
\vspace{0.3cm}
\onward

We give two simple applications of the above results.

\vspace{0.5cm}
\noindent {\it 5.1 WZW Braid Matrices}
\vspace{0.3cm}

The WZW braid matrices are matrices of non-planar tetrahedra.\cite{witten}
With the edge orientations in the definition
of the WZW braid matrix in terms of Chern-Simons
graphs chosen to make the connection with the bases defined
in section {\it 2.1} transparent, consider the braid
matrices specified by
\eq
   \fork{a}{b}{b}{a}{s} = \sum_{t}  B_{st} \left[ \begin{array}{cc}
                               b & \rho(b) \\
                            a & \rho(a)
                       \end{array}
                             \right]
       \eta^{b \rho(b)}_{0}      \tfork{a}{b}{b}{a}{t} ~~,
\label{brmatdef}
\en

\noindent which braid $\phi_b$ and $\phi_{\rho(b)}$ in the WZW
correlation functions $\vev{\phi_{a} \phi_{b} \phi_{\rho(b)} \phi_{\rho(a)}}$.
The relation to Chern-Simons tetrahedra is simply that
\eq
\begin{array}{ccl}
       B_{st}\left[ \begin{array}{cc}
                               b & \rho(b) \\
                            a & \rho(a)
                       \end{array}
                             \right]  & = &
\eta^{b\rho(b)}_{0}
(\chi_{q}(a)\chi_{q}(b))^{-1}  ~
 \npovtet{a}{a}{s}{b}{t}{b} \\
  ~ & ~  &  ~  \\
  ~ & ~ &  ~  \\
  ~  & = &
  (\chi_{q}(a)\chi_{q}(b))^{-1}
\cF_{b}^{-1} \eta^{b\rho(b)}_{0} \varrho^{ab}_{s} \varrho^{a\rho(b)}_{t}
                                \tetra{s}{t}{a}{b}{b}{a}  \\
  ~ & ~  &  ~  \\
  ~ & ~  &  ~
\end{array}
\label{onemore}
\en
\vspace{1.0cm}

\noindent This shows that the WZW braid matrices singled out by
eq.~\ref{brmatdef} are expressible in terms of
link-type tetrahedra.
Since the planar tetrahedra appearing in eq.~\ref{onemore} are
exactly those of $F_{st}$ (with $d=a$ and $c=b$)
\eq
     B_{st}\left[ \begin{array}{cc}
                               b & \rho(b) \\
                            a & \rho(a)
                       \end{array}
                             \right]  =
\cF_b^{-1}
(\eta^{b\rho(b)}_{0} \varrho^{ab}_{s} \varrho^{a\rho(b)}_{t}   F_{st}) ~.
\en
Using this correspondence and eq.~\ref{dualexres} we find that
\eq
 \begin{array}{ccl}
\ds \sum_{t}       B_{s^{\prime}t}\left[ \begin{array}{cc}
                               b & \rho(b) \\
                            a & \rho(a)
                       \end{array}
                             \right]
{}~B_{\sigma^{\Delta^{ab}_{s}}(\ts~\!)\sigma^{\Delta^{a}_{bt}}(\tt~\!)}
                      \left[ \begin{array}{cc}
                               \tb & \rho(\tb) \\
                            \ta & \rho(\ta)
                       \end{array}
                             \right]
& = &      e^{-2 \pi i (h(b) + h(\tb))}
\varrho^{ab}_{s^\prime} (\varrho^{ab}_{s})^{-1}
\ds    \sum_{t} F_{s^{\prime}t}  ~F_{s t}  \\
{}~ & = &
   e^{\pi i r(b)} e^{\pi i \Phi(b,b)} \delta_{s^{\prime}s}
\end{array}
\en
where $\Phi(b,b)$ is defined in eq.~\ref{fivethree}.
This complements the results in refs.~\mite{nrs} \&~\mite{nacschnit}
on WZW braid and fusion matrix dualities.
There, however, WZW braid matrices that braid
the pair of $\phi_{b}$ fields in
$\vev{\phi_{a} \phi_{b}\phi_{b} \phi_{d}}$ were considered.
These latter braid matrices are proportional
to a special class of non-link-type tetrahedra.

Similarly, the relation between the (link-type) WZW braid matrices that
differ by cominimal equivalence is found from eq.~\ref{exacres} to be
\eq
       B_{\sigma(s) \sigma(t)} \left[ \begin{array}{cc}
                               b &  \rho(b) \\
                           \sigma(a) & \rho(\sigma(a))
                       \end{array}
                             \right]
      =
                        B_{st}\left[ \begin{array}{cc}
                               b & \rho(b) \\
                            a & \rho(a)
                       \end{array}
                             \right]
\en

\vspace{0.8cm}
\noindent {\it 5.2 Quantum Group $6j$-Symbols and WZW Fusion Matrices}
\vspace{0.3cm}

Given appropriate normalizations,
the expectation values of
planar tetrahedra equal\cite{witten} the values of quantum $6j$-symbols
(of $\cU_{q}(\gn)$)
evaluated at the roots of unity $q= \exp(2\pi i/(K+ \overline{g}))$.
(In addition, the WZW fusion matrices are also directly matrices
of planar tetrahedra.)
Therefore sections $3$ and $4$ immediately yield
identities for these quantities. Given the standard
relation\cite{gaume,chairzhu}
\eq
      \fork{a}{b}{c}{d}{s} ~=~ \sum_{u}
           \left\{ \begin{array}{ccc}  a & b & s \\
                                c & d & u \end{array} \right\}_{q}
              \fusfork{a}{b}{c}{d}{u}
\en
the correspondence is
\eq
     \left\{ \begin{array}{ccc}  a & b & s \\
                                c & d & u \end{array} \right\}_{q}  =
{\punknot{u} \over \smduubaryon{a}{u}{d} \smddubaryon{b}{u}{c} }
                 \tetra{s}{u}{a}{b}{d}{c}  = G_{su}
\label{sixj}
\en
\vspace{1.3cm}

\noindent
The results of section $3$ yield the
transformation properties of quantum $6j$-symbols under cominimal
equivalence. For example,
\eq
      \left\{ \begin{array}{ccc}  \sigma(a) & b & \sigma(s) \\
          \sigma(c) & d & \sigma(u) \end{array} \right\}_{q} =
   \pm  \left\{ \begin{array}{ccc}  a & b & s \\
                                c & d & u \end{array} \right\}_{q}
\en

Similarly, the results of section $4$ show that there is
a rank-level duality between the quantum $6j$-symbols
of $\cU_{q}(\gn)$ and $\cU_{q}(\gk)$ for $q$ the common
root of unity $q=\exp(2\pi i /(K + \overline{g}))$.
\eq
     \left\{ \begin{array}{ccc}  a & b & s \\
                                c & d & u \end{array} \right\}_{q} =
\pm
 \left\{ \begin{array}{ccc}  \ta & \tb & \sigma^{\Delta^{ab}_{s}}(\ts) \\
       \tc & \td & \sigma^{\Delta^{a}_{du}}(\tu)
                        \end{array} \right\}_{q}
\en
The $\pm$ signs appearing in these two equations
depend on the phase conventions of
the $6j$-symbols (which are inherited from the residual vertex
normalization conventions of the planar tetrahedra via eq.~\ref{sixj}).
Exact identities can also be constructed in the special case corresponding
to link-type tetrahedra.

\vspace{3.0cm}
\noindent {\bf 6. Discrete Symmetries for All
                 Chern-Simons Observables}
\vspace{0.8cm}
\onward

Arbitrary planar graphs can be reduced to sums of products of planar
tetrahedra.\cite{martin}  The overall sign of a non-link-type graph depends
on the arbitrary normalizations of the graph vertices,
and the tetrahedra that appear in its reduction
can all have expectation values with
normalization dependent signs. However, if we fix the normalization of
all vertices of the original graph, then, since the new vertices that appear
in the reduction process come in pairs, the pattern of relative signs
between terms
in the sum is {\em not} normalization dependent. It is not clear what
determines these relative signs or how to calculate them.
The situation does not change for
link-type graphs. Such a graph has an unknown, intrinsic overall sign (since
each vertex appears an even number of times in the graph, so that
a change of normalization does not change the sign of the graph),
and, in addition, the relative signs between terms in
a tetrahedral decomposition are not known.
Without a way of calculating these signs the algorithm in
ref.~\mite{martin} is ineffective for general graphs.

To make this problem concrete consider the graph
\eq
G(a,b,\{s_{i}\}) = \gfour{a}{b}{s_{1}}{s_{2}}{s_{3}}{s_{4}}
             {\lower 1.5cm \hbox{~}}
\label{fourgraph}
\en
\vspace{1.5cm}

\noindent which has the reduction into generic tetrahedra
\eq
 \begin{array}{l}
      G(a,b,\{s_{i}\}) =
\eta^{a\rho(a)}_{0} \eta^{b\rho(b)}_{0} (\chi_{q}(a)\chi_q(b))^{-1}
 (\chi_q(s_{1})\chi_q(s_{2})\chi_q(s_{3})\chi_q(s_{4}))^{-\half} \\
\ds \!\!\!\!\!\!\!\!\!\!\!\!\!\!\!\!\!\!\!\!\!\!\!\!\!
                  \sum_{{t\atop ijkl}} {1\over \chi_q(t)}
  ~\!\!\!\!\!
     \cpastetra{b}{t}{s_{1}}{a}{s_{2}}{a} {\sct \cK^{s_{1}}_{s_{2} t}(j)}
  \! \cpbstetra{a}{t}{b}{s_{1}}{b}{s_{2}} {\sct \cK^{b}_{bt}(l)}
  \! \cpdstetra{a}{t}{b}{s_{3}}{b}{s_{4}} {\sct \cK^{bt}_{b}(l)}
 \! \cpcstetra{b}{t}{s_{3}}{a}{s_{4}}{a} {\sct \cK^{s_{3} t}_{s_{4}}(k)}    \\
  ~             \\
{}~        \\
{}~
\end{array}
\label{tetdecomp}
\en

\noindent The vertices not in common between
each term in the sum (\ie, those that do not appear in the original
graph) come in dual pairs so that
a change of residual vertex normalization does not change the relative sign
of the terms. Since this  is a link-type graph the vertices of the
original graph also appear in dual pairs so that its overall sign
is also clearly normalization independent.
While the non-linear identities (\ref{inveq} \&~\ref{keyeq})
do establish relations between the signs of certain tetrahedra, the
tetrahedra that appear here are not in the same set of basis change
coefficients and so are not related in this way.

Using the fusion rule identity in~\ref{propeq} one can show
that to an arbitrary
graph in a level $K$ $\gn$ theory there corresponds a class of
topologically identical graphs obtained by uniformly replacing
the representations along oriented, closed loops by cominimal equivalents.
Similarly, using eq.~\ref{fusdual}, one can also show
that in the level $N$ $\gk$ theory
there are dual graphs with cominimal equivalents of
transposes of the $\gnk$ representations along the edges.
In this latter case, however, the fusion rule identities only establish that
to each vertex
of one graph there is a dual vertex and one might ask whether these can be
pieced together consistently.
This is always possible since the integers $\delta$
(defined in eq.~\ref{smdeltadef}),
which measure the absence of exact box conservation across intermediate
channels, sum to zero around
closed loops. Then using the same pattern of reduction to tetrahedra for both
graphs and the symmetry results for these tetrahedra, we see that cominimally
equivalent graphs are identical functions of cominimally
equivalent tetrahedra, and that dual graphs are identical
functions of dual tetrahedra.  However, we only know that
the absolute values of these (real) tetrahedra are the same.
Therefore the above sign ambiguity only allows one
to conjecture that the results of sections 3 and 4 generalize
to arbitrary graphs.

The first part of this section is devoted to examining some exceptional
cases of graphs which {\em can} be calculated by the above
algorithm, providing further evidence for the above conjecture.
We then find linear equations for any
link-type graph. These play the same role for
general graphs that the analogous linear equations did
for tetrahedra. In the last part of this section
a graph-independent argument yields
the exact transformation property of knots and links under
cominimal equivalence.  An immediate consequence is a demonstration
of the transformation property of link-type graphs under
cominimal equivalence.  We obtain (but do not prove)
the analogous transformation identities for graphs, knots, and links under
rank-level duality.

\vspace{0.6cm}
\noindent {\it 6.1 Calculable link-type graphs}
\vspace{0.3cm}

The graph in eq.~\ref{fourgraph} has the alternate reduction
entirely in terms of link-type tetrahedra
\eq
  \begin{array}{c}
      G(a,b,\{s_{i}\}) = (\chi_{q}(a) \chi_q(b) )^{2}
\varrho^{ab}_{s_{1}} \varrho^{ab}_{s_{4}}
(\varrho^{ab}_{s_{2}} \varrho^{ab}_{s_{3}} )^{-1} \times   \\[0.3cm]
\ds   \sum_{t} {\chi_{q}(t)}^{-1}
(\eta^{b\rho(b)}_{0}
      \varrho^{ab}_{s_{1}} F_{s_{1} t} \varrho^{a\rho(b)}_{t})^{*}
(\eta^{b\rho(b)}_{0}
      \varrho^{ab}_{s_{2}} F_{s_{2} t} \varrho^{a\rho(b)}_{t})
(\eta^{b\rho(b)}_{0}
      \varrho^{ab}_{s_{3}} F_{s_{3} t} \varrho^{a\rho(b)}_{t})
(\eta^{b\rho(b)}_{0}
      \varrho^{ab}_{s_{4}} F_{s_{4} t} \varrho^{a\rho(b)}_{t})^{*}
 \end{array}
\en
where $F_{s_{i} t}$ is exactly that displayed in eq.~\ref{tetradisp} with
$c=b$ and $d=a$.
The braid eigenvalues all come from
insertions of factors of unity in the form of
products of braid eigenvalues, such as
$\varrho^{a\rho(b)}_{t} ({\varrho^{a\rho(b)}_{t}})^{*}$.
In this case the strategy outlined above is successful, and
eq.~\ref{exacres} leads to
\eq
   (\varrho^{\sigma(a) b}_{\sigma(s_{1})}
              \varrho^{\sigma(a)b}_{\sigma(s_{4})})^{-1}
\varrho^{\sigma(a) b}_{\sigma(s_{2})}
 \varrho^{\sigma(a) b}_{\sigma(s_{3})}
     ~~G(\sigma(a),b,\{\sigma(s_{i})\})
=    (\varrho^{a b}_{s_{1}} \varrho^{a b}_{s_{4}})^{-1}
   \varrho^{a b}_{s_{2}} \varrho^{a b}_{s_{3}}
{}~~G(a,b,\{s_{i}\})
\label{sixfour}
\en
Similarly, eq.~\ref{dualexres} yields
\eq
  \begin{array}{l}
(\varrho^{a b}_{s_{1}} \varrho^{a b}_{s_{4}})^{-1}
\varrho^{a b}_{s_{2}} \varrho^{a b}_{s_{3}}
{}~~G(a,b,\{s_{i}\}) \\[0.3cm]
{}~~~~~~~~=
\varrho^{\ta\tb}_{\sigma^{\Delta^{ab}_{s_{1}}}(\ts_{1})}
\varrho^{\ta\tb}_{\sigma^{\Delta^{ab}_{s_{4}}}(\ts_{4})}
(\varrho^{\ta\tb}_{\sigma^{\Delta^{ab}_{s_{2}}}(\ts_{2})}
  \varrho^{\ta\tb}_{\sigma^{\Delta^{ab}_{s_{3}}}(\ts_{3})})^{-1}
                    ~~G(\ta,\tb,\{ \sigma^{\Delta^{ab}_{s_{i}}}(\ts_{i}) \})
\end{array}
\label{sixfive}
\en
The entire complex part of the (combined) phase from both sides
of each of these equations cancels, yielding the result that the related
graphs have the same expectation values up to a (calculable) sign .
Although other examples of graphs with such reductions can be found, it
does not seem possible in general to reduce general link-type
graphs into sums of products of {\em link-type} tetrahedra.

A general link-type graph can have several types of intermediate
channel edges which differ from those occurring in eq.~\ref{fourgraph}. In all,
there are four types of such channels.
Depending on the orientations of the edges adjacent to an
intermediate channel edge, one can have an $S$ or $T$ channel.
In addition, depending on whether the pair of representations adjacent to
one vertex are permuted at the opposite vertex, one can have a
{\em twist} (with permutation) or {\em parallel} (without permutation)
edge.  The four intermediate channel edges in eq.~\ref{fourgraph}
are $S$-channel twist edges. Link-type tetrahedra exhibit both
$S$-channel and $T$-channel twist edges, but parallel edges are not
possible.
A simple example with parallel edges is the graph
\eq
         G(a,b,\{s_{1},s_{2},t\}) = \bigraph{b}{a}{s_{1}}{s_{2}}{t}
\label{big}
\en
\vspace{1.2cm}

\noindent The reduction of this graph to tetrahedra
\eq
\begin{array}{rl}
\ds G(a,b,\{s_{1},s_{2},t\}) = &
\ds {\eta^{a\rho(a)}_{0}
                \over \sqrt{\chi_q(a) \chi_q(b) \chi_q(t)} }
\sum_{i}
    \cpstetra{s_{1}}{t}{a}{b}{b}{a} {\sct ~\cK_{bt}^{a}(i)}
           \cprtstetra{s_{2}}{t}{b}{a}{a}{b} {\sct \cK^{bt}_{a}(i)}  ~~, \\
  ~   \\
  ~
\end{array}
\en
can be written in the form
\eq
\begin{array}{l}
\eta^{a\rho(a)}_{0} \varrho^{ab}_{s_{1}} (\varrho^{ab}_{s_{2}})^{-1}
G(a,b,\{s_{1},s_{2},t\}) =  \\
   \ds
{}~~~~~~~(\chi_{q}(a)\chi_{q}(b))^{{3\over 2}}  \chi_{q}(t)^{-\half}
 \sum_{i}
(\eta^{b\rho(b)}_{0} \varrho^{ab}_{s_{1}} F_{s_{1} t}(i)
                                               \varrho^{a\rho(b)}_{t})
(\eta^{b\rho(b)}_{0} \varrho^{ab}_{s_{2}} F_{s_{2} t}(i)
                                               \varrho^{a\rho(b)}_{t})^{*}
\end{array}
\en
where the sum is over the ${N_{a\rho(b)}}^{t}$
types of the coupling $\cK_{bt}^{a}$ (or its dual) appearing
at the indicated vertices. This example isolates
a further problem, if the fusion multiplicity is greater than two.
If the multiplicity is exactly two, then the two couplings
(generally) correspond
to symmetric and anti-symmetric combinations and the corresponding
braid eigenvalues differ by a sign so that no degeneracy need occur.
If, however,  ${N_{a\rho(b)}}^{t} \geq 3$ for some $t$,
a degeneracy will necessarily occur,
in the sense that the braid eigenvalues occur with multiplicity.
To apply the results of sections 3 and 4 to this case (as well, in
fact, to the previous case) we are implicitly assuming that an
orthogonal basis
of the degenerate couplings can be chosen so that the tetrahedra
that only differ by such couplings are exactly equal.
In the case of low multiplicity or where this is
possible, we find that
\eq
\begin{array}{l}
\eta^{\sigma(a) \rho(\sigma(a))}_{0}    \varrho^{\sigma(a)b}_{\sigma(s_{1})}
    (\varrho^{\sigma(a)b}_{\sigma(s_{2})} )^{-1}
G(\sigma(a),b,\{ \sigma(s_{1}),\sigma(s_{2}),\sigma(t) \} ) =~~~~~~~~ \\
{}~~~~~~~~~~~\eta^{a\rho(a)}_{0}
\varrho^{ab}_{s_{1}} (\varrho^{ab}_{s_{2}})^{-1} G(a,b,\{s_{1},s_{2},t\})
\end{array}
\label{sixeight}
\en
In addition,
\eq
\begin{array}{l}
\eta^{a\rho(a)}_{0}
   \varrho^{ab}_{s_{1}} (\varrho^{ab}_{s_{2}})^{-1} G(a,b,\{s_{1},s_{2},t\})
=  \\[0.3cm]
{}~~~~~~~~~\eta^{\ta\rho(\ta)}_{0}
(\varrho^{\ta\tb}_{\sigma^{\Delta^{ab}_{s_{1}}}(\ts_{1})})^{-1}
  \varrho^{\ta\tb}_{\sigma^{\Delta^{ab}_{s_{2}}}(\ts_{2})}
G(\ta,\tb,\{\sigma^{\Delta^{ab}_{s_{1}}}(\ts_{1}),
     \sigma^{\Delta^{ab}_{s_{2}}}(\ts_{2}),\sigma^{\Delta^{a}_{bt}}(\tt~\!) \})
\end{array}
\label{sixnine}
\en

\noindent This example shows that
no braid eigenvalue pre-factor appears for the parallel $T$-channel.
(The same result holds for a parallel $S$-channel.)

\vspace{0.6cm}
\noindent {\it 6.2 Linear skein relations for link-type graphs}
\vspace{0.3cm}

Let us suppose that we have (presumably non-linear) equations
that constrain the expectation values of a set of general graphs of
a given topology.  Then they should involve exactly the same
sign ambiguity as the non-linear equations for tetrahedra.
Therefore, for the graphs of this topology with edge representations
and orientations chosen so
that the graphs can support knots or links, their overall signs will
be intrinsic but undetermined by these equations.
The number of undetermined signs in the array of graphs
$G(s_{1}, s_{2}, \ldots; t_{3}, \ldots)$, indexed by
the intermediate channels, is
$\sum_{i} f_{i}$. Here the sum is over the edges labeled
as intermediate channels (in the case of graphs that support
two component links, this is just $({\rm \# crossings}) \times f$).
We now show how to  construct (at least)
this number of inhomogeneous linear equations.
For each  intermediate channel edge there will be two representations $a$ and
$b$ at each vertex. From one vertex trace out a Wilson line for $a$ so that
it follows the graph edges and eventually comes back to the
other vertex of the intermediate channel edge.  Then do the same for the
representation $b$, but in such a way that whenever a
strand of $a$ must be crossed the $b$ Wilson line goes
uniformly over (or under) the $a$ Wilson line.  If there remain
other external edges (bearing representations $a_{m}$)
not traversed by this procedure trace out unlinked unknots until
all edges have been traversed.
The indicated (two-vertex)
graph is just a fancy way of specifying a baryon (multiplied
\setcounter{footnote}{1}
by a braid eigenvalue and, perhaps, a
product of unknot expectation values).\footnote{Actually,
one can let the $a$ strand cross the
$b$ strand in any way that leaves the graph topologically equivalent
to a baryon; in which case further equations result.}
 On the
other hand at every crossing we can insert an $S$- or $T$-channel spectral
decomposition and obtain the baryon as a sum of graphs from the
array $G(s_{1}, s_{2}, \ldots; t_{3}, \ldots)$
(all of the same topology and with the same representations on the
external legs).

For example, using the graph in eq.~\ref{fourgraph}, we can draw
\eq
                  \fourlineq{a}{b}{s_{2}}
\label{fourlin}
\en
\vspace{1.7cm}

\noindent This graph leads to the equations
\eq
  \begin{array}{l}
(\varrho^{ab}_{s_{2}})^{2} \sqrt{\chi_{q}(s_{2})}
(\chi_q(a)\chi_q(b))^2 \\[0.2cm]
\ds  = \sum_{s_{1},s_{3},s_{4}}
\sqrt{\chi_{q}(s_{1}) \chi_{q}(s_{3}) \chi_{q}(s_{4}) }
{}~(\varrho^{ab}_{s_{1}})^{2}
{}~~[(\varrho^{ab}_{s_{1}} \varrho^{ab}_{s_{4}})^{-1}
\varrho^{ab}_{s_{2}} \varrho^{ab}_{s_{3}}
  ~~G(a,b,\{s_{1}, s_{2}, s_{3}, s_{4} \})]
\end{array}
\label{sixeleven}
\en
They have been written in a way that isolates (in square brackets)
a special product
of braid eigenvalues with the graph; it is exactly
the special quantity in eqs.~\ref{sixfour} and~\ref{sixfive}.
These equations, with the unknowns considered to be the special products,
are exact invariants of both cominimal equivalence and
rank-level duality, since the
coefficients are squares of eigenvalues. This
equation is a precise analog of the linear equations (\ref{lineq})
found for link-type tetrahedra.
In addition, other patterns of crossings
lead to further equations. For example,
the mirror image of the graph in~\ref{fourlin} yields the
complex conjugate of eq.~\ref{sixeleven}.

Similarly, using the graph in eq.~\ref{big}, we find that
the non-planar graph
\eq
           \bilinkeq{b}{a}{s_{1}}
\en
\vspace{0.8cm}

\noindent leads to the equation
\eq
\sqrt{\chi_{q}(s_{1}) } (\chi_q(a)\chi_q(b))^{\frac{3}{2}}
= \sum_{t,s_{2}}  \sqrt{\chi_{q}(t) \chi_{q}(s_{2})}
[\eta^{a\rho(a)}_{0}
   \varrho^{ab}_{s_{1}} (\varrho^{ab}_{s_{2}})^{-1} G(a,b,\{s_{1},s_{2}, t \})]
\label{exam2}
\en
This equation has been written in a way that isolates
the exact invariant of the discrete symmetries that appears in
eqs.~\ref{sixeight} and~\ref{sixnine}.

For arbitrary graphs the result is a set of equations,  only slightly different
for each of the four types of intermediate channels appearing in the
graph. For each pair of representations $a$ and $b$ which join at
(at least) one vertex in the
graph, a subset of the following equations holds. Letting
$\cC$ denote the number of intermediate edges of $G$, and
$\cW$ the sum of the crossing signs of the (two-vertex)
graph constructed above, the $S$-channel twist, $T$-channel
twist, and parallel equations are
\eq
\begin{array}{l}
\ds \left( \prod_{m} \chi_{q}(a_{m}) \right)
             (\varrho^{ab}_{s_{l}})^{-\cW} \sqrt{\chi_{q}(s_{l})} =
               (\chi_{q}(a)\chi_{q}(b))^{-\half \cC} ~\zeta~\times  \\
\ds \sum_{\{r_{i}, t_{j}, s_{k} (k\neq l) \}}
\left( \prod_{i,j,k\neq l} \sqrt{\chi_{q}(r_{i})\chi_{q}(t_{j})\chi_{q}(s_{k})}
(\varrho_{t_{j}})^{w(j)}   (\varrho_{s_{k}})^{-w(k)}  \right)
G(a,b,\ldots,\{r_{i}, t_{j}, s_{k},\ldots \})  \\[0.9cm]
\ds \left( \prod_{m} \chi_{q}(a_{m}) \right)
(\varrho^{a\rho(b)}_{t_{l}})^{\cW} \sqrt{\chi_{q}(t_{l})} =
                (\chi_{q}(a)\chi_{q}(b))^{-\half \cC}  ~\zeta~\times \\
\ds \sum_{\{r_{i}, t_{j} (j\neq l), s_{k}\} }
\left( \prod_{i,j\neq l,k} \sqrt{\chi_{q}(r_{i})\chi_{q}(t_{j})\chi_{q}(s_{k})}
(\varrho_{t_{j}})^{w(j)}   (\varrho_{s_{k}})^{-w(k)} \right)
G(a,b,\ldots,\{r_{i}, t_{j}, s_{k},\ldots \}) \\[0.9cm]
\ds \left( \prod_{m} \chi_{q}(a_{m}) \right)
\sqrt{\chi_{q}(t_{l})}  =
              (\chi_{q}(a)\chi_{q}(b))^{-\half \cC} ~\zeta~\times \\
\ds \sum_{\{r_{i} (i\neq l), t_{j}, s_{k}\} }
\left( \prod_{i\neq l,j,k} \sqrt{\chi_{q}(r_{i})\chi_{q}(t_{j})\chi_{q}(s_{k})}
(\varrho_{t_{j}})^{w(j)}   (\varrho_{s_{k}})^{-w(k)} \right)
G(a,b,\ldots,\{r_{i}, t_{j}, s_{k},\ldots \})
\end{array}
\label{genlineq}
\en
where, in each case, the index $i$ labels parallel channels (either
$S$ or $T$ type), the index $j$ labels twist $T$-channels, and
the index $k$ labels twist $S$-channels. The representations
$a_{m}$ correspond to paths on the graph that yield unknots.
In the third equation
the crossing signs $w(j) = \pm 1$ for $T$-channel crossings
and $w(k) = \pm 1$ for $S$-channel crossings satisfy the constraint
that $\cW= \sum_{j} w(j) + \sum_{k} w(k) = 0$.
(The above construction implies that $\cW = \pm 1$ in the
first and second equations.)
The omitted superscripts on the braid eigenvalues come from one of the
pairs $\{a_{m},a\},\{a_{m},b\},\{a_{m_1},a_{m_2}\},\{a,b\}$
according to the edges adjacent to the intermediate channel in question.
The sign $\zeta$ is given by
\eq
\zeta = \left(\prod_{S} \eta^{a\rho(a)}_{0} \eta^{b\rho(b)}_{0} \right)
          \left( \prod_{T} \eta^{a\rho(a)}_{0} \right),
\label{zetdef}
\en
where the products  are over all $S$ and $T$ channel intermediate
edges, respectively, and $a$ and $b$ are the adjacent edge representations
of the given intermediate channel.

These equations provide as many constraints
(with the complex conjugates twice as many) as
undetermined signs for the family of graphs indexed by the
representations appearing on the intermediate edges.
Since we have, with the index $i$ running over all intermediate
edges,
 $\prod_{i} f_{i}$
graphs (for example,
$f^{({\rm \# crossings})}$ graphs, for graphs supporting two component
links), these linear relations are insufficient in general to determine
the graph expectation values.

They have, nevertheless, the same structure as the
analogous equations for tetrahedra in~eq.~\ref{lineq} and
can be written in a similar manner to isolate a quantity
that just depends on the squares of eigenvalues and on
products of $q$-dimensions.
If $f=2$, they give a complete set of equations for
the graphs in~\ref{fourgraph} and~\ref{big} and
permit verification of the identities
in eqs.~\ref{sixfour},~\ref{sixfive},~\ref{sixeight}, and~\ref{sixnine}
without appeal to any results about tetrahedra.
This direct, linear
approach to these identities does not require,
in the case of fusion multiplicities, the
existence of a special choice of
basis for which tetrahedra of degenerate channels are equal.
(Of course this problem recurs for the graphs themselves.)

In conclusion, it seems reasonably certain that
\eq
\begin{array}{c}
\left( \ds \prod_{j,k} (\varrho^{\sigma(a) b}_{\sigma(s_{k})})^{-w(k)}
(\varrho^{\sigma(a) \rho(b)}_{\sigma(t_{j})})^{w(j)}   \right)
\zeta^{\sigma}
G(\sigma(a),b,\ldots,\{\sigma(r_{i}), \sigma(t_{j}), \sigma(s_{k}),\ldots \})
=               \\[0.5cm]
 ~\left( \ds \prod_{j,k} (\varrho^{a b}_{s_{k}})^{-w(k)}
(\varrho^{a \rho(b)}_{t_{j}})^{w(j)}  \right)
\zeta G(a,b,\ldots,\{r_{i}, t_{j},s_{k},\ldots \})
\end{array}
\label{fincograph}
\en
for any choice of the signs $w(j)$ and $w(k)$ as long as
$\sum_{j} w(j) + \sum_{k} w(k) = 0$, since these quantities
satisfy some form of the relevant equations in~\ref{genlineq}
which are exactly invariant under cominimal
equivalence. The signs $\zeta$ and
$\zeta^{\sigma}$ ($\widetilde{\zeta}$ below)
are defined by eq.~\ref{zetdef}, interpreted
with reference to the appropriate graphs.

Similarly, one expects
\eq
\begin{array}{l}
\left( \ds \prod_{j,k} (\varrho^{a b}_{s_{k}})^{-w(k)}
(\varrho^{a \rho(b)}_{t_{j}})^{w(j)}  \right)
\zeta G(a,b,\ldots,\{r_{i}, t_{j},s_{k},\ldots \}) =  \\[0.4cm]
{}~~~~~~~ \left(
\ds \prod_{j,k} (\varrho^{\ta \tb}_{\sigma^{\Delta^{ab}_{s}}(\ts_{k})})^{w(k)}
(\varrho^{\ta \rho(\tb)}_{\sigma^{\Delta^{a}_{bt}}(\tt_{j})})^{-w(j)} \right)
\widetilde{\zeta} G(\ta,\tb,\ldots,\{\sigma^{\Delta^{ab}_{r}}( \trr_{i}),
\sigma^{\Delta^{a}_{bt}}(\tt_{j}),\sigma^{\Delta^{ab}_{s}}(\ts_{k}),\ldots \})
\end{array}
\label{findugraph}
\en
to hold under the same condition on $w(j)$ and $w(k)$.

We have shown that eqs.~\ref{fincograph} and~\ref{findugraph}
actually do hold for various particular graphs. In addition, the
linear constraints that exist for any link-type graph also support
the expectation that these equations will hold for all link-type
graphs.

\vspace{0.6cm}
\noindent {\it 6.3 Symmetries for Links and Knots}
\vspace{0.3cm}

Using eq.~\ref{comgcfdim} we find that the characteristic
polynomial of the square of the braid matrix $B_{\sigma(a)b}$
\eq
  \prod_{s} (B_{\sigma(a) b}^{2} - (\varrho^{\sigma(a)b}_{\sigma(s)})^{2}) = 0
\label{charpoly}
\en
transforms to
\eq
   \prod_{s} ( (e^{-\pi i q(b)} B_{\sigma(a)b})^{2} - (\varrho^{ab}_{s})^{2})
=0
\en
which is exactly the characteristic equation for the square
of the braid matrix $B_{ab}$,
\eq
\prod_{s} ( B_{ab}^{2} - (\varrho^{ab}_{s})^{2}) =0  ~~~.
\label{charpolygen}
\en
This implies that a link $\cL(\sigma(a),b,\ldots)$
with $\sigma(a)$ on an unknotted component
and the same link with $a$ replacing $\sigma(a)$,
$\cL(a,b,\ldots)$,
multiplied by a phase $\exp(\pm i\pi q(b))$ for each
($\mp$ signed) crossing for each component that the $a$ component crosses,
satisfy identical skein relations.
Similarly, using eq.~\ref{confopdual} the characteristic
polynomial of the square of the $\gnk$ braid matrix $B_{ab}$
can be written (with $\Phi(a,b)$ defined in eq.~\ref{fivethree})
\eq
\prod_{s} \left(  (e^{- \pi i \Phi(a,b)} B_{ab})^{2} -
(\varrho^{\ta\tb}_{\sigma^{\Delta^{ab}_{s}}(\ts)})^{-2} \right) = 0
\en
which is exactly the characteristic equation for the square of the
braid matrix $B_{\ta\tb}^{-1}$ in the $\gkn$ theory.
This implies that
a linking of unknots $\cL(a,b,\ldots)$ in a $\gnk$ theory will satisfy
the same skein relation as the mirror image link in the $\gkn$ theory
multiplied by a phase $e^{\pm \pi i \Phi(a,b)}$ for each
($\pm$ signed) crossing of components $a$ and $b$ (and
this holds for each pair of components).
Similar statements result from
comparison of the characteristic
polynomial of the  knot-type braid matrix
$B_{\sigma(a) \sigma(a)}$
\eq
 \prod_{s} (B_{\sigma(a) \sigma(a)} -
           \varrho^{\sigma(a)\sigma(a)}_{\sigma^2(s)})    = 0
\label{knotpoly}
\en
with that of $B_{aa}$ (by using
eq.~\ref{braidfin} which relates the braid eigenvalues).
A comparison of the characteristic
polynomial for $B_{aa}$ in a $\gnk$ theory with
that for $B_{\ta \ta}$ in a $\gkn$ theory requires the use of
eq.~\ref{exdualbr} (which we have proved in many but not all cases).
Since not all knots or links can be untied with the
skein relations corresponding to characteristic
polynomials (such as~\ref{charpolygen} or~\ref{knotpoly}), these
results alone would only permit the comparison of a restricted
class of knots and links.

In the case of the Dynkin diagram symmetries a
cabling argument yields a proof of the
exact connection between any link and its cominimal equivalents
(eq.~\ref{comlinres} below).
To obtain the analogous
rank-level link relation (eq.~\ref{duallinres}) we will need to
examine the case of general knots and
links by means of the reduction to planar graphs followed by an appeal to
eq.~\ref{findugraph}.  This will provide a complete proof
of eq.~\ref{duallinres} only for special classes of
links (\ie, those built on certain special graphs such
as those studied in sections {\it 6.1} and {\it 6.2}).

Consider an arbitrary link $\cL(\{a_{i}\}) $ with representations
$a_{i}$ on the link components and the cominimally equivalent
link $\cL(\sigma(a_{i}), \{a_{j},j\neq i \}) $.
The exact relation between
these links is obtained as follows.
Given any link $\cL$, with a specified component $\cK$,
let $\{ \cL_{s}, \phi_{s} \in \phi_{a} \cdot \phi_{b} \}$ be the set of links
with representations $s$ on that component, and
let $\cK_{a} \cdot \cL_{b}$ be the link with an
untwisted, two-cable of the component $\cK$ in place of
the original component. Then
\eq
        \vev{ \cK_{a} \cdot \cL_{b}} = \sum_{s} \vev{ \cL_{s}} ~~,
\en
always holds. This fact and the fusion rule for
cominimal representations,
$\phi_{\sigma(0)} \cdot \phi_{a} = \phi_{\sigma(a)}$, permits
replacement of the Wilson line with $\sigma(a_{i})$ with the two-cable
of Wilson lines with $a_{i}$ and $\sigma(0)$ on the two
(everywhere locally parallel) lines, in order to obtain
\eq
\vev{\cL(\sigma(a_{i}),\{ a_{j}, j\neq i \}) } =
  \vev{ \cK_{\sigma(0)} \cdot \cL(\{ a_{i}\} )  } ~.
\label{singterm}
\en
Then the one-term skein relation based on the
characteristic polynomial (of the same fusion rule)
\eq
           B_{\sigma(0) r} =
(\varrho^{\sigma(0) r}_{\sigma(r)})^{2} B^{-1}_{\sigma(0) r}
 = e^{2\pi i q(r)} B^{-1}_{\sigma(0) r}
\label{singtermskein}
\en
for $r= a,b,\ldots$, can be used
to lift up the $\sigma(0)$ component, $\cK_{\sigma(0)}$,
detaching it from the rest of the link, at the cost of one factor of the
braid eigenvalue in  eq.~\ref{singtermskein} (or its inverse)
for every negative (positive) {\em under-crossing} by $\cK_{\sigma(0)}$ of
any other component, to obtain
\eq
  \vev{\cL(\sigma(a_{i}),\{a_{j},j\neq i\}) } =
e^{-i \pi \sum_{j\neq i} q(a_{j}) w(i,j)}  e^{-2\pi i q(a_{i}) w(i,i)}
\vev{ \cK_{\sigma(0)} }     \vev{\cL(\{ a_{i} \}) } ~.
\en
where $w(i,j)$ is the sum of the crossing
signs\cite{kauffman} between the components $i$ and $j$.
Then each crossing in the knot $\cK_{\sigma(0)}$
can be replaced (at the cost of further braid
eigenvalue factors) with an un-crossing  by means of the skein relation
corresponding to the characteristic polynomial
\eq
B_{\sigma(0) \sigma(0)} = \varrho^{\sigma(0) \sigma(0)}_{\sigma^2(0)} I ~~.
\en
Since the expectation value of a $\sigma(0)$ unknot equals unity
(eq.~\ref{comgqdim}),
\eq
\vev{ \cK_{\sigma(0)} } =
(\varrho^{\sigma(0) \sigma(0)}_{\sigma^{2}(0)})^{-w(i,i)} =
e^{-i\pi {p + 1 \over p} (\kappa | \sigma(0))  w(i,i)}
\en
Therefore,
\eq
 \vev{\cL(\sigma(a_{i}), \{a_{j}, j\neq i\})} =
{}~~e{\lower 0.35 cm
\hbox{$^{\sct -i\pi [({\ds \sum_{j\neq i}} w(i,j) q(a_{j})) + 2 w(i,i) q(a_{i})
+ w(i,i) {p+1\over p} (\kappa | \sigma(0)) ]}$}}
\vev{\cL(\{a_{j}\}) }
\label{comlinres}
\en
The explicit values of $(\kappa | \sigma(0))$
are listed in eq.~\ref{sigintval},
and $p$ is the order of $\sigma$.
We can now use~\ref{comlinres} to obtain further constraints on
a set of graphs of a given type.  There will turn out to
be enough (linear) equations to
completely determine the expectation value of
any link-type graph in terms of the links
they support. While we do not know these expectation values independently
(so that we cannot use these equations to calculate the graphs) we do
know their transformation properties (eq.~\ref{comlinres})
and so can obtain the transformation
property of graphs given in eq.~\ref{fincograph}.

First one can insert a spectral decomposition for each crossing
in an arbitrary link $\cL_{a,b,\ldots}$
to get a representation of the link as a sum over planar graphs.
At each crossing one can choose to insert an $S$-channel spectral
decomposition so that
\eq
 \ds \cL =
{(\eta^{a\rho(a)}_{0} \eta^{b\rho(b)}_{0})^{\cC} \over
(\chi_{q}(a) \chi_{q}(b))^{\half \cC} }
\sum_{\{s_{i}\}} \left( \prod_{i} \sqrt{\chi_{q}(s_{i})}
\varrho^{a_{i} b_{i}}_{s_{i}} \right)
              G(a_{i}, b_{i}, \ldots, \{s_{i},\ldots\})
\label{linkdecomp}
\en
with $\cC = \sum |w(i)|$ denoting the number of crossings of
the link (alternately, the number of intermediate-channel edges in
each graph).
Here the indicated graphs only have $S$-channel twist-type
intermediate edges.
(While all faces of $G$ with just two edges can be immediately
excised, this raises the powers of the braid eigenvalues
appearing in~\ref{linkdecomp} and potentially introduces a variety
of the four channel types, which complicates the argument somewhat.)
For example, the family of links
\eq
          \cL(a,b;\{n_{i}\}) = \twist{a}{b}
\label{sixtwensix}
\en
\vspace{3.3cm}

\noindent (where each $n_{i}$ equals
the sum of the crossing signs of the neighboring braiding)
has the decomposition
$$
          \vev{ \cL(a,b;\{n_i\})} =
\sum_{s_{1} s_{2} s_{3} s_{4}}
\left( \prod_{i=1}^{4}  \sqrt{{\chi_{q}(s_{i})\over \chi_{q}(a) \chi_{q}(b)}}
(\varrho^{ab}_{s_{i}})^{-n_{i}}   \right)
     \gfour{a}{b}{s_{1}}{s_{2}}{s_{3}}{s_{4}}
$$
\vspace{1.5cm}
\eq
{}~
\label{sixtwenseven}
\en
If $a\neq b$, then
each $n_{i}$ must be odd so that the sum of the
crossing signs between two different components, $\sum_{i} n_{i}$,
is  even. The general cabling result yields, in this case,
the equations
\eq
                 \cL(\sigma(a),b;\{n_i\}) =
         e^{-i\pi q(b) (\sum_{i} n_{i})} ~\cL(a,b;\{n_i\})
\label{cabres}
\en
which  can be thought of as  a set of
$f^4$ linear homogeneous
equations (parameterized by the crossing signs $n_{i}$) for
the $f^4$ quantities
$G(\sigma(a),b;\{\sigma(s_{i})\})- G(a,b;\{s_{i}\})$, multiplied
by certain braid eigenvalues.
While the $n_{i}$ run over all odd integers, the braid polynomial
skein relations relate links with different numbers of crossing signs.
Given that no unforeseen degeneracies occur this set of equations
can be solved only by requiring that eq.~\ref{fincograph} holds.

Assuming the graph identity in eq.~\ref{findugraph} yields,
via eq.~\ref{sixtwenseven},
\eq
                      \cL(a,b;\{n_{i} \})_{\gnk} =
      e^{-\pi i \Phi(a,b) \sum_{i} n_{i} }
         \cL(\ta,\tb;\{-n_{i}\})_{\gkn}
\label{rlcase}
\en
as the expected general result for the link in eq.~\ref{sixtwensix}
(assuming that $a\neq b$ so that $\sum_{i} n_{i}$ is even).

Similarly, the class of links
\eq
    \cL(a,b;\{n_{i}\}) =   ~\bilink{b}{a}
\en
\vspace{3.0cm}

\noindent has a decomposition as a sum of graphs of
the type in eq.~\ref{big} via one $T$ channel and
two $S$ channel insertions so that eq.~\ref{comlinres} for
these links (which is identical to eq.~\ref{cabres})
leads to the graph identity in eq.~\ref{exam2}.
Again, assumption of eq.~\ref{findugraph} for the underlying graph again leads
to eq.~\ref{rlcase}.

In general, the rank-level-duality identity for
graphs (eq.~\ref{findugraph}) implies
the link identity
\eq
  \begin{array}{c}
     \cL( \{ a_{i} \} )_{\sst \gnk} =
e^{\pi i \sum_{i} w(i,i) r(a_{i})}  e^{-\pi i \sum_{i,j} w(i,j) \Phi(a_i,a_j)}
\widetilde{\cL}(\{\ta_{i}\})_{\sst \gkn}\\[0.3cm]
\Phi(a_i,a_j) = \cases{ r(a_{i}) r(a_{j})/NK  &  $\sunk$ \cr
                          0                & $\spnk$ \& $\asodnk$ }
 \end{array}
\label{duallinres}
\en
where $\widetilde{\cL}$ is the mirror image link of $\cL$.
On the basis of the concrete results, and the structure of the
known constraint equations,
we expect that eq.~\ref{duallinres}
holds for all knots and links.

\vspace{0.8cm}
\noindent {\bf 7. Conclusion}
\vspace{0.3cm}
\onward

In order to study the exact
symmetries of arbitrary Chern-Simons observables
we need a systematic reduction of all such observables to known
quantities.
A previously proposed algorithm involving the
reduction of such observables to tetrahedra is ineffective
due to the presence of undetermined signs that appear in these
reductions.  We have found an
extension of this algorithm that permits the examination of the symmetries of
tetrahedra and certain other Chern-Simons observables.
Using this, we have derived the exact form of cominimal equivalence and
rank-level duality for tetrahedra. (This result does not depend
on any choice of a system of permutation signs.)
For arbitrary link-type graphs (including tetrahedra)
we find a set of linear equations; these equations suggest the
general form of cominimal equivalence and rank-level duality for
arbitrary link-type graphs. In the case of cominimal equivalence this is
confirmed by an argument based on an independent result for
links.
For rank-level duality we only show that the expected graph
result implies the expected link identities (and vice versa). In both
cases we exhibit several non-trivial examples consistent
with these identities.

For knots these identities require precise control over the permutation
signs and conformal dimensions appearing in the braid
eigenvalues. Study of these quantities led to an exact formula for the
permutation signs in the multiplicity free case (ref.~\mite{newp}) and
the examination of the simple current charges
led to the identification of the simple classical origin
of these charges (section $3$).

The most pressing problem raised by this work is to find
an effective way of calculating arbitrary (or even
just link-type) graphs in a systematic
way. The reduction of a knot, link, or graph to a sum of
products of tetrahedra (planar or non-planar) is akin to
evaluating a lattice partition function, but with an important
difference: the Boltzmann weights of a (unitary) lattice partition function
are positive definite, but the tetrahedra are not.
\newpage

\vspace{0.9cm}
\noindent {\bf Technical Note: Plethysm, Permutation Signs, and Baryons}
\vspace{0.4cm}
\renewcommand{\theequation}{T.\arabic{equation}}
\onward

\vspace{0.3cm}
\noindent {\it Young Tableaux and Dynkin Indices}
\vspace{0.3cm}

The representations of $\sun$, $\spn$,
and $\ason$ that appear on the
components and edges of Wilson links and graphs are
referred to primarily via
Young tableaux. For $\sun$, $\spn$, and $\asodn$, the tableau row
lengths are given in
terms of the Dynkin indices of the highest weight by
\eq
    \row_{i} =
  \cases{\ds \sum_{j=i}^{{\rm rank}\{G\}} a_{j} & for G =$\sun$ \& $\spn$ \cr
                            ~                    & ~ \cr
         \ds \half a_{n} + \sum_{j=i}^{n-1} a_{j} & for $\asodn$ ~~.}
\en
For ${\rm so}(2n)$ the natural labels
\eq
     \row_{i} =
\cases{ \ds \half ( a_{n} + a_{n-1} )
              + \sum_{j=i}^{n-2} a_{j} & for $1\leq i\leq n-1$ \cr
          ~     & ~ \cr
        \half ( a_{n} - a_{n-1} )      & for $i=n$}
\en
correspond to tableau row lengths for $i=1,\ldots, n-1$.
The final tableau row length is
given by $|\row_{n} |$. If $\row_{n}\neq 0$, then the
representation is characterized by its tableau and the
number $\nu \in \{ 0,1\}$ defined by
$(-1)^{\nu} = {\rm sgn}( a_{n} - a_{n-1})$. The tableau for a
spin-tensor $\{ \psi; a\}$ with tensor part $a$
is formed by adjoining a column of $n$ half boxes to the
ordinary tableau for $a$.

For $\sun$ and $\ason$ we refer to {\em reduced} tableaux.
A tableau is reduced if $\row_{N} = 0$ for $\sun$ and if
$\col_{1} \leq N - \col_{1}$ for $\ason$ (the $\col_{i}$ are
column lengths). While the tableaux defined in the
paragraph above are all reduced,
unreduced tableaux appear in the standard procedures for computing
tensor products using Young tableaux.
In addition, for $\ason$, the
{\it associate} tableau $\alpha(a)$ of a tableau $a$ only
differs from $a$ in that $\col_{1}(\alpha(a))
= N - \col_{1}(a)$. If $\alpha(a) = a$ ~(requiring $N=2n$), the representation
$a$ is {\it self-associate}.  Since $\row_{n}$ is non-vanishing
for these representations, specification of self-associate
representations requires the sign $\nu$ in addition to a tableau.
All spinors of ${\rm SO}(2n)$ are self-associate.

Implicit in this paper is the  assumption that the
tableaux appearing in a level $K$ $\gn$ theory label
integrable representations of $\gnk$.
This means that $\row_{1} \leq K$ for
$\sun$ and $\spn$ and $\row_{1} + \row_{2} \leq K$ for $\ason$.

\vspace{0.3cm}
\noindent {\it Baryon Normalization and Permutation Sign Constraints}
\vspace{0.3cm}

The crossing constraint $\eta^{ab}_{c} = \eta^{\rho(c) b}_{\rho(a)}$
in eq.~\ref{chernsign}
follows from comparison of the  standard untwisting
\eq
    B_{ab} \openy{a}{b}{c} =
   \gbrtwis{b}{a}{c} = \varrho^{ab}_{c} \openy{b}{a}{c}
{\lower 1.5cm \hbox{~}}
\en
\vspace{0.6cm}

\noindent with the alternate untwisting
\eq
     \gbrtwis{b}{a}{c} = \yfancy{b}{a}{c} =
    \cF_{a} (\varrho^{\rho(c) a}_{\rho(b)})^{-1} \openy{b}{a}{c}
\en
\vspace{1.3cm}

\noindent where $\cF_{a} = q^{\half Q(a)}$ is the framing factor
incurred in undoing the (positive crossing-sign) self-crossing.

We adopt the standard normalization of baryons
\eq
      \cpbaryon{a}{\rho(c)}{b} =
     \sqrt{\chi_{q}(a) \chi_{q}(b) \chi_{q}(\rho(c))}
\label{standnorm}
\en
\vspace{1.0cm}

This normalization and the crossing constraint
implies a definite relation between any coupling and its dual and leads
to the conjugation constraint $\eta^{ab}_{c} =
\eta^{\rho(a)\rho(b)}_{\rho(c)}$ in eq.~\ref{chernsign}.
Given a consistent choice of a system of
permutation signs there is one sign $\omega$ left for the set
of four couplings $\cK^{ab}_{c}, \cK^{ba}_{c}, \cK_{ab}^{c},
\cK_{ba}^{c}$.

In the remainder of this subsection we discuss the origin of the
fusion constraint in eq.~\ref{chernsign}.
Examination of the $U$-channel spectral decomposition of the identity
shows that
\eq
        \curid{b}{~}{a}{~} = A(a,b)  \curchan{b}{b}{a}{a}{0}
\label{Adef}
\en
\vspace{1.0cm}

\noindent where $A(a,b)= \pm 1$. Then
\eq
         \rcurid{b}{~}{a}{~} =
        A(a,b) \eta^{a\rho(a)}_{0}  \rcurchan{b}{b}{a}{a}{0}
              {\lower 1.0cm \hbox{~}}
\label{identrem}
\en
\vspace{1.0cm}

\noindent If $a=b$ then the sign $A(a,a)$ is intrinsic in the sense that
it does not depend on the residual normalization of any couplings,
but if $a\neq b$ then $A(a,b)$ depends on the residual
normalization of the couplings $\cK^{a}_{0a}$ and $\cK^{0b}_{b}$.
It is natural (though perhaps not necessary) to set the
normalization of $\cK_{\rho(a)}^{0\rho(a)}$, which couples
the ingoing states $\ket{0} \otimes \ket{\rho(a),i}$ to
the outgoing states $\ket{\rho(a),i}$, equal
to the normalization of $\cK_{a}^{0a}$. This implies that
\eq
          \idrop{~~~a} ~~~= \idrop{\rho(a)} ~,
\label{tdrop}
\en
\vspace{0.7cm}

\noindent which, from eqs.~\ref{Adef} and~\ref{identrem}, leads to the
equality $A(\rho(a),b) = A(a,b)$.

Similarly, we find that, necessarily,
\eq
        \curchan{a}{c}{b}{d}{t} = R_{t} \dcurchan{a}{c}{b}{d}{\rho(t)}
\label{arrowreversal}
\en
\vspace{1.0cm}

\noindent with $R_{t} = R_{\rho(t)} = \pm 1$. The couplings that
appear on the left- and right-hand side of eq.~\ref{arrowreversal}
are related by the action of the charge conjugation
operator $\cC^{\rho(t) t} : t \longrightarrow \rho(t)$.
For example, $\cK_{b}^{\rho(t) d} = \cC^{\rho(t) t} (\cK_{t b}^{d})^{T_1}$
(where $T_1$ is the partial transpose that effectively raises the
indices labeling the states of $t$).
That the right hand side of eq.~\ref{arrowreversal}
can be constructed from the
left by an odd number of applications of this operator makes it
reasonable that a non-trivial sign could appear in eq.~\ref{arrowreversal}.
In this light it is also reasonable that while $R_{t}$
depends on $t$ it does not depend on the other representations
in eq.~\ref{arrowreversal}.
Using these results we will now show that $A(a,a) =1$ and that
$R_{t} = \eta^{t\rho(t)}_{0}$.

Using eq.~\ref{identrem} we find that
\eq
 \sptetra{0}{0}{a}{a}{a}{a} = A(a,a) \eta^{a\rho(a)}_{0} \duubaryon{a}{0}{a}
\en
\vspace{1cm}

\noindent Since $R_{a}^2=1$ eq.~\ref{arrowreversal} leads to
\eq
    \sptetra{0}{~0}{a}{a}{a}{a} =
\spotetra{0}{~0}{a}{a}{\!\! \rho(a)}{~\rho(a)}
  = A(\rho(a),a) \baryon{\rho(a)}{0}{a}
\en
\vspace{1cm}

\noindent These two decompositions,  eq.~\ref{tdrop},
and eq.~\ref{standnorm}  imply that
\eq
       \duubaryon{a}{0}{a} ~=~ \eta^{a\rho(a)}_{0} \chi_{q}(a)
\label{crux}
\en
\vspace{1.0cm}

\noindent which from eq.~\ref{identrem} yields $A(a,a)=1$.
If $\rho(a)=a$, then eq.~\ref{crux} holds necessarily,
\ie, with any choice of residual couplings.
The important point here is simply that (almost
necessarily) not all baryons can be normalized to be positive.
Since this is the case we have adopted eq.~\ref{tdrop} in order
to put all representations on the same footing.
Similar manipulations starting from tetrahedra with only one edge
carrying the identity, such as
\eq
           \tetra{c}{0}{a}{b}{a}{b},
\en
\vspace{1.0cm}

\noindent show that for any $a$, $b$, and $c$, with $c\in a\otimes b$
\eq
      \udubaryon{a}{c}{b} = \eta^{c\rho(c)}_{0} \baryon{a}{\rho(c)}{b}
              = \eta^{c\rho(c)}_{0} \chi_{q}(a) \chi_{q}(b) \chi_{q}(c)~~~
\label{arrbarynorm}
\en
\vspace{1.3cm}

\noindent This result and
eq.~\ref{arrowreversal} immediately imply that
$R_{c} = \eta^{c\rho(c)}_{0}$.
These baryon signs are responsible for the permutation
signs in the dual basis equations in section {\it 2.1}.
By reversing all the arrows in a baryon, eq.~\ref{arrbarynorm}
implies the fusion constraint
$\eta^{a\rho(a)}_{0} \eta^{b\rho(b)}_{0} =
\eta^{c\rho(c)}_{0}$ that appears in eq.~\ref{chernsign}.

In this way we obtain the three constraints on the signs in
eq.~\ref{chernsign} from the standard baryon
normalization and the apparently innocuous choice of vertical framing.

\vspace{0.3cm}
\noindent {\it The Natural Permutation Signs}
\vspace{0.3cm}

In the traditional
quantum group construction of link and graph invariants\cite{reshq}
finite-dimensional
$\cR$-matrices act diagonally on matrices of
$q$-Clebsch-Gordan coefficients, with the diagonal elements given
by eq.~\ref{breigdef}, except that the non-crossing
symmetric $\epsilon^{ab}_{c}$ appears in
place of the crossing-symmetric $\eta^{ab}_{c}$.
This difference reflects the fact that
the diagram calculus inspired by quantum groups\cite{reshq}
differs from that appropriate to Chern-Simons theory by
singling out a particular direction (a `time' direction).
In this sense the Chern-Simons graphical calculus is a Lagrangian
version of the (`Hamiltonian') quantum group graphical
calculus. Due to the close correspondence between these two
approaches we expect a simple relation between
$\eta^{ab}_{c}$ and $\epsilon^{ab}_{c}$, which, in fact,
is the case (eq.~\ref{cstoclass}).
The quantum group permutation signs can be deduced from
(if $a\neq b$ defined via)
the matrices of $q$-Clebsch-Gordan coefficients
$\left(\cC_{ab}^{c}(q) \right)_{ij}^{k}$ that
appear in the tensor product decompositions
\eq
 \ket{c,k} = \sum_{ij}
\left(\cC_{ab}^{c}(q) \right)_{ij}^{k} \ket{a,i} \otimes \ket{b,j} ~~,
\label{cgdecomp}
\en
by means of the identity
\eq
{\cal P}^{ab}   \cC^{ab}_{c}(q)
= \epsilon^{ab}_{c} \cC^{ba}_{c}(q^{-1}) ~~~.
\label{classperm}
\en
(Iteration of eq.~\ref{classperm} shows immediately that
$\epsilon^{ab}_{c} = \epsilon^{ba}_{c}$.)
The classical
limit ($q\rightarrow 1$) shows that $\epsilon^{ab}_{c}$ is simply
the (ordinary) group theory permutation sign.
While the intrinsic signs $\epsilon^{aa}_{c}$ can just as well
be calculated from the classical Clebsch-Gordan coefficients
by setting $q=1$ in eq.~\ref{classperm}, in the case
of $\epsilon^{ab}_{c}$ with $a\neq b$ the powers of $q$ keep track of a
natural ordering of states.
For example, the fact that in ${\rm SU}(3)$ one has
\setlength{\unitlength}{0.15cm}
$\epsilon^{~\!\!\young2000001 ~\young2000001}_{~
{\raise 0.2cm \hbox{$\young3100002$}}} = -1$
may be deduced by applying eq.~\ref{classperm}  to  the embedding
of the highest weight state of
\setlength{\unitlength}{0.18cm} {\raise 0.15cm \hbox{$\young3100002$}} in
$\young2000001 \otimes \young2000001$
\eq
    \ket{21} = \omega \sqrt{ { [2]_{q} \over [4]_{q}} }
(q^{\half} \ket{20} \otimes \ket{01} - q^{-\half} \ket{01} \otimes \ket{20} )
{}~~,
\label{decomaa}
\en
where the states are labeled by Dynkin indices and $\omega =\pm 1$
remains unfixed after imposing $\vev{21 | 21} = 1$.
The permutation sign is {\em intrinsic} in the sense that
it originates in the relative sign between
the leading term and its permutation in such decompositions
and is independent of $\omega$.
The same procedure,
when applied to the embeddings of the highest weight state of
\setlength{\unitlength}{0.18cm}
{\raise 0.12cm \hbox{$\young2100002$}} ~in the tensor
product
$\young2000001 \otimes \young1000001$ ~and in its permutation~
$\young1000001 \otimes \young2000001$, respectively,
\eq
 \begin{array}{ccc}
    \ket{11} & = & \omega_{1} {1 \over \sqrt{[3]_{q}}}
( \sqrt{[2]_{q}} q^{{1 \over 4}} \ket{20} \otimes \ket{-1 1}
          - q^{-\half} \ket{01} \otimes \ket{10} ) \\
    \ket{11} & = & \omega_{2} {1 \over \sqrt{[3]_{q}}}
(q^{\half} \ket{10} \otimes \ket{01}
  - \sqrt{[2]_{q}} q^{-{1 \over 4}} \ket{-1 1} \otimes \ket{20}  )
\end{array}
\label{decomab}
\en
yields \setlength{\unitlength}{0.15cm}
$\epsilon^{~\!\!\young2000001
{}~\young1000001}_{~{\raise 0.25cm \hbox{$\young2100002$} } } =
- \omega_{1} \omega_{2}$.
We have defined the normalization sign to be the sign of
the highest power of $q$.
Given this it seems natural to take a uniform
sign of the square root when imposing $\vev{11 | 11} =1$ so
that $\omega_{1} = \omega_{2}$. The permutation sign
then indicates the relative sign between the leading state
(as ordered by powers of $q$) and its permutation (with the
inverse power of $q$).

In all cases, once a system of permutation
signs is chosen, we have exactly one sign choice $\omega$ remaining
for each $c \in a \otimes b$.
This corresponds to the residual normalization remaining for Chern-Simons
vertices.

These natural---or structural---permutation signs can be obtained without
having to compute the couplings $\cC_{ab}^{c}(q)$.
The explicit formula eq.~\ref{classepdef} for the
multiplicity-free case is derived in ref.~\mite{newp} and allows rapid
evaluation of $\epsilon^{ab}_{c}$ in these cases.
While this is useful for many purposes,
a spurious dependence on row lengths sometimes appears which would
actually disappear if one knew how to impose the condition
of no multiplicities in general.

Since the fusion ring is a quotient of the tensor ring by a certain
ideal, the terms $\phi_{c}$ remaining in the fusion of
$\phi_{a}$ and  $\phi_{b}$ inherit the permutation sign from the
tensor ring. This is unambiguous if $c$ appears in the tensor
product $a \otimes b$ with multiplicity one, or if the fusion
multiplicity equals the tensor multiplicity. There remains a problem
if the fusion ring multiplicity is less than the tensor ring
multiplicity (but not zero), which we call the problem of
{\em reduced} fusion multiplicities. In such cases,
the known algorithms\cite{wcalc} for computing the fusion
product do not indicate whether symmetric
or anti-symmetric terms are removed from the product; they only
yield the sum of symmetric and anti-symmetric multiplicities
\eq
{N_{ab}}^{c} =   {N_{ab}^{+}}^{c} +   {N_{ab}^{-}}^{c}    ~~.
\label{lastlab}
\en
We have found two ways to obtain
information about ${N_{ab}^{\pm}}^{c}$
(via Chern-Simons theory itself). First,
by expanding a singly-twisted unknot (or its mirror image)
with a spectral decomposition of the crossing,
two equations for the difference
${N_{ab}^{+}}^{c} - {N_{ab}^{-}}^{c}$ are
obtained. While this (allied with the single-multiplicity
formula~\ref{cstoclass}) is effective
in many situations, it seems possible that more than two
separate cases of reduced multiplicities could appear in a fusion product.
Second, the result in eq.~\ref{comlinres}, which embodies the
exact symmetry under cominimal equivalence,
often correlates reduced multiplicities
in one fusion product with non-reduced multiplicities
in cominimal fusion products.

\vspace{0.3cm}
\noindent {\it Cominimal Equivalence and Permutation Signs}
\vspace{0.3cm}

The application of this
Chern-Simons argument (in eq.~\ref{sigtref}-\ref{braidfin}) uses this
latter approach (\ie, that via eq.~\ref{comlinres}) to demonstrate
the braid eigenvalue relations directly in almost all
cases, including that of reduced multiplicities.
In addition, the identities in eqs.~\ref{eppropeq}
and~\ref{sunsign} for $\sun$,
and eq.~\ref{orthsign} for the remaining groups,
follow in the case of no multiplicity directly from the level formula.
One reason that this is possible is that the leading sign
in the case of multiplicities (\ie, that spurious sign given by the
level formula to all copies of a representation in the case of
multiplicity) satisfies the single multiplicity equations without
further constraint so that no spurious
row length dependencies appear in these single multiplicity formulae.

In contrast, in the similar situation for rank-level duality, the analogous
level formula relation contains a complicated dependence on row lengths.
This dependence would disappear if one knew in general how to impose
the condition of no multiplicity.  It is at this point that
the complementary approach via plethysm comes to the rescue
since it is not restricted to the multiplicity-free case.

\vspace{0.3cm}
\noindent {\it The Rank-level Duality of Permutation Signs from Plethysm}
\vspace{0.3cm}

The Littlewood-Richardson product of tableau characters is denoted by
$$
{\rm char}(a) \cdot {\rm char}(b) = \sum {\rm char}(c) ~~.
$$
If the character of any tableau whose first column
length exceeds a given integer $N$ ($N\geq 2$) is set to zero
identically, then this product
is just the tensor product of (purely covariant tableaux of) $\gln$.
If, in addition, the character of any
tableau whose first column length equals $N$ is identified with
the character of the tableau obtained by removing this first
column of length $N$, then this product is just the tensor product
of $\sun$. The product of characters needed for $\spn$ and $\son$
is defined in terms of the Littlewood-Richardson product by
\eq
      {\rm char}(a) \times {\rm char}(b)
   = \sum_{d} {\rm char}( (a/d) \cdot (b/d) )  ~~.
\label{conpro}
\en
Here $(a/d)$ denotes the sum of all tableaux
$a_{d}$ such that $a_{d} \cdot d$ contains
$a$. Define $\Gamma(a_{d}) = r(d)$ (\ie,
the number of boxes in the tableau $d$).  The tensor product
of $\spn$ and $\son$ (tensor) representations is obtained from
eq.~\ref{conpro} by imposing certain character identities
(that are more involved than in the $\sun$ case\cite{king}).
Then, for $c \in a\otimes b$,
the quantity $\Gamma(c)$ gives the number of contractions
of tensor indices
needed to obtain $c$ in the tensor product  $a \otimes b$.

The permutation signs for $\sun$, $\spn$, and
$\sodn$ may all be obtained from $\gln$, via
\eq
\begin{array}{lcl}
 {\epsilon^{ab}_{c}}_{\sst \sun} & = & {\epsilon^{ab}_{c}}_{\sst \gln}  \\
 {\epsilon^{ab}_{c}}_{\sst \spn} & = & e^{i\pi \Gamma_{c}}
{\epsilon^{a_{d}b_{d}}_{c}}_{\sst \gln}  \\
 {\epsilon^{ab}_{c}}_{\sst \sodn} & = &
   {\epsilon^{a_{d}b_{d}}_{c}}_{\sst \gln}
\end{array}
\label{glreduce}
\en
where $a_{d} \in (a/d)$, $b_{d} \in (b/d)$ and $c \in a_{d} \cdot b_{d}$.

Under rank-level duality a representation
$c$ in the decomposition of $a \otimes a$ is often paired by transposition
(\ref{transmap}) with $\tc \in \ta \otimes \ta$ (the case
$\Delta^{aa}_{c} = 0$ in eq.~\ref{fusdual}). For all these
cases we can use a standard result of
the calculus of plethysm to show that\cite{nrs}
\eq
        \epsilon^{aa}_{s} \epsilon^{\ta\ta}_{\ts} = e^{i\pi r(a)}  ~~{\rm
GL}(N)
\label{remarkable}
\en
Here we give the proof of this result.
We will denote the operation of plethysm between two tableaux by a star:
\eq
               a * \mu
\en
denotes the plethysm of the tableau $a $ by
the tableau $\mu$.
\setlength{\unitlength}{0.2cm}
We will be interested in
the case $\mu = \young2000001 $ (which corresponds to the
symmetric product of $a$) and $\mu= \young1100002$ (the anti-symmetric
product).  A classic theorem of
plethysm states that:

\vspace{0.5cm}
\noindent {\it Theorem} ~~(p. 54 of ref.~\mite{plethyatom})

   Given that  $a * \mu = \sum s$,

   ~~if $r(a)$ is even  then $\tilde{a} * \mu = \sum \tilde{s}$, while

   ~~if $r(a)$ is odd~  then $\tilde{a} * \tilde{\mu} = \sum \tilde{s}$.

\vspace{0.5cm}
\noindent {\it Proof of Eq.}~\ref{remarkable}:
Consider the product $\epsilon^{aa}_{s}
\epsilon^{\ta ~\!\ta}_{\ts}$.  First assume that $s$ is in
the symmetric
part of $a\otimes a$ so that $\mu=\young2000001$. Then, if
$r(a)$ is even the above theorem states that $\ts$ is also in
the symmetric part of $\ta \otimes \ta$.
Therefore $\epsilon^{aa}_{s} \epsilon^{\tilde{a}\tilde{a}}_{\ts}
  = 1 = e^{i\pi r(a)}$,
since $r(a)$ is even. If $r(a)$ is odd then the above theorem
states that $\ts$ is in the anti-symmetric part of $\ta\otimes \ta$,
so that $\epsilon^{aa}_{s} \epsilon^{\tilde{a}\tilde{a}}_{\ts}
= -1 = e^{i\pi r(a)}$. Similarly, assuming that $s$ is in the
anti-symmetric part, so that
$\mu=\young1100002$, means that $\ts$ is also in the anti-symmetric
part if $r(a)$ is even, but is in the symmetric part if
$r(a)$ is odd and in either case
$\epsilon^{aa}_{s} \epsilon^{\tilde{a}\tilde{a}}_{\ts} =e^{i\pi r(a)}$.
Therefore, eq.~\ref{remarkable} holds for all $a$.

For $\spnk$ the fusion rule identity
${N_{ab}}^{c} = {N_{\ta \tb}}^{\tc}$, eq.~\ref{glreduce}, and
the $\gln$ result (\ref{remarkable}) yield the rank-level
permutation sign transformation in the intrinsic case $a=b$,
except in the case of reduced multiplicities.
With $\Gamma(s)$ denoting the number of contractions in the
tensor product,
\eq
    (\epsilon^{aa}_{s})_{\spnk} (\epsilon^{\ta \ta}_{\ts})_{\spkn} =
(\epsilon^{a_{d} a_{d} }_{s})_{\gln} ~e^{i\pi \Gamma(s)}
{}~(\epsilon^{\widetilde{a_{d}} \widetilde{a_{d}}}_{\ts})_{\glk}
{}~e^{i\pi \Gamma(\ts) }
\en
where $s \in a_{d} \cdot a_{d}$ and $\ts \in \ta_{d} \cdot \ta_{d}$.
Using the fact that  $\Gamma(s) = \Gamma(\ts)$ and
eq.~\ref{remarkable} we find that
\eq
  (\epsilon^{aa}_{s})_{\spnk} (\epsilon^{\ta \ta}_{\ts})_{\spkn} =
     e^{i\pi r(a_{d})} = e^{i\pi(r(a) - r(d))}
\en
But $r(d) = \Gamma(s)$, and we find that
\eq
      (\epsilon^{aa}_{s})_{\spnk} (\epsilon^{\ta \ta}_{\ts})_{\spkn} =
          e^{i\pi (r(a) - \Gamma(s))}
\label{spfinep}
\en
holds in all cases (given that $s$ and $\ts$ do not appear with reduced
multiplicities).  Note that eq.~\ref{remarkable} implies that if the
tensor and fusion multiplicities are equal then
$$
{N_{aa}^{+}}^{c} - {N_{aa}^{-}}^{c} = e^{i\pi r(a)}
 ({N_{\ta~\!\ta}^{+}}^{\tc} - {N_{\ta~\! \ta}^{-}}^{\tc})
$$
so that ${N_{aa}^{+}}^{c} = {N_{\ta~\!\ta}^{\pm}}^{\tc}$ as
$r(a)$ is even or odd. Then we can  extend the definition
of the transposition map so that eq.~\ref{spfinep} holds both
in the single and full multiplicity cases.

The identical result follows for the
$\sodnk$ fusion rule by similar reasoning
as follows. From the fusion rule identity
${N_{ab}}^{c} = {N_{\ta \tb}}^{\sigma^{\Delta^{ab}_{c}}(\tc~\!)}$ we
know that $\epsilon^{aa}_{s}$ is paired with  $\epsilon^{\ta \ta}_{\ts}$
if $\Delta^{aa}_{s} = 0$ mod $2$ and with $\epsilon^{\ta \ta}_{\sigma(\ts~\!)}$
if $\Delta^{aa}_{s} = 1$ mod $2$.
In the first case
\eq
    (\epsilon^{aa}_{s})_{\asodnk} (\epsilon^{\ta \ta}_{\ts})_{\asodkn} =
(\epsilon^{a_{d} a_{d} }_{s})_{\gln}
{}~(\epsilon^{\widetilde{a_{d}} \widetilde{a_{d}}}_{\ts})_{\glk}
\en
and exactly the same arguments show that
\eq
      (\epsilon^{aa}_{s})_{\asodnk} (\epsilon^{\ta \ta}_{\ts})_{\asodkn} =
          e^{i\pi (r(a) - \Gamma(s))}
\label{sofinep}
\en
holds (apart from the case of reduced multiplicities).
If $\Delta^{aa}_{s} = 2r(a) - r(s) = 1$ mod $2$, then $r(s)$ is
odd. This can only occur if the reduction rule
has been used in producing $r(s)$ via eq.~\ref{conpro}.
Since $\epsilon^{aa}_{s} =  \epsilon^{aa}_{\alpha(s)}$
and $\widetilde{\alpha(s)} = \sigma(\tilde{s})$, where
$\alpha(s)$ is the associate tableau of the representation $s$,
we have
\eq
 (\epsilon^{aa}_{s})_{\asodnk} (\epsilon^{\ta \ta}_{\sigma(\ts~\!)})_{\asodkn}
=
(\epsilon^{a_{d}a_{d}}_{\alpha(s)})_{\gln}
(\epsilon^{\ta_{d} \ta_{d}}_{\widetilde{\alpha(s)}})_{\glk}
= e^{i\pi r(a_{d})}
\en
where $\alpha(s) \in a_{d} \otimes a_{d}$ and
      $\widetilde{\alpha(s)} \in \ta_{d} \otimes \ta_{d}$.
Then $r(a_{d}) = r(a) - \Gamma(s)$, so that
\eq
      (\epsilon^{aa}_{s})_{\asodnk}
(\epsilon^{\ta \ta}_{\sigma^{\Delta^{aa}_{s}}(\ts~\!)})_{\asodkn} =
          e^{i\pi (r(a) - \Gamma(s))}
\en
holds in all cases (except those involving reduced multiplicities).

If $\Delta^{aa}_{s} = 0$ then eq.~\ref{remarkable} is exactly the
$\sun$ result for this case.
If $\Delta^{aa}_{s} \neq 0$ but $\Omega^{aa}(s) = 0$, which
means that the {\em unreduced} tableau $s^u$
has $\row_{1}^{u} \leq K$, then  the tableau for
$\sigma^{\Delta^{aa}_{s}}(\ts)$ is just that for
$\widetilde{ s^{u}}$ so that
\eq
 (\epsilon^{aa}_{s})_{\sunk}
(\epsilon^{\ta \ta}_{\sigma^{\Delta^{aa}_{s}}(\ts~\!)})_{\sukn} =
 (\epsilon^{aa}_{s^{u}})_{\gln}
(\epsilon^{\ta \ta}_{\widetilde{s^{u}}})_{\glk} = e^{i\pi r(a)}
         ~~~({\rm for~}\Omega^{aa}(s) = 0)
\en
holds in all (non-reduced multiplicity) cases.
A consideration of various specific cases suggests that,
in fact,
\eq
        \epsilon^{aa}_{s} \epsilon^{\ta\ta}_{\sigma^{\Delta^{aa}_{s}}(\ts~\!)}
   = e^{i\pi (r(a) + \Omega^{aa} (s))}
\en
for $\sunk$.

Since $\eta^{aa}_{s} = \epsilon^{aa}_{s}$ all these results
hold directly for the Chern-Simons (single and full multiplicity)
permutation signs as well.

While all these tensor ring permutation sign results are
inherited by the fusion ring in the indicated situations, to
obtain results for the reduced multiplicity case for all
groups and the $\Omega^{aa}(s)\neq 0$ cases for $\sun$ we must
appeal to the Chern-Simons knot-based arguments. For
cominimal equivalence the cabling argument yields a complete
confirmation that the single multiplicity case extends
to all cases with multiplicity (whether reduced or not).
For rank-level duality we can only use the figure-eight and
its complex conjugate to get two equations for the
reduced multiplicity signs. While this only
implies that the single multiplicity case
necessarily extends to fusion products with
at most two reduced multiplicity terms, this represents an
infinite class of non-trivial examples, on which to base the general
result.

\end{document}